\documentclass{ws-ijmpa}
\usepackage[super,compress]{cite}
\usepackage{graphicx}
\usepackage{slashed}  
\usepackage{multirow,array,bm,bbm,esint}
\usepackage{makecell}
\usepackage{epigraph}
\usepackage{amsmath}
\usepackage{mathtools}
\usepackage{comment}

\usepackage{makeidx}         
\usepackage{graphicx}        
\usepackage{multicol}        
\usepackage[bottom]{footmisc}

\newcommand{\nn}{\nonumber}
\newcommand{\bec}{\begin{center}}
\newcommand{\eec}{\end{center}}
\newcommand{\beq}{\begin{equation}}
\newcommand{\eeq}{\end{equation}}
\newcommand{\bea}{\begin{eqnarray}}
\newcommand{\eea}{\end{eqnarray}}

\newcommand{\hf}{\frac{1}{2}}

\newcommand{\cD}{{\cal D}}

\newcommand{\cN}{{\cal N}}
\newcommand{\cO}{{\cal O}}

\newcommand{\Tr}{{\rm Tr}}

\newcommand{\psib}{{\overline{\psi}}}

\newcommand{\B}{{\mathcal{B}}}
\newcommand{\Q}{Q}
\newcommand{\mcS}{\mathcal{S}}

\newcommand{\Qb}{\overline{Q}}

\usepackage{xcolor}

\begin{document}

\markboth{Anosh Joseph and Arpith Kumar}
{Complex Langevin Simulations of Supersymmetric Theories}

%
\catchline{}{}{}{}{}
%

\title{Complex Langevin Simulations of Supersymmetric Theories}

\author{ANOSH JOSEPH}

\address{National Institute for Theoretical and Computational Sciences, \\
School of Physics, and Mandelstam Institute for Theoretical Physics, \\
University of the Witwatersrand, Johannesburg, Wits 2050, South Africa \\
anosh.joseph@wits.ac.za}

\author{ARPITH KUMAR}

\address{Key Laboratory of Quark and Lepton Physics (MOE) and Institute of Particle Physics, \\
Central China Normal University, Wuhan 430079, China \\
arpithk@ccnu.edu.cn}

\maketitle

\begin{abstract}

This review explores the Complex Langevin Method (CLM), a stochastic quantization technique designed to address the sign problem in quantum field theories with complex actions. 
Beginning with foundational principles, the review examines the applications of CLM across a range of models, including zero- and two-dimensional systems, supersymmetric quantum mechanics, and the IKKT matrix model, a candidate for non-perturbative string theory. 
Key advancements, such as stabilization techniques and mass deformations, are highlighted as solutions to challenges like numerical instability and singular drift terms. 
The review emphasizes the capacity of CLM to simulate complex systems and reveal non-perturbative phenomena, positioning it as a powerful tool for exploring quantum field theory and string theory. 
Future directions, including higher-dimensional applications and benchmarking against quantum simulations, underscore the potential of CLM to advance both theoretical understanding and computational methodologies.

\end{abstract}

\ccode{PACS numbers: 02.70.Ss; 11.15.Tk; 12.60.Jv; 11.25.Tq; 11.15.Ha; 11.30.Pb}

\tableofcontents

\section{Introduction}

A fundamental tool for investigating many non-perturbative aspects of quantum field theory (QFT) is the Monte Carlo simulation of a lattice-regularized version of the field theory path integral.
The core idea behind path integral Monte Carlo is to generate field configurations with probability weights derived from the exponential of the negative of the action (in Euclidean spacetime). 
The path integral is then computed by statistically averaging this ensemble of importance-sampled field configurations. 
However, when the action is complex, such as in finite-temperature QCD with baryon/quark chemical potential, QCD with a theta term, Chern-Simons gauge theories, or chiral gauge theories, the fermion determinant can also become complex. 
This introduces the well-known {\it sign problem} or {\it complex phase problem}, which undermines the reliability of conventional Monte Carlo methods.

In the context of string theory, for instance, the IKKT matrix model -- viewed as a promising non-perturbative formulation of superstring theory -- also suffers from a complex fermion operator. 
These complex entities within the theory pose significant challenges for simulation algorithms based on path integral Monte Carlo.

Several methods have been developed to address complex actions in quantum field theories, including analytical continuation, Taylor series expansion, and more recently, approaches that involve the complexification of integration variables, such as the Lefschetz thimble method and the complex Langevin method (CLM).

The CLM is a natural generalization of the real Langevin method, which extends stochastic quantization techniques to scenarios where the action is complex. 
It overcomes the sign problem by defining a stochastic process with complexified field variables, allowing for the calculation of observables from this process. 
In recent years, the CLM has been applied with notable success in a range of models.

This review presents a detailed exploration of stochastic quantization and the CLM, with a particular focus on its application to supersymmetric field theories. 
The central idea behind stochastic quantization is that the expectation values of observables emerge as equilibrium values of a stochastic process. 
In the context of Langevin dynamics, this is achieved by evolving the system in a fictitious time direction, referred to as {\it Langevin time}, subject to stochastic noise. 
For complex actions, the fields are naturally complexified during the Langevin evolution, which presents new challenges in terms of proving convergence to the correct equilibrium distribution.

This review is structured across nine sections. 
Section \ref{chap:Lattice_regularized_path_integrals} introduces the reader to the basics of quantum field theory (QFT), with a particular emphasis on the path integral formalism and Monte Carlo methods. 
It examines various strategies to overcome the sign problem, focusing on complex Langevin dynamics and providing a broader context through non-perturbative methods. 
Section \ref{sec:Sign_problem_in_complex_actions_and_methods_to_cure_it} delves into the sign problem, a critical challenge in simulating quantum field theories with complex actions. 
In traditional path integral Monte Carlo methods, a real Euclidean action allows configurations to be weighted probabilistically using a positive Boltzmann factor. 
However, in systems with complex actions -- such as finite-temperature QCD with chemical potentials, topological terms, or theories with fermions -- the Boltzmann weight becomes complex, precluding its interpretation as a probability measure. 
This leads to oscillatory integrals, exponential noise, and unreliable simulations, particularly as the system size increases. 
This section sets the stage for exploring specific applications and advancements of the complex Langevin method in addressing the sign problem. 
In Section \ref{sec:A_brief_review_of_complex_Langevin_method}, we review the complex Langevin method. 
Originating in the 1980s from Parisi and Wu's work, the method extends Langevin dynamics -- traditionally used for real actions -- to systems with complexified fields. 
CLM evolves field configurations in a fictitious Langevin time under stochastic noise, using a complexified version of the Langevin equation to simulate equilibrium distributions.
The method involves treating fields as complex variables and solving their dynamics using stochastic differential equations. 
Observables are computed as equilibrium averages over these complexified configurations, with the evolution governed by a Fokker-Planck equation adapted for complex distributions. 
This allows CLM to handle the oscillatory nature of integrals in systems with complex actions, bypassing the sign problem encountered in conventional Monte Carlo methods. 
CLM’s ability to manage complex actions and severe sign problems makes it a promising tool for simulating strongly coupled quantum systems and supersymmetric models, as further explored in subsequent sections. 
Section \ref{sec:Complex_Langevin_simulations_of_zero-dimensional_models} focuses on applying the CLM to zero-dimensional quantum field theory models, serving as a simpler setting to study fundamental issues such as spontaneous supersymmetry (SUSY) breaking and the reliability of CLM. 
These models allow detailed exploration of the method's dynamics and provide a testing ground for addressing the sign problem in more complex systems. 
This section highlights the effectiveness of CLM in tackling complex actions and its robustness in simpler settings, establishing a foundation for its application to higher-dimensional and more intricate models in subsequent sections. 
Section \ref{sec:Complex_Langevin_method_for_SUSY_quantum_mechanics} explores the application of the CLM to supersymmetric quantum mechanics (SUSY QM), focusing on its ability to address the sign problem in systems with complex actions and to investigate SUSY-breaking dynamics. 
The section builds on foundational studies in zero-dimensional models and extends CLM's use to one-dimensional supersymmetric systems. 
In Section \ref{sec:Complex_Langevin_simulations_of_two-dimensional_models} we examine the application of the CLM to two-dimensional quantum field theory models, focusing on scalar field theories and supersymmetric systems. 
This expansion from zero- and one-dimensional settings allows the study of more complex interactions and dynamics while addressing the notorious sign problem. 
It establishes the efficacy of CLM in simulating two-dimensional quantum systems, showcasing its ability to handle complex interactions and overcome computational challenges in these higher-dimensional settings. 
Section \ref{sec:Complex_Langevin_analysis_of_the_IKKT_model} focuses on the application of the CLM to the IKKT (Ishibashi-Kawai-Kitazawa-Tsuchiya) matrix model, a prominent candidate for a non-perturbative formulation of superstring theory. 
This section explores how CLM addresses the complex fermion determinant and other computational challenges inherent to the IKKT model. 
This section also demonstrates the viability of CLM in tackling the computationally demanding IKKT model, offering a promising tool for exploring the non-perturbative regime of string theory and quantum gravity. 
In Sec. \ref{sec:Conclusions_and_future_directions} we end the review by providing conclusions and future directions.

\section{Lattice regularized path integrals}
\label{chap:Lattice_regularized_path_integrals}

Quantum field theory (QFT), a theoretical framework that merges the principles of special relativity and quantum mechanics, offers profound insights into the fundamental forces of nature. 
Specifically, the electromagnetic, weak, and strong interactions between elementary particles are described through quantum electrodynamics, quantum flavourdynamics, and quantum chromodynamics (QCD), respectively. 
The Standard Model of particle physics unifies these three interactions, excluding gravity, and has been remarkably successful in explaining a wide range of experimentally discovered particles.

While perturbative methods in QFT have produced highly accurate results for weakly interacting systems -- such as the anomalous magnetic moment of the electron, first computed by Schwinger in 1948 -- these techniques have significant limitations. 
Perturbation theory, though powerful, is ultimately an asymptotic expansion with a divergent sum at higher orders. 
More critically, it fails entirely in strongly coupled theories, like QCD at low energies, where non-Abelian gauge symmetry and asymptotic freedom lead to the confinement of quarks and gluons within hadrons. 
Confinement, along with phenomena such as chiral symmetry breaking, cannot be captured by perturbative approaches, necessitating a non-perturbative framework. 
Lattice regularized path integrals provide such a framework, enabling the study of these complex, non-perturbative aspects of quantum field theories.

\subsection{Path integral in Euclidean spacetime}

Feynman path integrals are functional integrals over the space of all possible quantum mechanical trajectories that satisfy specific boundary conditions. 
One of the most effective ways to extract non-perturbative physics is through the exact evaluation of these path integrals. 
To illustrate, let us consider a real scalar field, $\phi(\vec{x}, t)$, in four-dimensional Minkowski spacetime. 
The action, $S \equiv S[\phi(\vec{x}, t)]$, of this theory is the spacetime integral of the Lagrangian density, $\mathcal{L} \equiv \mathcal{L}[\phi, \partial_{\mu}\phi]$, given by
\bea
S &=& \int_0^T dt \int_\Omega d^3x~ \mathcal{L} \nn \\
&=& \int_0^T dt \int_\Omega d^3x~ \left( \hf \partial_\mu \phi \partial^\mu \phi - V(\phi) \right),
\eea
where $T$ represents the time interval, $\Omega$ is the finite spatial volume, and the potential term is $V(\phi)$.

In the path integral formulation, propagators -- such as position space Green's functions -- are central to deriving the physical properties of a system. 
The propagator for transitioning from a field configuration $\phi_1 \equiv \phi_1(\vec{x})$ to $\phi_2 \equiv \phi_2(\vec{x})$ over a time interval $T$, denoted as $G(\phi_2, \phi_1; T)$, is given by the probability amplitude
\bea
G(\phi_2, \phi_1; T) &=& \langle \phi_2(\vec{x}) | e^{-iHt} | \phi_1(\vec{x}) \rangle \nn \\
&\equiv& \underbrace{\int_{\phi(\vec{x}, 0) \equiv \phi_1(\vec{x})}^{\phi(\vec{x}, T) \equiv \phi_2(\vec{x})} \mathcal{D} \phi}_{_{ \substack{\text{over the space of} \\ \text{continuous trajectories}}}}~ e^{iS[\phi(\vec{x},t)]}.
\eea
Here, $H$ is the Hamiltonian, and the integrand has a highly oscillatory nature, expressed as $\exp(iS[\phi])$. 
The oscillatory behavior presents challenges for defining a sensible measure on the set of paths. 
To manage this, path integrals are often treated in \textit{Euclidean time} via analytical continuation.

To understand this, we can draw an analogy with statistical mechanics, considering a canonical system at inverse temperature $\beta$, with partition function
\beq
Z(\beta) = {\rm Tr} \left(  e^{-\beta H} \right) = \sum_{n}e^{-\beta E_n}. 
\eeq
Using the basis independence of the trace operation, the partition function in position space is written as
\beq
Z(\beta) = \int  \mathcal{D} \phi (\vec{x})~ \langle \phi(\vec{x}) | e^{-\beta H} | \phi(\vec{x}) \rangle. 
\eeq
By performing a {\it Wick rotation} to Euclidean time, $t \to - i \tau$, we express
\bea
\langle \phi(\vec{x}) | e^{-\beta H} | \phi(\vec{x}) \rangle &=& \int_{\phi(\vec{x}, 0) \equiv \phi(\vec{x})}^{\phi(\vec{x}, \beta) \equiv \phi(\vec{x})} \mathcal{D} \phi~ e^{-S_{E}[\phi(\vec{x}, \tau)]} \nn \\
&=& G(\phi(\vec{x}), \phi(\vec{x}); - i \beta),
\eea
where $S_E [\phi(\vec{x}, \tau)]$ is the Euclidean action, and the Euclidean Lagrangian density $\mathcal{L}_E$ is given by
\bea
S_E[\phi(\vec{x}, \tau)] &=& \int_0^\beta d\tau \int_\Omega d^3x~ \mathcal{L}_E(\phi, \partial_\mu \phi) \nn \\
&=& \int_0^\beta d\tau \int_\Omega d^3x~ \left( \hf \partial_\mu \phi \partial^\mu \phi + V(\phi) \right).
\eea
This leads to a partition function in Euclidean space
\beq
Z(\beta) = \oint_{PBC} \mathcal{D} \phi ~ e^{-S_E[\phi(\vec{x}, \tau)]},
\eeq
with periodic boundary conditions (PBC), $\phi(\vec{x}, 0) = \phi(\vec{x}, \beta)$. 
The observables in Euclidean QFTs can now be computed as
\beq
\langle \mathcal{O}(\phi) \rangle = \frac{1}{Z(\beta)} \int \mathcal{D} \phi~ \mathcal{O}(\phi)~ e^{-S_E[\phi]},
\eeq 
and analytically continued back to real-time dynamics via inverse Wick rotation. 
While this formulation improves mathematical behavior, the path integral remains infinite-dimensional, making exact computation impractical. 
However, lattice regularization and numerical simulations provide viable approaches to obtaining physically meaningful results in such cases.

\subsection{Lattice discretization}

Lattice regularization is a powerful method for quantizing field theories and addressing divergences by discretizing spacetime into a lattice and computing Euclidean path integrals. 
This approach replaces continuous field variables with those defined on discrete lattice points. 
Interestingly, the use of lattices in physics predates the formal development of field variables, having been employed in condensed matter systems to model phenomena like electron behavior in crystalline structures. 
In 1974, Wilson made a seminal contribution by applying lattice regularization to gauge field theories, particularly in the context of QCD. 
His work not only introduced lattice gauge theory but also led to the discovery of quark confinement at strong coupling within QCD, a breakthrough that significantly enhanced our understanding of the strong nuclear force \cite{Wilson:1974sk}.

The basic approach to discretizing a four-dimensional spacetime on a lattice involves dividing spacetime into discrete points, creating a grid-like structure:
\bea
\int_0^\beta d\tau \int_\Omega d^3x &\rightarrow& a^4 \sum_n, \\
\phi(\vec{x}, t) &\rightarrow& \phi_n, \\
\partial_\mu \phi(\vec{x}, t) &\rightarrow& \sum_\mu \frac{\left(\phi_{n + e_\mu} - \phi_n \right)}{a}.
\eea
In these equations, $n$ denotes the four-dimensional lattice points, and $e_\mu$ represents the unit vectors in each direction. 
The field variable at a lattice point is $\phi_n$, and $a$ is the lattice spacing. 
This discretization replaces the continuum integrals over time and space with sums over lattice sites, where $\beta = N_\tau a$ and $\Omega = (N_x a)^3$. 
Here, $\beta$ and $\Omega$ denote the temporal extent and spatial volume of the continuum theory, respectively, and $N_\tau$ and $N_x$ represent the number of lattice sites in the temporal and spatial directions. 
It is also possible to use different lattice spacings for different directions.

The original infinite-dimensional continuous path integral can be represented by a finite-dimensional lattice regularized form in the continuum limit. 
Specifically,
\beq
Z = \int \mathcal{D} \phi~ e^{- S[\phi]} \equiv \lim_{\substack{a \to 0 \\ N_\tau \to \infty \\ N_x \to \infty}} \int \left( \prod_n d\phi_n \right)~e^{- S_{lat}[\phi_n]},
\eeq
where $S_{lat}[\phi_n]$ denotes the lattice action. 
Lattice regularization is a crucial method in modern theoretical physics, enabling the computation of physical observables through numerical techniques. 
By discretizing spacetime coordinates, the lattice framework facilitates practical simulations and calculations, making it an invaluable tool for exploring and analyzing quantum field theories.

Traditionally, path integrals in Euclidean space are evaluated using the Monte Carlo method, specifically through importance sampling \cite{Metropolis:1953am}. 
This approach transforms the calculation of a Euclidean QFT into a problem of simulating a statistical system. 
For a real-valued action $S[\phi(\vec{x}, t)]$, a set of field configurations $\{ \phi_r \}$ are generated by treating the normalized Boltzmann factor $ \exp(-S_{lat}[\phi_r]) / Z $ as a probability weight. 
The expectation values of observables are then computed as averages over a large number of sampled configurations $(N_{\{\phi_r\}})$. 
Mathematically, this process is represented as:
\bea
{\langle \mathcal{O}(\phi) \rangle}_E &=& \frac{1}{Z} \int \mathcal{D} \phi~ \mathcal{O}(\phi)~ e^{-S[\phi]} \\
&\equiv& \lim_{\substack{a \to 0 \\ N_\tau \to \infty \\ N_x \to \infty}} \int \left( \prod_r d\phi_r \right)~\mathcal{O}(\phi_r) e^{-S_{lat}[\phi_r]} \\
&\approx& \frac{1}{N_{\{\phi_r \}}} \sum_{\{\phi_r \}} \mathcal{O}(\phi_r).
\eea
However, it is crucial to assess whether this method remains valid when the Euclidean action is complex.

\section{Sign problem in complex actions and methods to cure it}
\label{sec:Sign_problem_in_complex_actions_and_methods_to_cure_it}

\subsection{Sign problem and complex actions}

Path integral Monte Carlo methods are effective for handling real-valued and non-negative actions, but many physically significant systems present complex-valued actions. 
Indeed, such systems are so prevalent that they are arguably more common than exceptions. 
Notable examples of systems with complex actions include:

\begin{itemize}
\item {\it Field theories in Minkowski space}: For field theories formulated in Minkowski space, addressing the challenges posed by complex actions is crucial. 
Although a Wick rotation typically converts the Minkowski action into a manageable form in Euclidean space, there is value in developing techniques that tackle the complexities inherent in the Minkowski signature itself \cite{deAguiar:2010ue, Anzaki:2014hba}.

\item {\it Systems with chemical potential}: Low-energy studies of strongly interacting systems at finite chemical potential pose significant challenges. Investigations at non-zero chemical potential are crucial to comprehend the behavor of QCD transition and locating the critical endpoint \footnote{For physical quark masses, the critical endpoint marks a second order phase transition boundary on the QCD phase diagram between smooth crossover pseudo-phase transitions at low chemical potentials and true first order phase transitions at high chemical potentials.}. 
In general, for SU$(N)$ gauge theories with $N \ge 3$, in the presence of a non-zero chemical potential, the Dirac operator loses key symmetries (such as $\gamma^5$-hermiticity) leading to complex / non-positive fermion determinant, complicating the effective action governing these systems \cite{Barbour:1986jf,Fodor:2001au,deForcrand:2009zkb,Aarts:2010gr, Schmalzbauer:2016pbg, Nagata:2017pgc}.

\item {\it Theories with external charges}: Gauge theories with external static charges treat these charges as fixed color sources rather than dynamical quarks. 
They do not appear in the fermion determinant but instead as sources in the action, complicating the Euclidean path integral. 
For instance, computing the string tension and the charge separation with static-quark potential is vital for understanding quark confinement, a fundamental aspect of strong interactions \cite{Ambjorn:1986fz, Blum:1995cb, Greensite:2003bk}.

\item {\it Systems with fermions}: Simulating fermions in quantum field theories is challenging due to the lack of a real-number representation in the conventional path integral formalism \cite{Kogut:1974ag, Nielsen:1981hk}. 
Methods often rely on the fermion determinant or Pfaffian, which introduces complex actions and negative probabilities. 
Addressing these challenges requires advanced techniques and contributes to a deeper understanding of fundamental particle physics and applications in condensed matter physics \cite{Li:2018wnz, Wang:2015vha, Chen:2003vy}.

\item {\it Effective actions in the presence of topological $\theta$-term or Chern-Simons theories}: Topological terms like the $\theta$-term, Chern-Simons term, and Wess-Zumino-Witten (WZW) term impact low-energy theories significantly. 
For instance, Chern-Simons gauge theory generalizes compact Lie groups to complex Lie groups, and these terms lead to complex phases that complicate the Boltzmann factor \cite{Witten:1989ip, Gukov:2003na}. 
The $\mathcal{CP}$ problem in theories with strongly interacting matter arises from the introduction of a topological charge through $\mathcal{CP}$-violating angle $\theta$. Although expected to be extremely small ($\sim 10^{-10}$), non-zero $\theta$ offers a rich structure and introduces additional complex phases in the theory \cite{Veneziano:1979ec, Vicari:2008jw, Bongiovanni:2014rna, Hirasawa:2020bnl, Matsumoto:2021zjf}.

\item {\it Non-equilibrium physics of quantum many-body systems}: Investigating the non-equilibrium dynamics of quantum many-body systems presents a significant research challenge. 
Quantum Monte Carlo (QMC) methods, which are highly effective for equilibrium studies, face severe sign problems when extended to non-equilibrium scenarios due to oscillatory integrals in real-time formulations \cite{Troyer:2004ge, Polkovnikov:2010yn}. 
This complexity escalates exponentially with simulation time, making such studies computationally demanding but essential for understanding quantum phase transitions and the behavior of strongly correlated materials under external perturbations \cite{Eisert:2014jea, Cohen:2015}.

\item {\it Condensed matter systems of strongly correlated electrons}: In condensed matter physics, several systems present complex actions that challenge both our theoretical understanding and computational techniques. 
A prominent example is the repulsive Hubbard model on bipartite or triangular lattices, where doping introduces a significant sign problem \cite{Loh:1990zz, Blanc:2015, Yamamoto:2015ura}. 
Other intriguing systems, such as spin-polarized electron gases, frustrated magnetic configurations, and the Shastry-Sutherland antiferromagnetic spin model, also face similar complexities. 
These systems reveal fascinating physical phenomena, prompting researchers to develop innovative computational methods and theoretical frameworks to decode their complexities. 
Such efforts hold the promise of uncovering exotic phases of matter and advancing the fields of materials science and technology \cite{Huffman:2013mla, Wessel_2018, D_Emidio_2020, Pan:2022fgf}.
\end{itemize}

Field theories with complex actions present significant challenges for non-perturbative analysis due to the non-positive or generally complex nature of the Boltzmann factor. 
While the partition function remains well-defined, the Boltzmann factor cannot be interpreted as a probability weight, leading to the notorious \textit{sign problem} in path integral Monte Carlo simulations. 
This issue is known by various names across different areas of physics, including the \textit{numerical sign problem}, \textit{complex phase problem}, \textit{complex action problem}, and \textit{negative sign problem}. 
Recognized as {\it NP-hard}, the sign problem is one of the most significant and infamous challenges in modern computational physics, severely hindering the accurate analysis of both equilibrium and non-equilibrium behaviors in diverse, cutting-edge physical systems.

The sign problem poses a major challenge for applying importance sampling, as it complicates or even precludes the selection of a positive-definite probability distribution. 
One of the most direct approaches to address this issue is the re-weighting procedure \cite{Ferrenberg:1988yz}, which integrates the complex phase of the Boltzmann weight into the observable calculations. 
Specifically, we can express the complex Boltzmann weight $e^{- S[\phi]}$ as $| e^{-S[\phi]}| e^{i \omega}$, where $e^{i \omega}$ represents the complex phase. 
Sampling is then performed using the magnitude of the weight, or phase-quenched weight \( |e^{-S[\phi]}| \), as the probability measure.

The expectation values of observables are given by
\bea
{\langle \mathcal{O}(\phi) \rangle} = \frac{1}{Z} \int \mathcal{D} \phi~ \mathcal{O}(\phi)~ e^{-S[\phi]} = \frac{\int \mathcal{D} \phi~ \mathcal{O}(\phi) ~|e^{-S[\phi]}| e^{i \omega}} {\int \mathcal{D} \phi ~|e^{-S[\phi]}| e^{i \omega}}. 
\eea
We can multiply both the numerator and the denominator by the phase-quenched partition function $Z_{\rm pq} = \int \mathcal{D} \phi~|e^{-S[\phi]}|$.
Then,
\bea
{\langle \mathcal{O}(\phi) \rangle} &=& \frac{ \int \mathcal{D} \phi~ \mathcal{O}(\phi) ~|e^{-S[\phi]}| e^{i \omega}}{\int \mathcal{D} \phi~|e^{-S[\phi]}|} \times \frac{\int \mathcal{D} \phi~ |e^{-S[\phi]}|}{\int \mathcal{D} \phi~|e^{-S[\phi]}| e^{i \omega}} \\
&=& \frac{{\langle \mathcal{O}(\phi) e^{i \omega} \rangle}_{Z_{\rm pq}}}{{\langle e^{i \omega} \rangle}_{Z_{\rm pq}}},
\eea
where ${\langle \cdot \rangle}_{Z_{\rm pq}}$ denotes the expectation values with respect to the phase-quenched weight $|e^{-S[\phi]}|$. 
This re-weighting approach effectively transforms the problem of computing observables with a complex Boltzmann weight into one where the weight is real and positive.

Although the procedure appears elegant, its practical implementation is hindered by the highly oscillatory nature of $e^{i \omega}$. 
Both the numerator and denominator of the re-weighted expression become extremely small and decay exponentially as the physical size of the spacetime lattice increases. 
The severity of the sign problem is quantified by the expectation value of the complex phase:
\bea
\langle e^{i \omega} \rangle_{Z_{\rm pq}} = \frac{Z}{Z_{\rm pq}} = e^{- \Omega \Delta f},
\eea
where $\Omega$ is the volume of the spacetime lattice, and $\Delta f = f - f_{\rm pq}$ represents the difference in free energy densities between the original and phase-quenched theories. 
Here, $Z_{\rm pq}$ denotes a bosonic ensemble with sums over non-negative real numbers, while $Z$ is a path integral that incorporates the phase. 
Consequently, $\Delta f$, the difference in free energy density between the bosonic and fermionic systems, is necessarily positive. 
For further details, see Refs. \cite{Chandrasekharan:1999cm, Chandrasekharan:1999ys, Alford:2001ug, Berger:2019odf}.

The small value of $\langle e^{i \omega} \rangle$ is evident since $Z \leq Z_{\rm pq}$ and it vanishes as $\Omega \to \infty$. 
Furthermore, the statistical uncertainty $\sigma$ in Monte Carlo simulations, which decreases as ${N_{\phi}}^{-1/2}$ with $N_\phi$ samples, is overshadowed by the exponentially decaying nature of $\langle e^{i \omega} \rangle_{Z_{\rm pq}}$:
\beq
\frac{\sigma}{{\langle e^{i \omega} \rangle}_{Z_{\rm pq}}} = \frac{e^{\Omega \Delta f}}{\sqrt{N_{\phi}}}.
\eeq

This equation highlights the formidable challenge of addressing the sign problem using straightforward methods like re-weighting. 
The sign problem represents an exponential computational barrier, manifesting as memory constraints for non-stochastic methods and as a signal-to-noise problem in statistical methods.

\subsection{Methods to tame the sign problem}

Various strategies have been suggested for tackling systems with severe sign problems. 
While some of these approaches are recent and still evolving, each presents its own set of advantages and challenges. 
Nonetheless, they hold considerable promise for addressing the sign problem effectively. 
Below are a few notable and successful methods.

\begin{itemize}
\item {\it Meron cluster and fermion bag approaches}: The Meron cluster and fermion bag methods are notable advancements in addressing the fermion sign problem in quantum Monte Carlo simulations. 
The Meron cluster approach, used in various systems within the Hubbard model and for relativistic fermions, involves decomposing fermion world lines into clusters, each contributing independently to the fermion permutation sign. 
A {\it meron} is a cluster whose flip alters the sign. 
By expressing the partition function as a gas of clusters in the zero-meron sector, this method mitigates the sign problem by shifting contributions from $\pm$1 to non-negative values of 0 and 1 \cite{Chandrasekharan:1999cm}.

The fermion bag approach builds on the Meron cluster concept by segmenting fermion degrees of freedom into smaller, entangled regions called fermion bags, which are treated as largely independent. 
Each bag is analyzed by summing over all quantum fluctuations within it, ensuring that the bag weights remain positive for efficient Monte Carlo simulation. 
Applied to continuous-time Hamiltonian formulations, this method has been particularly useful for describing interacting two-dimensional massless Hamiltonian staggered fermions. 
By dividing the system into fermion bags, the approach simplifies updates and observables calculations in Monte Carlo simulations \cite{Chandrasekharan:2009wc}. 

Both methods not only tackle the sign problem but also offer innovative insights into fermionic systems, illustrating that the partition function of fermionic systems can be framed in terms of classical statistical mechanics models of clusters or fermion bags \cite{Chandrasekharan:2010iy, Huffman:2017swn}.

\item {\it Majorana fermions algorithm}: The Majorana Quantum Monte Carlo (MQMC) algorithm represents a significant breakthrough in computational physics, particularly in tackling the fermion sign problem in interacting fermion models \cite{Li:2014tla}. 
MQMC is especially effective for simulating spin-less fermion models on bipartite lattices at half-filling. 
The technique involves expressing complex fermion operators $c_i$ using Majorana fermions $\gamma_i^1$ and $\gamma_i^2$, with the relation $c_i = \frac{1}{2}(\gamma_i^1 + i\gamma_i^2)$ \cite{Liu:2015mxb}. 
This Majorana representation enables a novel formulation of the Hamiltonian, incorporating hopping integrals $t_{ij}$ and density interactions $V_{ij}$, which facilitates Hubbard-Stratonovich transformations within Majorana hopping channels \cite{Li:2017sbk}.

A key advantage of MQMC is its capacity to make the Boltzmann weight positive-definite, effectively addressing the fermion sign problem. 
This is achieved through a time-reversal transformation that standardizes the Hamiltonians of different Majorana species \cite{Li:2018wnz}. 
Ensuring the positive definiteness of the Boltzmann weight is crucial for the stability and precision of QMC simulations. 
MQMC is also adaptable to finite-temperature simulations and projector algorithms, which are often more effective for studying ground-state properties. 
Projector MQMC computes expectation values of operators in the ground state, thereby enhancing the versatility and effectiveness of the method \cite{Wang:2015vha}.

Moreover, the development of MQMC is guided by a broader principle that leverages the inherent Lie group and Lie algebra structures in fermionic QMC simulations. 
This principle, which includes time-reversal symmetry and the application of the split orthogonal group \( O(n, n) \), expands the range of fermionic models that can be simulated without encountering the sign problem. 
This advancement paves the way for exploring physical phenomena in systems that were previously intractable due to computational constraints \cite{Wei:2016sgb, Chen:2003vy}.

\item {\it Density of states}: The density of states (DOS) method plays a pivotal role in computational physics, particularly in Monte Carlo simulations and lattice QCD studies. 
This approach involves calculating the density of states \(\rho(E)\), which indicates the number of states available at each energy level and is essential for determining the system's thermodynamic properties \cite{Wang:2000fzi, Bazavov:2012ex, Langfeld:2012ah}. 
In Monte Carlo simulations, the DOS method entails performing random walks within specific energy ranges to estimate \(\rho(E)\) efficiently, which is particularly useful for large systems. 
This technique provides direct insights into free energy and entropy across different temperatures and is effective for investigating both first- and second-order phase transitions \cite{Ejiri:2007ga, Fodor:2007vv}.

In the context of lattice QCD, the DOS method helps address the complex action problem at non-zero chemical potentials by integrating thermodynamic observables over a fixed parameter, such as the plaquette expectation value. 
Accurate calculations of physical observables depend on the density of states derived from the constrained partition function. 
Nonetheless, this method is computationally intensive and often plagued by significant corrections due to finite lattice spacing and size constraints \cite{Springer:2021liy}.

\item {\it Taylor expansion method}: The Taylor expansion method is a widely accepted framework in lattice QCD for studying fermions at finite temperature and non-zero baryon density. 
Expanding thermodynamic observables in a Taylor series around zero chemical potential, $\mu = 0$, circumvents the sign problem associated with real chemical potentials. 
This approach involves calculating derivatives of the pressure to $\mu$ at $\mu = 0$, thereby constructing an approximate expression valid for small $\mu/T$ ratios \cite{Gottlieb:1988cq, Gavai:2001fr, Allton:2002zi}. 
Although it systematically includes higher-order terms to enhance accuracy, the reliability of the method diminishes at larger chemical potentials due to series truncation and convergence issues. 
Moreover, calculating higher-order coefficients becomes increasingly complex, limiting its applicability near the anticipated critical endpoint in the QCD phase diagram. 
Despite these limitations, the Taylor expansion method remains one of the most effective tools for mapping the QCD phase structure at low to moderate baryon densities and determining thermodynamic properties through quark susceptibilities \cite{deForcrand:2002hgr, Fodor:2002km, Karsch:2003jg, HotQCD:2012fhj}.

\item {\it Imaginary chemical potential method}: The imaginary chemical potential method is a key technique in lattice QCD for exploring fermions at finite temperature and baryon density. 
This method addresses the sign problem encountered with real chemical potentials by simulating the partition function at an imaginary chemical potential, $\mu = i\nu$, where the fermion determinant remains real and positive under certain conditions, thus allowing for standard Monte Carlo simulations \cite{Alford:1998sd, Fodor:2001au, DElia:2002tig, deForcrand:2003vyj}. 
The grand canonical partition function $Z(\mu)$ at imaginary $\mu$ is linked to the canonical partition function $Z_N$ at fixed baryon number $N$ through a Fourier transform, and the results are subsequently analytically continued to real $\mu$ to obtain physical observables. 
However, even at imaginary $\mu$, the theory has a rich phase structure, exhibiting Roberge-Weiss symmetry \cite{Roberge:1986mm}. 
Notably, introducing the chemical potential as a phase restricts its magnitude and renders the Fourier integrands highly oscillatory for large baryon numbers, thereby resulting in significant numerical challenges. 
Nonetheless, the imaginary chemical potential method has been effectively applied in QCD, providing valuable insights into phase transitions \cite{DElia:2004ani, DElia:2007bkz, Karbstein:2006er, Philipsen:2012nu, Cea:2012ev, Bonati:2014kpa, Wu:2018oed, Guenther:2017hnx, Borsanyi:2020fev}.

\item {\it Complexification of space}:
\begin{itemize}
\item {\it Path optimization method}: The Path Optimization Method (POM) represents a significant advancement in addressing the sign problem in Monte Carlo simulations for complex actions, commonly encountered in quantum many-body theories and lattice QCD at finite densities \cite{Ohnishi:2017zxh, Mori:2017pne}. 
POM operates by variationally optimizing the integral path in the complex plane to improve the average phase factor. 
It begins with a trial function to parameterize the integral path and uses a cost function that measures the severity of the sign problem. 
This approach is distinct from other methods, such as the Lefschetz-thimble method or the complex Langevin method, because it does not require pre-determining the fixed points of the action \cite{Ohnishi:2018jjw, Kashiwa:2018vxr, Bursa:2018ykf}. 

POM has proven effective in a one-variable toy model, where it outperforms the complex Langevin method in scenarios with severe phase oscillations and frequent cancellations. 
In this model, which involves a one-dimensional integral, the phase oscillations under certain conditions can cause significant cancellations. 
POM’s optimized path closely approximates the Lefschetz thimble at stationary points of the action, allowing for accurate calculations of observables in regions plagued by substantial sign problems. 
Additionally, POM's adaptability to complex systems, including its use of neural networks for optimization, indicates its broad potential for various complex actions \cite{Ohnishi:2017zxh, Mori:2017pne, Bursa:2018ykf, Ohnishi:2018jjw, Kashiwa:2018vxr, Mori:2019rql, Mori:2019tux, Kashiwa:2019lkv, Detmold:2020ncp}.

\item {\it Complex Langevin method}: Originally introduced in the early 1980s, this method extends conventional Langevin equations to the complex plane \cite{Klauder:1983sp, Parisi:1983mgm}. 
It involves complex stochastic quantization, a technique for quantizing systems with complex actions \cite{Klauder:1983nn, Klauder:1983zm}. 
This approach addresses the sign problem and offers new perspectives on the dynamics of quantum systems.
In the next section we will review the complex Langevin method.

\item {\it Lefschetz Thimble method}: The thimble approach involves deforming the integration domain into the complex plane to ensure that the imaginary part of the action remains constant along each thimble, thereby suppressing oscillations \cite{Cristoforetti:2012su, Fujii:2013sra}. 
Based on Morse theory, each thimble corresponds to the steepest ascent path originating from a critical point of the action. 
This method replaces the integration over real fields with integration over these complex thimbles. 
Thus, by summing over the integrals of all relevant thimbles, one recovers the integral over the original real domain.

The main challenges of this method include identifying the thimbles that contribute to the integral and performing the integration over these complex manifolds. 
A key advantage of the Lefschetz thimble method is its treatment of the phase factor, which is a major source of the sign problem. 
By isolating and controlling the fluctuations of the phase factor, the method improves the efficiency of Monte Carlo sampling. 
Additionally, it allows for systematic approximations by considering a finite number of thimbles, which is particularly useful when a single thimble dominates the contribution to the path integral. 
The Lefschetz thimble method has been successfully applied to various models with severe sign problems, offering new insights and computational techniques across diverse fields, from QCD to condensed matter physics \cite{DiRenzo:2015foa, Tanizaki:2015rda, Fujii:2015vha, Alexandru:2015xva}.
\end{itemize}
\end{itemize}

\section{A brief review of complex Langevin method} 
\label{sec:A_brief_review_of_complex_Langevin_method}

The complex Langevin method is designed to address the \textit{sign problem} by extending the concept of stochastic quantization from systems with real actions to those with complex actions \cite{Klauder:1983nn, Klauder:1983zm, Klauder:1983sp, Parisi:1983mgm}. 
This section will introduce the fundamental ideas behind complex Langevin dynamics and stochastic quantization. 

The two primary techniques for quantizing field theories are canonical and path integral quantization. 
In the 1980s, Parisi and Wu established a link between Euclidean field theories and statistical systems interacting with a heat bath \cite{Parisi:1980ys}, proposing an alternative quantization method based on stochastic differential equations, known as Langevin equations. 
This approach, referred to as stochastic quantization, treats Euclidean field theory as the equilibrium state of a statistical system driven by a stochastic process. 
The expectation values of observables in the original theory, which involve complex weights, can be calculated by measuring the observables of the complexified variables generated by the Langevin process and evaluating their expectation values over sufficiently long simulation times.

\subsection{Stochastic process: basic concepts}

A stochastic process describes the evolution of a random variable over stochastic time $\theta$. 
To illustrate the mathematical framework of these processes, we can examine Brownian motion, or the Wiener process, in one dimension. 
The erratic motion of a particle suspended in a liquid is governed by the following stochastic differential equation:
\beq
\label{eqn:stoch-process}
m \dot{v}(\theta) = - \gamma {v}(\theta) + \eta(\theta),
\eeq 
where $m$ is the particle's mass, $\gamma$ represents the friction coefficient due to the liquid's viscosity, and the dot indicates a derivative with respect to $\theta$. 
The term $\eta(\theta)$ represents the stochastic contributions -- random forces from the surrounding liquid particles. 
This equation is known as the Langevin equation for free Brownian motion, assuming no external potentials like gravity or spring. 
The Langevin equation models how a random variable evolves under the influence of random forces. 
By multiplying the equation by an integrating factor $e^{\gamma \theta / m}$, we can solve it, yielding:
\beq
{v}(\theta) = {v}(0) e^{-\gamma \theta / m} + \frac{1}{m} \int_0^\theta d\theta' \eta(\theta') e^{- \gamma (\theta - \theta') / m}.
\eeq

To compute physical quantities like position, velocity, and their correlation functions, we must account for the properties of the noise term $\eta(\theta)$. 
Let us consider the simplest scenario where $\eta(\theta)$ follows a Gaussian distribution over time and satisfies the condition:
\beq
\langle \eta (\theta) \rangle = 0,~~
~ \langle \eta (\theta) \eta (\theta') \rangle = \alpha \sigma^2 \gamma^2  \delta(\theta - \theta'),
\eeq
where $\sigma^2$ is the variance, and $\alpha$ governs the strength of the noise correlations. 
Using these properties, we can derive the following results:
\bea
\langle x(\theta) \rangle &=& x(0) + \frac{m}{\gamma} \dot{x}(0) \left(1 - e^{- \gamma \theta / m} \right), \\
\langle {v}(\theta) \rangle &=& {v}(0) e^{- \gamma \theta / m}, \\
\langle {v}(\theta){v}(\theta') \rangle &=& \langle v(\theta) \rangle \langle v(\theta')\rangle + \frac{\alpha \sigma^2 \gamma}{m} \left(e^{- \gamma (\theta - \theta') / m} - e^{- \gamma (\theta + \theta') / m} \right).
\eea
For very large times, $\theta = \theta' \to \infty$, $\langle v^2(\theta) \rangle$ approaches $\frac{\alpha \sigma^2 \gamma}{2m} = \frac{k_B T}{m}$, which is consistent with the equipartition theorem. 
To calculate higher-order correlation functions, we require the higher moments of the probability distribution of $\eta(\theta)$. 
Instead of directly specifying all moments, it is more convenient to use the probability distribution itself. 
The generalized functional probability distribution for the noise is given by:
\beq
P[ \eta ] \propto \exp \left( - \int_{-\infty}^{\infty} d \theta ~\frac{\eta^2(\theta)}{2 \alpha \sigma^2 \gamma^2} \right).
\eeq

\subsection{The prescription of Parisi and Wu}

Let us consider a real scalar field $\phi(x)$ in $d$-dimensions with a real Euclidean action $S[\phi(x)]$. 
With the help of the path integral
\beq
\langle \mathcal{O}(\phi)\rangle = \frac{1}{Z}{\int \mathcal{D} \phi(x)~ \mathcal{O}(\phi) e^{-S[\phi(x)]}};~~ Z = \int \mathcal{D} \phi(x)~ e^{-S[\phi(x)]},
\eeq
we can compute the expectation values of physical observables $\mathcal{O}(\phi)$. 
In stochastic quantization, these expectation values are found as the equilibrium values of a stochastic process. 
The system is updated according to Langevin dynamics in a fictitious Langevin time \(\theta\), driven by a Gaussian noise \cite{Parisi:1980ys}. 
At any Langevin time $\theta$, the evolution of the field is described by:
\beq
\label{eqn:real-Langevin}
\frac{d\phi(x; \theta)}{d\theta} = - \frac{\partial S[\phi(x; \theta)]}{\partial \phi(x; \theta)} + \eta(x; \theta),
\eeq 
where $\eta(x; \theta)$ is a Gaussian noise that satisfies the following properties:
\beq
\langle \eta (x; \theta) \rangle = 0,~~
\langle \eta (x; \theta) \eta (x; \theta') \rangle = 2 \delta(x - x') \delta(\theta - \theta').
\eeq
This equation governs the stochastic evolution of the scalar field in Langevin time, allowing the system to evolve towards equilibrium where physical observables can be computed. 
Notice that the noise has a variance of 2, chosen such that the diffusion coefficient in the Fokker-Planck equation matches the standard convention. 
This choice ensures that the equilibrium distribution of the Fokker-Planck equation corresponds to the Boltzmann factor.

Let us denote $\phi_{\eta}(x; \theta)$ as the solution of the Langevin equation which is obtained as the equilibrium field configuration. 
At large Langevin time $\Theta$, the Langevin time average of the observable $\mathcal{O}(\phi_{\eta}(x; \theta))$ is expected to approach expectation value given by the path integral:
\beq
\frac{1}{\Theta} \int_{\theta_0}^{\theta_0 + \Theta} d\theta ~\mathcal{O}(\phi_{\eta}(x; \theta)) \xrightarrow[]{\Theta \to \infty} \langle \mathcal{O}(\phi(x))\rangle.
\eeq
Since the solution $\phi_{\eta}(x; \theta)$ depends on the noise $\eta(x; \theta)$, different noise realizations generate a probability distribution $P(\phi; \theta)$ for the field configurations at Langevin time $\theta$. 
Therefore, the noise-averaged expectation value of $\mathcal{O}(\phi_{\eta})$ can be expressed as:
\beq
{\langle \mathcal{O}(\phi_{\eta}(x; \theta))\rangle}_{\eta} = \int \mathcal{D} \phi(x)~ P(\phi; \theta) \mathcal{O}(\phi(x; \theta)), 
\eeq   
where the left-hand side represents the noise-averaged expectation value. 
The evolution of the probability distribution $P(\phi; \theta)$ follows from the Langevin dynamics, satisfying the Fokker-Planck equation:
\beq
\frac{\partial P(\phi; \theta)}{\partial \theta} = - H_{\rm FP} P(\phi; \theta);~~ P(\phi; 0) =  \delta(\phi - \phi_0),
\eeq
where $H_{\rm FP}$ is the Fokker-Planck Hamiltonian, given by:
\beq
H_{\rm FP} = \int d^d x~\frac{\partial}{\partial \phi} \left( \frac{\partial}{\partial \phi} + \frac{\partial S}{\partial \phi} \right),
\eeq
and $\phi_0$ is the initial field configuration. 
This equation governs how the probability distribution evolves towards equilibrium over Langevin time.

Applying a similarity transformation to the probability distribution $P(\phi; \theta)$, we define a new distribution $\widetilde{P}(\phi; \theta)$ as
\beq
\widetilde{P}(\phi; \theta) = e^{S[\phi(x; \theta)] / 2} P(\phi; \theta).
\eeq
This transforms the Fokker-Planck equation into
\beq
\frac{\partial \widetilde{P}{(\phi; \theta)}}{\partial t} = - \widetilde{H}_{\rm FP} \widetilde{P}{(\phi; \theta)},
\eeq
where the new Hamiltonian $\widetilde{H}_{\rm FP}$ is given by
\bea
\widetilde{H}_{\rm FP} &=& e^{S[\phi]/2} ~H_{\rm FP}~ e^{- S[\phi]/2} \nn \\
&=&  \int d^d x~\left( -\frac{\partial}{\partial \phi} + \hf\frac{\partial S}{\partial \phi} \right)\left( \frac{\partial}{\partial \phi} - \hf\frac{\partial S}{\partial \phi} \right).
\eea
For a real action $S[\phi]$, the operator $\widetilde{H}_{\rm FP}$ becomes Hermitian and positive semi-definite. 
The Langevin process, in the long-time limit, makes the field distribution $\widetilde{P}{(\phi; \theta)}$ to approach the expected Boltzmann distribution $e^{-S[\phi]}$, ensuring that the dynamics correctly reproduce the desired path integral
\beq
\lim_{\theta\to \infty}{P}{(\phi; \theta)} \propto e^{-S[\phi]}.
\eeq
This guarantees that the real Langevin dynamics converge as intended. 

\subsection{Generalization to complex actions}

Shortly after Parisi and Wu's introduction of stochastic quantization, it became evident that this concept could be extended to complex actions. 
In 1983, both Klauder and Parisi independently proposed a method to simulate a complex measure involving an entire holomorphic action on a real manifold, $\mathcal{M}$ \cite{Klauder:1983nn, Klauder:1983zm, Klauder:1983sp, Parisi:1983mgm}. 
They established a stochastic process on the complexified version of the manifold, $\mathcal{M}_c$, ensuring that the expectation values of holomorphic observables $\mathcal{O}$ obtained from this process would match those computed with the original complex measure. 
For such complex actions, the solution to the Langevin equation naturally becomes complex as well. 
Consequently, the field variables need to be extended into the complex plane. 
In the simplest case (zero-dimensional), this is expressed as:
\beq
\phi(\theta) = \phi_r(\theta) + i \phi_i(\theta),
\eeq 
where $\phi_r$ and $\phi_i$ represent the real and imaginary parts of $\phi(\theta)$, respectively. 
The complex Langevin equation resembles Eq. \eqref{eqn:real-Langevin}, the equation for the real Langevin
\beq
\frac{d\phi(\theta)}{d\theta} = K (\phi; \theta) + \eta(\theta); ~~~~ K (\phi; \theta) = - \frac{\partial S(\phi(\theta))}{\partial \phi(\theta)},
\eeq
but now all quantities are complex, except for the stochastic noise $\eta(\theta)$, which remains real. 
For a more detailed analysis involving complex noise, see Ref. [\refcite{Aarts:2013uza}].

The relaxation dynamics of the complex Langevin equation can be understood by considering a complex-valued density $\rho(\phi_r; \theta)$ defined on $\mathcal{M}$, parameterized by the real variable $\phi_r$. 
This density evolves according to
\beq
\frac{\partial }{\partial \theta} \rho(\phi_r; \theta) = L^T_0 \rho(\phi_r; \theta); ~~ \rho(\phi_r; 0) = \delta(\phi_r - \phi_{r_0}), 
\eeq 
where $L_0^T$ is the complex Fokker-Planck operator given by
\beq
L^T_0 \equiv \nabla_r \left[\nabla_r - \nabla_r S(\phi_r)\right], 
\eeq
with $\nabla_r = \frac{\partial}{\partial \phi_r}$. 

For a more general case involving any $\phi_{i_0} \in \mathcal{M}$, the generalized complex Fokker-Planck operator is
\beq
L^T_{c0} \equiv \nabla_r \left[\nabla_r - \nabla_r S(\phi_r + i \phi_{i_0}) \right],
\eeq
which acts on a complex-valued density (or measure) on $\mathcal{M}$, again parameterized by the real variable $\phi_r$ \cite{Klauder:1983zm}. 
However, these operators do not preserve positivity, which restricts their probabilistic interpretation.

In complex Langevin dynamics, although the trajectories may extend into complex directions, the underlying process remains a real stochastic process. 
This can be demonstrated by decomposing the process into its real and imaginary components:
\beq
\frac{d\phi_r}{d\theta} = K_r + \eta(\theta), \quad \frac{d\phi_i}{d\theta} = K_i,
\eeq
where
\beq
K_r = - \text{Re} \left[ \frac{\partial S(\phi_r + i \phi_i)}{\partial \phi_r} \right], \quad K_i = - \text{Im} \left[ \frac{\partial S(\phi_r + i \phi_i)}{\partial \phi_i} \right].
\eeq
For this real stochastic process, we can define a real and positive definite probability density $P(\phi_r, \phi_i; \theta)$ on $\mathcal{M}_c$. 
This density evolves according to the Fokker-Planck equation:
\bea
\label{eqn:clm-holomorphic-L}
\frac{\partial}{\partial \theta} P(\phi_r, \phi_i; \theta) = L^T P(\phi_r, \phi_i; \theta); ~~ P(\phi_r, \phi_i; 0) = \delta(\phi_r - \phi_{r_0}) \delta(\phi_i), 
\eea
where the real Fokker-Planck operator $L$ is given by
\beq
L^T \equiv \nabla_r \left[\nabla_r - K_r \right] - \nabla_i K_i,
\eeq
with $\nabla_r = \frac{\partial}{\partial \phi_r}$ and $\nabla_i = \frac{\partial}{\partial \phi_i}$.

The crucial question is whether the real and complex Langevin evolutions yield consistent results for the expectation values of holomorphic observables $\mathcal{O}$. 
Specifically, we need to verify if
\beq
\langle \mathcal{O} \rangle_{P(\theta)} = \frac{\int d\phi_r d\phi_i ~ \mathcal{O}(\phi_r + i \phi_i) P(\phi_r, \phi_i; \theta)}{\int d\phi_r d\phi_i ~ P(\phi_r,\phi_i; \theta)}
\eeq
and
\beq
\langle \mathcal{O} \rangle_{\rho(\theta)} = \frac{\int d\phi_r ~ \mathcal{O}(\phi_r) \rho(\phi_r; \theta)}{\int d\phi_r~ \rho(\phi_r; \theta)}
\eeq
remain equal if they agree initially at $\theta = 0$. 
To establish this, we aim to demonstrate:
\beq
\langle \mathcal{O} \rangle_{P(\theta)} = \langle \mathcal{O}\rangle_{\rho(\theta)}, 
\eeq
given the initial condition:
\beq
\label{eqn:clm-density-initial}
P(\phi_r, \phi_i; 0) =  \rho(\phi_r; 0) \delta(\phi_i - \phi_{i_0}).
\eeq 

The question of whether real and complex Langevin evolutions yield identical expectation values for holomorphic observables was formally addressed in Refs. [\refcite{Aarts:2009uq,Aarts:2011ax}]. 
The authors demonstrated that this equivalence holds specifically for holomorphic observables, provided the action and its gradient are holomorphic functions of the complex field $\phi$. 
This is an important point since only holomorphic observables allow for the extension of the complex Langevin operator $L^T_{c0}$ to the analytic continuation in the complexified space $\mathcal{M}_c$. 

The complex Langevin operator $\widetilde{L}$ is defined as:
\beq
\widetilde{L}^T \equiv \nabla_\phi \left[\nabla_\phi -  \nabla_\phi S(\phi)\right].
\eeq
The action of this operator on these analytic continuations aligns with that of the real Fokker-Planck operator $L$, with the difference $L - \widetilde{L}$ vanishing due to the Cauchy-Riemann (CR) equations.

To clarify, the Langevin operator $\widetilde{L}$ is derived as an analytic continuation of the complex Fokker-Planck operator $L_{c0}$, which acts on the real manifold $\mathcal{M}$. 
The operators $\widetilde{L}$ and $L$ both act on functions defined on the complexified space $\mathcal{M}_c$, but they agree only for holomorphic functions.

To analyze the time evolution of holomorphic observables using the complex Langevin method, we focus on how these observables evolve over time rather than densities, leveraging the Cauchy-Riemann (CR) equations. 
We can use the operator $\widetilde{L}$ or $L$ interchangeably to evolve holomorphic observables. 

We have
\bea 
\partial_{\theta} \mathcal{O}(\phi; \theta) =  {L} \mathcal{O}(\phi; \theta),
\eea
with the initial condition given by $\mathcal{O}(\phi; 0) = \mathcal{O}(\phi)$. 
Once we solve this differential equation, formally, we get
\beq
\label{eqn:clm-LO-evolve}
\mathcal{O}(\phi; \theta) = \exp[\theta {L}]~ \mathcal{O}(\phi).
\eeq

To justify the method, we need to establish that, in the large Langevin time limit $\Theta = \theta \to \infty$, the expectation values of the real probability density $\langle \mathcal{O} \rangle_{P(\Theta)}$ and the complex density $\langle \mathcal{O} \rangle_{\rho(\Theta)}$ are equal. 
To do this, we define a quantity \( F(\Theta, \theta) \) for \( 0 \le \theta \le \Theta \) as follows:
\beq
F(\Theta, \theta) \equiv \int d\phi_r d\phi_i~P(\phi_r, \phi_i; \Theta - \theta)~ \mathcal{O}(\phi_r + i \phi_i; \theta).
\eeq
This quantity \( F(\Theta, \theta) \) interpolates between the expectation values:
\beq
F(\Theta, 0) = \langle \mathcal{O} \rangle_{P(\Theta)}
\eeq
and
\beq
F(\Theta, \Theta) = \langle \mathcal{O} \rangle_{\rho(\Theta)}.
\eeq
The first equality is straightforward. 
The second can be understood as follows:
\bea
F(\Theta, \Theta) &=& \int d\phi_r d\phi_i~P(\phi_r, \phi_i; 0)~ \mathcal{O}(\phi_r + i \phi_i; \Theta) \nn \\
&=& \int d\phi_r d\phi_i~P(\phi_r, \phi_i; 0)~ \exp[\Theta L]\mathcal{O}(\phi_r + i \phi_i; 0).
\eea
In the above, we employed the property in Eq. \eqref{eqn:clm-LO-evolve} to evolve the observable.

We can use the initial condition, given in Eq. \eqref{eqn:clm-density-initial}, to get
\bea
F(\Theta, \Theta) &=& \int d\phi_r d\phi_i ~\rho(\phi_r; 0) \delta(\phi_i - \phi_{i_0}) \exp[\Theta L] \mathcal{O}(\phi_r + i \phi_i; 0) \nn \\
&=& \int d\phi_r ~\rho(\phi_r; 0) \exp[\Theta L_{c0}] \mathcal{O}(\phi_r + i \phi_{i_0}; 0) \nn \\
&=& \int d\phi_r d\phi_i ~\mathcal{O}(\phi_r + i \phi_i; 0) \exp[\Theta L^T_{c0}] \rho(\phi_r; 0) \nn \\
&=& \langle \mathcal{O} \rangle_{\rho(\Theta)}.
\eea
In the above, we used integration by parts in $\phi_r$, and neglected the boundary terms at $\infty$ and the poles coming from the drift term. 
Note that when $F(\Theta, \theta)$ is independent of $\theta$ these boundary terms vanish. 
To understand this, let us consider the derivative of $F(\Theta, \theta)$:
\bea
\label{eqn:boundary-terms}
\frac{\partial}{\partial \theta}F(\Theta, \theta) &=& \int d\phi_r d\phi_i~  P(\phi_r, \phi_i; \Theta - \theta) \widetilde{L}\mathcal{O}(\phi_r + i \phi_i; \theta) \nn \\ 
&& - \int d\phi_r d\phi_i~ \left(L^T P(\phi_r, \phi_i; \Theta - \theta) \right) \mathcal{O}(\phi_r + i \phi_i; \theta). 
\eea

At large Langevin time, in equilibrium, the condition for the stationarity of observables in the stochastic process is:
\beq
\label{eqn:clm-cc}
\lim_{\theta \to \infty} \partial_\theta \langle \mathcal{O} \rangle = \langle \widetilde{L} \mathcal{O} \rangle \equiv \int d\phi_r d\phi_i~ P(\phi_r, \phi_i; \infty) \widetilde{L} \mathcal{O}(\phi_r + i \phi_i; 0) = 0.
\eeq
Note that the above expression resembles the Schwinger-Dyson equations. This is known as the {\it consistency condition}. 
Sometimes, the consistency condition, along with additional criteria, is sufficient to test the correctness of the equilibrium measure in the complex Langevin method \cite{Aarts:2011ax}.

\subsubsection{Correctness criteria for the complex Langevin method}

After a sufficiently large Langevin time, the probability distribution associated with the field configurations should ideally approach the Boltzmann factor. 
This convergence is well-established for real actions. 
However, no exact mathematical proof guarantees convergence for complex actions. 
Due to this unresolved issue, the complex Langevin method has historically garnered limited attention from physicists. 
The formal arguments discussed earlier highlight several significant mathematical questions:
\begin{itemize}
\item {\it Langevin / Fokker-Planck Operators}: The question of whether the Langevin / Fokker-Planck operators $L$, $L_{c0}$, and $\widetilde{L}$ can be exponentiated remains unresolved. 
Specifically, the existence of a unique stochastic process and the time evolutions generated by these operators are not fully established. 
In practical terms, this issue is mitigated by using an {\it adaptive step size}, as discussed in the following sections \cite{Aarts:2011ax}.

\item {\it Convergence to an Equilibrium Measure}: There is no mathematical proof ensuring that a positive density converges to the equilibrium measure. 
This issue is related to the spectrum of Langevin operators \cite{Weingarten:2002xs}. 
Klauder and Peterson (1985) noted the {\it conspicuous absence of a general theorem} for non-self-adjoint operators \cite{Klauder:1985kq, Klauder:1985ks}.

\item {\it Boundary Terms}: The validity of various integrations by parts, which facilitate the shifting of time evolution between measures and observables, remains uncertain \cite{Aarts:2008wh}. 
One must consider whether boundary terms need to be accounted for in this context.
\end{itemize}

Despite the lack of rigorous mathematical proof, physicists have continued to develop the complex Langevin method with practical success. 
The method has demonstrated its utility in numerous influential studies \cite{Gausterer:1985sm, Klauder:1985kq, Karsch:1985cb, Horowitz:1986dt, Flower:1986hv, Ambjorn:1986fz, Haymaker:1987ey, Catterall:1990qn, Kieu:1994xg, Gausterer_1998, Adami:2000fs, Berges:2005yt, Berges:2007nr}. 
Recently, there has been renewed interest in proving the convergence of the probability distribution $P(\phi_r, \phi_i; \theta)$ to the complex measure $\rho = e^{-S[\phi]}$. 
Several correctness criteria have been proposed to assess the reliability of the method:
\begin{itemize}
\item {\it Langevin-Operator Criterion}: Revived in 2009 by Aarts, Seiler, and Stamatescu \cite{Aarts:2009uq}, this criterion requires that the Langevin operator acting on observables should vanish, i.e.,
\beq
\langle \widetilde{L} \mathcal{O} \rangle = 0.
\eeq
We can use this as a reliability criterion in numerical simulations to ensure the convergence of the distribution $P(\phi_r, \phi_i; \theta)$ to the equilibrium measure. 
Note that this criterion needs to be satisfied for a complete set of observables, $\mathcal{O}[\phi]$ in an aptly chosen basis \cite{Aarts:2011ax}, resulting in an infinite series of identities similar to Schwinger-Dyson equations.

\item {\it Probability Drift Criterion}: Introduced in 2016 by Nagata, Nishimura, and Shimasaki \cite{Nagata:2015uga}, this criterion assesses the magnitude of the drift term in the probability distribution $P(\phi_r, \phi_i; \theta)$. 
The presence or absence of boundary terms is linked to the growth of the holomorphic function and the associated drift term. 
We can define a magnitudes $u$ as the absolute value of the gradient of the action:
\beq
u \equiv \left| \frac{\partial S[\phi]}{\partial \phi} \right |.
\eeq
For reliable simulations, we need an exponential or faster decay of the probability of the drift term, $P(u)$ at larger values of $u$.
\end{itemize}

\subsection{Simulations using complex Langevin method}

Let us briefly look at the complex Langevin method from the perspective of numerical simulations. 
In the case of zero dimensions, for simplicity, the complex Langevin equation is given by
\beq
\frac{d\phi(\theta)}{d\theta} = K(\phi; \theta) + \eta(\theta); \quad K(\phi; \theta) = -\frac{\partial S(\phi(\theta))}{\partial \phi(\theta)},
\eeq
where $\eta(\theta)$ represents a real continuous noise function with a variance of 2. 
To implement this equation numerically, we must discretize the continuous Langevin equation. 
A straightforward discretization, where the Langevin time is represented as $\theta = n \epsilon$ (an integer multiple of the time step size $\epsilon$), results in the following discrete update rule:
\beq
\phi_{\theta + \epsilon} = \phi_\theta - \frac{\partial S[\phi]}{\partial \phi} \Big|_{\phi_\theta} \epsilon + \eta^d_\theta \epsilon,
\eeq
where $\eta^d_\theta$ denotes the discretized noise term.

Although the discretization of the complex Langevin equation may seem straightforward at first glance, handling the noise function is somewhat intricate. 
To illustrate this, consider a continuous noise function $\eta(\theta)$ with variance $\sigma^2$. 
The relationship between this continuous noise and the discretized noise $\eta^d_\theta$ is given by
\bea
\langle \eta (\theta ) \eta (\theta') \rangle = \sigma^2 \delta(\theta - \theta') ~~~~~ 
&\xrightleftharpoons[\rm continuous]{\rm discretized} & ~~~~~~~\langle \eta^d_\theta \eta^d_{\theta'} \rangle = c \delta_{\theta \theta'} \\
\Big \downarrow {\substack{\text{integrate} \\ \text{over continuous } \theta} }~~~~~~~~~~&&~~~~~~~~~~ \Big \downarrow {\substack{\text{sum} \\ \text{over discretized } \theta} } \nn \\
\int d\theta ~\langle \eta (\theta ) \eta (\theta') \rangle = \sigma^2 ~~~~~~~~ &\xrightleftharpoons[\rm continuous]{\rm discretized}& ~~~~~\epsilon \sum_{n = 1}^\infty \langle \eta^d_\theta \eta^d_{\theta'}  \rangle = \sigma^2. 
\eea
From this procedure, we can determine the constant $c$:
\beq
\epsilon \sum_{n = 1}^\infty c \delta_{\theta \theta'} = \sigma^2 {\implies} c = \frac{\sigma^2}{\epsilon}.
\eeq 

Thus, the discretized Gaussian noise $\eta^d_\theta$ satisfies
\beq
\langle \eta^d_\theta \rangle = 0, \quad \langle \eta^d_\theta \eta^d_{\theta'} \rangle = {\sigma^2} \delta_{\theta \theta'}/{\epsilon}.
\eeq
By rescaling the noise
\beq
\eta^d_\theta \to {\eta_{\theta} } / {\sqrt{\epsilon}}
\eeq
the dependence on the Langevin step-size $\epsilon$ can be eliminated. 
For a variance of $\sigma^2 = 2$, ensuring consistency between Fokker-Planck equation's equilibrium solution and the Boltzmann distribution, the discretized complex Langevin equation becomes
\beq
\label{eqn:disc-clm}
\phi_{\theta + \epsilon} = \phi_{\theta } - \frac{\partial S[\phi]}{\partial \phi} \Big |_{\phi_\theta}  {\epsilon} + \eta_{\theta} \sqrt{ \epsilon},
\eeq
where the discretized Gaussian noise $\eta_\theta$ satisfies
\beq
\langle \eta_\theta \rangle = 0, \quad \langle \eta_\theta \eta_{\theta'} \rangle = 2 \delta_{\theta \theta'}.
\eeq

\begin{figure}[htp]
\begin{center}
\includegraphics[width =0.6\textwidth]{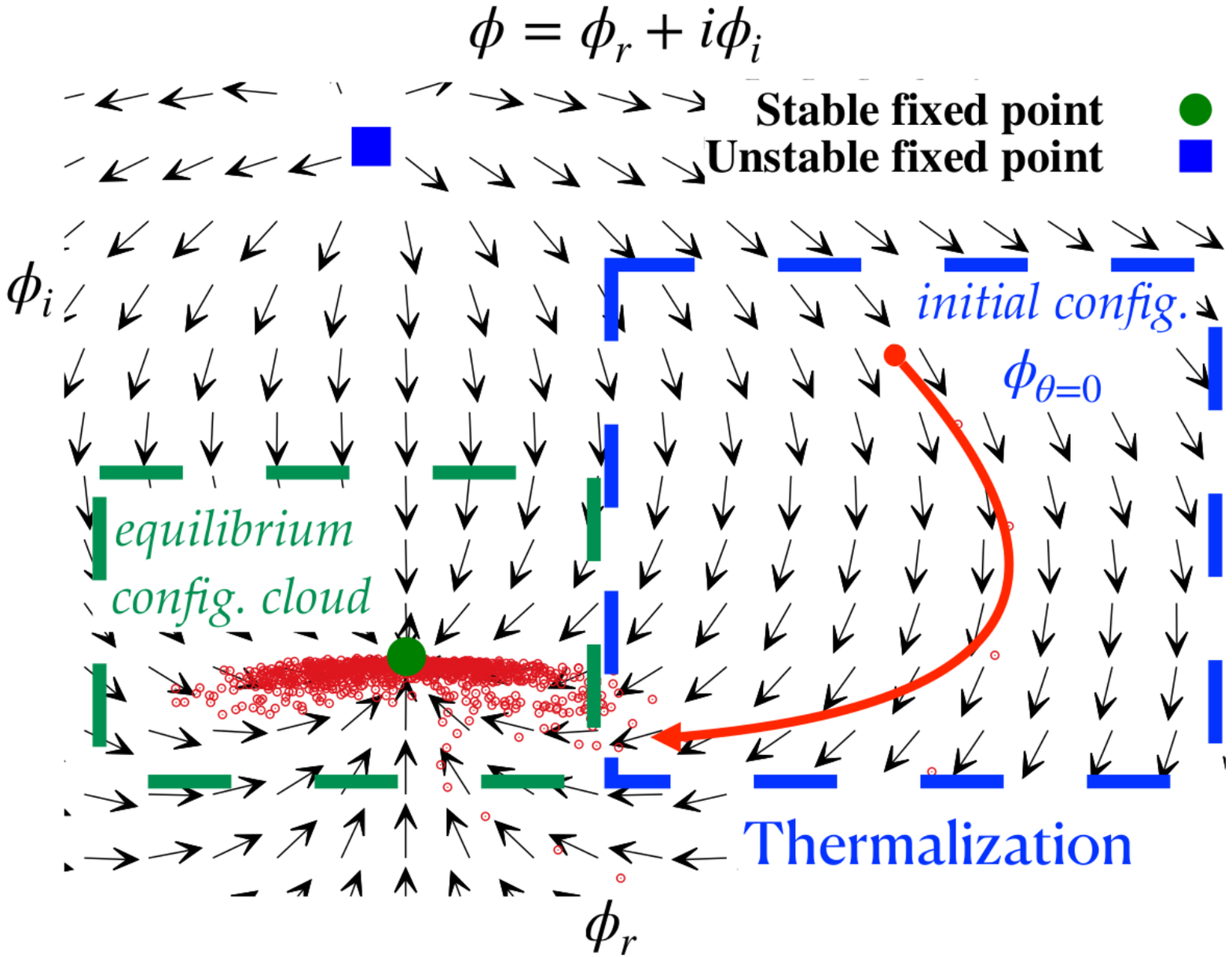}
\end{center}
\caption{A schematic representation of the evolution of the complex field following the complex Langevin dynamics.}
\label{fig:schematic-clm}	
\end{figure}

Finally, we make use of the discretized version of the complex Langevin equation, given in Eq. \eqref{eqn:disc-clm}, to update the field configurations of the theory. 
In this process, the fields evolve as the Langevin time progresses, following the Langevin dynamics, guided by the drift term $-\frac{\partial S[\phi]}{\partial \phi} \big|_{\phi_{\theta}}$. 

Figure \ref{fig:schematic-clm} illustrates the evolution of a complex field under the Langevin dynamics for a theory with a complex scalar field. 
Starting from an initial field configuration, $\phi_0$ at Langevin time $\theta = 0$, the field evolves under the influence of the drift, with arrows representing the direction of this drift. 
As the field progresses, it moves towards the neighborhood of a stable fixed point. 
Due to the stochastic nature of the process, the field does not simply converge to this fixed point but instead forms a cloud around it. 

After sufficient thermalization, at a large Langevin time $\theta$, this cloud of field configurations approximates a real probability distribution $P(\phi_r, \phi_i; \theta)$, which is expected to be the equilibrium solution of the Fokker-Planck equation. 
After the equilibrium is achieved, the noise averaged expectation values of the observables are obtained using the probability distribution $P(\phi_r, \phi_i; \theta)$. 
Specifically,
\bea
\langle \mathcal{O} \rangle_{\eta} = \lim_{\theta \to \infty } \langle \mathcal{O (\phi(\theta))} \rangle_{\eta} &=& \lim_{\theta \to \infty } \int d\phi_r d\phi_i~\mathcal{O}(\phi_r + i \phi_i) P(\phi_r, \phi_i; \theta) \nn \\
&\approx& \frac{1}{N} \sum_{n = 0}^N \mathcal{O}(\phi_r + i \phi_i).
\eea

\subsubsection{Numerical challenges and techniques for stabilization}

The complex Langevin method gained significant attention when it was introduced in the 1980s due to its potential to address models with severe sign problems, as it does not depend on a probabilistic interpretation of the weight. 
Despite this promising start, early studies faced substantial challenges. 
Numerical instability and incorrect convergence posed significant hurdles, leading to difficulties in applying the method effectively \cite{Ambjorn:1985iw, Ambjorn:1986fz, Gausterer:1992jz, Gausterer:1998jw}. 
This section provides an overview of these numerical challenges and the stabilization techniques developed to address them. 
Specifically, the issues stem from the difficulty in achieving convergence of the probability distribution $P(\phi_r, \phi_i; \theta)$ to the equilibrium measure. 
The primary problems are {\it runaways}, where field configurations fail to converge even after prolonged Langevin time, and {\it convergence to an incorrect limit}, which represents a more severe numerical issue.

Recent advancements have led to a successful revival of the complex Langevin method, demonstrating its ability to produce accurate results even in the presence of severe sign problems. 
Although the classical flow often exhibits unstable fixed points, the introduction of stochastic noise helps to prevent these trajectories from becoming trapped, thereby maintaining stability in the dynamics. 

However, the complexification of fields introduces additional degrees of freedom that are typically unbounded, which can result in divergent trajectories and render numerical simulations unstable. 
This instability can occur particularly when field configurations approach unstable directions. 
Therefore, careful numerical integration of the Langevin equations is crucial.

To address these issues, several stabilization techniques can be employed, including \textit{adaptive step size}, \textit{gauge cooling}, and \textit{dynamical stabilization}. 
These methods help manage the challenges associated with unstable trajectories. 
A brief overview of each technique is provided below.

\subsubsection*{Adaptive step size}

Langevin trajectories can venture significantly into imaginary directions, and while using a small step size might mitigate this issue to some extent, it does not universally prevent instabilities. 
This approach can result in very slow evolution, requiring numerous updates to thoroughly explore the configuration space \cite{Aarts:2009dg}. 

To address these challenges, an adaptive step size algorithm is often implemented in the numerical integration of Langevin equations. 
In a basic implementation, the maximum drift value, $K_{\rm max}(\theta)$, is computed at each Langevin step. 
The step size for the subsequent evolution sweep is then determined by the formula:
\beq
\epsilon = \frac{\gamma}{K_{\rm max}(\theta)},
\eeq
where the parameter $\gamma$ is chosen according to the specifics of the model.

An enhanced implementation of the adaptive step size algorithm involves monitoring the maximum drift $K_{\rm max}^{(n)}$ at each discrete Langevin time step $n$, where $\theta = n \epsilon$. 
The maximum drift is defined as:
\beq
K_{\rm max}^{(n)} \equiv \max_{x} |K_{x}^{(n)}| = \max_{x} \sqrt{K_{x,R}^{2}(n) + K_{x,I}^{2}(n)}.
\eeq

The step size $\epsilon_n$ for the next Langevin evolution is then determined by placing an upper bound on the product $\epsilon K_{\rm max}$:
\beq
\epsilon_n = \frac{\bar{\epsilon}}{K_{\rm max}/K_{\rm max}^{(n)}},
\eeq
where $\bar{\epsilon}$ is the desired average step size, and $K_{\rm max}$ is either precomputed or determined during the thermalization phase. 
This approach ensures that the step size is adaptive to the local conditions of the simulation, becoming smaller near instabilities and larger in more stable regions.

Alternatively, the step size can be adjusted to maintain the product $\epsilon K_{\rm max}$ within a specified range relative to a reference value $\bar{K}$, specifically:
\beq
\frac{1}{p} \times \bar{K} \leq \epsilon K_{\rm max} \leq p \times \bar{K},
\eeq
where $p$ and $\bar{K}$ are predetermined constants. 
If $\epsilon K_{\rm max}$ falls outside this range, the step size is adjusted by a factor of $p$, with the process repeated as necessary \cite{Aarts:2009dg}. 
This method provides a more dynamic response to the behavior of the Langevin trajectory and does not require fine-tuning, provided that regions with inappropriate behavior are avoided.

\subsubsection*{Gauge cooling}

The complex nature of the action in complex Langevin dynamics can cause fields to drift into imaginary directions. 
For the case of Hermitian matrix fields $X_\mu$, the dynamics can cause them to venture into anti-Hermitian directions. 
These excursions can lead to an enlargement of the group space, for example, from SU($N$) to SL($N, \mathbb{C}$). 
The field configurations, when stray too far from SU($N$), we encounter the so-called excursion problem. 

A solution to this problem, proposed in Ref. [\refcite{Seiler:2012wz}] is called \textit{gauge cooling}. It introduces a \textit{Hermiticity norm} \cite{Nagata:2016vkn} to quantify deviations from Hermitian configurations. 
For example, for the IKKT matrix model, this norm is defined as:
\beq
\mathcal{N}_{\rm H} \equiv - \frac{1}{D N} \sum_\mu \text{tr} \left( \left[ X_\mu - X_\mu^\dagger \right]^2 \right),
\eeq
where $D$ is the number of matrix fields $X_\mu$. 
This norm measures how much the matrix fields deviate from Hermitian configurations. 

To address the excursions, the matrix fields $X_\mu$ are transformed under the enlarged gauge symmetry:
\bea
X_\mu \rightarrow g X_\mu g^{-1}, \quad g \in \text{SL}(N, \mathbb{C}),
\eea
where $g$ is chosen as $g = \text{e}^{-\alpha \delta \mathcal{N}_{\rm H}}$ with a tuning parameter $\alpha$, which is a real and positive, and
\bea
\delta \mathcal{N}_{\rm H} = \frac{1}{N} \sum_\mu \left[ X_\mu, X_\mu^\dagger \right].
\eea
The norm $\mathcal{N}_{\rm H}$ is not invariant under this gauge transformation. 
This non-invariance allows for iterative application of the gauge transformation to progressively minimize $\mathcal{N}_{\rm H}$ and bring the matrix fields closer to Hermitian directions. 
The procedure of gauge cooling has been shown to comply with the correctness criteria for complex Langevin method \cite{Nagata:2016vkn}.

\subsubsection*{Dynamical stabilization}

This method was recently introduced by Aarts, Attanasio, Jager, and Sexty. \cite{Aarts:2016qhx, Attanasio:2018rtq}. 
It involves decreasing the unitary norm by adding a custom drift term to the complex Langevin process, which vanishes in the continuum limit. 
Specifically, the drift term $K_x$ at lattice site $x$ is modified as follows:
\beq
K_x \to \widetilde{K}_x = K_x + i \alpha_{\rm DS} M_x,
\eeq
where $\alpha_{\rm DS}$ is a control parameter and $M_x$ acts only in the imaginary direction, orthogonal to the SU($N$) manifold, and grows with the unitary norm. 
Although this adjustment to the drift term is designed to vanish in the continuum limit, it is manually incorporated into the process. 
As a result, it does not satisfy the complex Langevin correctness criteria, making it challenging to assert that simulations using this dynamical stabilization method yield correct results.

\begin{figure}[t]
\begin{center}
\includegraphics[width = 0.7\textwidth, origin = c, angle = 0]{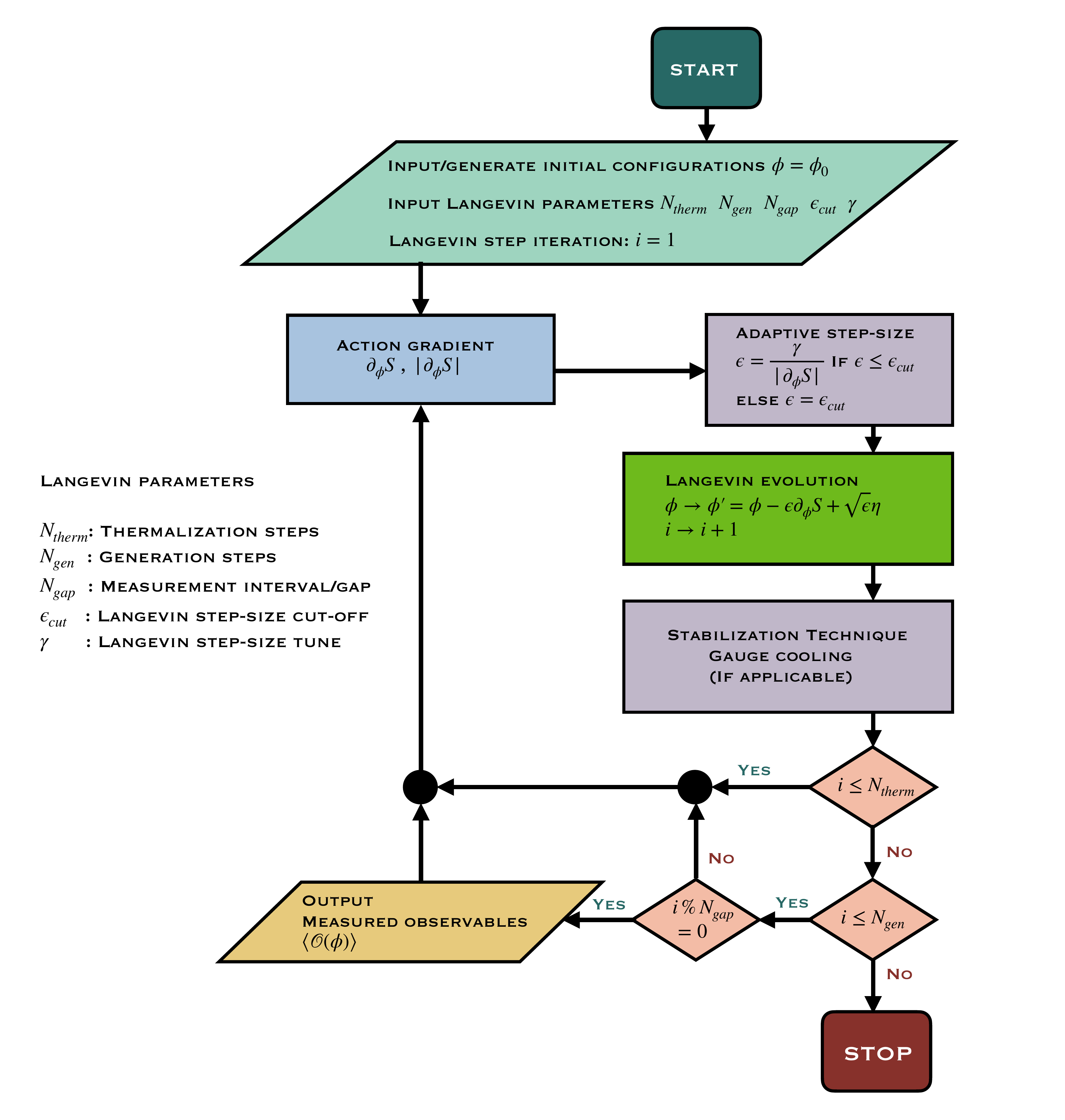}
\end{center}
\caption{Flowchart for the complex Langevin algorithm.}
\label{fig:flowchart-clm}	
\end{figure}

\subsubsection{Outline of the complex Langevin algorithm}

The flowchart in Fig. \ref{fig:flowchart-clm} outlines the implementation process for the complex Langevin method, incorporating adaptive step size and stabilization techniques. 
Below are the key components of the flowchart:
\begin{itemize}
\item {\it Initialization:} Start by initializing the field configurations, denoted as $\phi = \phi_0$. 
Set key Langevin parameters: $N_{\text{therm}}$ for the number of thermalization steps, $N_{\text{gen}}$ for the number of generation steps, $N_{\text{gap}}$ for the measurement interval, $\epsilon_{\text{cut}}$ for the step-size cut-off, and $\gamma$ for tuning the step size. 
Begin the Langevin iteration with $i = 1$.

\item {\it Langevin Evolution and Adaptive Step Size:} Compute the action gradient $\partial_\phi S$ to determine the adaptive step size $\epsilon$, given by $\epsilon = \frac{\gamma}{|\partial_\phi S|}$. 
If $\epsilon$ is less than or equal to $\epsilon_{\text{cut}}$, use $\epsilon$ as computed; otherwise, set $\epsilon$ to $\epsilon_{\text{cut}}$. 
Perform the Langevin evolution by updating the field configuration to $\phi' = \phi - \epsilon \partial_\phi S + \sqrt{\epsilon} \eta$, where $\eta$ represents stochastic noise. 
Increment the iteration counter $i$.

\item {\it Stabilization:} Apply a stabilization technique, such as gauge cooling, if applicable. 
Check whether the current step $i$ falls within the thermalization phase ($i \leq N_{\text{therm}}$) or the generation phase ($i \leq N_{\text{gen}}$), and if measurements should be taken at the current step ($i \% N_{\text{gap}} = 0$). 
If so, measure the observables $\langle \phi \rangle$. 
The procedure continues until the maximum number of generation steps is reached.
\end{itemize}

\subsection{Complex Langevin method: Recent studies and results}

This section briefly highlights some recent successful applications of the complex Langevin method. 

Since its inception, the complex Langevin method has undergone continuous improvements to address its primary shortcomings: {\it non-convergence} and the notorious {\it wrong convergence} caused by zeros in the probability density within the complex plane. 
The associated Fokker-Planck equation generally allows multiple equilibrium solutions in the space of distributions, see Ref. [\refcite{Salcedo:1993tj}]. 
While violations of the correctness criterion can help identify such scenarios, fundamentally understanding and resolving these issues remains a significant challenge.

The underlying problems are typically attributed to either the insufficient decay of the probability distribution in the complexified configuration space (at infinity or the poles of the drift force) or to a failure of ergodicity. 
These shortcomings result in boundary terms during integration by parts, as outlined in Eq. \eqref{eqn:boundary-terms}, and thereby lead to slow decay, which undermines the formal proof of correctness as discussed in Ref. [\refcite{Seiler:2020mkh}]. 
This was extensively studied in simple but pragmatic models, such as the U($1$) one-link model, where analytic results can be directly compared with numerical simulations, see Ref. [\refcite{Scherzer:2018hid}]. 
Further efforts to address these issues Ref. [\refcite{Scherzer:2018udt}] focused on explicitly computing boundary terms to test the convergence of the method and investigated the deconfinement phase transition in two-flavor Wilson fermion QCD. 
Moreover, examining the boundaries at infinity at the cost of measuring higher-order boundary term observables extended these investigations to various models, including one-plaquette models, the Polyakov chain model, the 3D XY model, and heavy dense QCD (Ref. [\refcite{Scherzer:2019lrh}]). 
Despite these advances, debates around the persistence of boundary-related issues continue as discussed in Ref. [\refcite{Seiler:2023kes}] and recent findings in Ref. [\refcite{Mandl:2024zvz}] advocated that boundary terms alone are insufficient as a correctness criterion, with additional complications arising from unwanted integration cycles. 
Refs. [\refcite{Okamoto:1988ru,Okano:1991tz,Salcedo:2018fvt}] established that introducing a suitable complex holomorphic kernel into the Langevin equation can effectively address boundary terms issues. 
However, the integration cycle complications persist, necessitating ongoing research to ensure correct convergence. 
See Ref. [\refcite{Hansen:2024kjm}] for investigations in this direction using toy models. 

Another recent promising strategy to mitigate incorrect convergence leverages insights from Lefschetz thimbles, employing weight regularization based on their associated structures to correct biases [\refcite{Boguslavski:2024yto,Boguslavski:2024zqf}]. 
A new parameter-free gauge cooling technique, the alternating descent method, was also proposed in Ref. [\refcite{Dong:2020mtk}] and successfully illustrated on the one-dimensional Polyakov loop model and 4D heavy quark QCD at finite chemical potential. 
The resurgence in interest has also driven significant advancements. 
The efficacy of adaptive step-size methods in managing unstable complex Langevin trajectories in lattice QCD was first demonstrated in the three-dimensional XY model, as detailed in Refs. [\refcite{Aarts:2010aq,Aarts:2010vk,Aarts:2012ft}]. 
Further optimizations and stabilizing techniques for simulations are discussed in Refs. [\refcite{Aarts:2011zn,Seiler:2012wz,Aarts:2017vrv,Attanasio:2018rtq}]. 
Ref. [\refcite{Attanasio:2017rxk,Attanasio:2018kwp}] provides an update on dynamical stabilization in the heavy quark limit and its application to the XY model at finite chemical potential. 

Complex Langevin simulations have significantly advanced our understanding of the real-time formulation of quantum field theories. 
The original motivations arose from systems that must be treated in real-time, such as out-of-equilibrium systems that cannot be formulated using Euclidean techniques. 
Ref. [\refcite{Berges:2005yt}] provides a detailed exploration of simulations of non-equilibrium quantum fields. 
Inspired by these successes, applications were extended to scalar and non-Abelian gauge fields, as discussed in Ref. [\refcite{Berges:2006xc}], while gauge theories in Minkowski spacetime, optimized for efficient updating, were examined in Ref. [\refcite{Berges:2007nr}]. 
Later, Ref. [\refcite{deAguiar:2010ue}] investigated finite-temperature field theory in the real-time formalism using stochastic quantization. 
Recently, Ref. [\refcite{Alvestad:2021hsi}] explored the potential of modern implicit solvers for stochastic partial differential equations in the context of real-time complex Langevin dynamics. 
The implicit solvers can be shown to be unconditionally asymptotically stable, preventing the occurrence of runaway trajectories, as long as the underlying complex Langevin dynamics remain finite. 
They also allow for simulations at comparatively large Langevin time steps, leading to lower computational cost. 
In Ref. [\refcite{Boguslavski:2022dee}], using the Schwinger-Keldysh formalism for SU($N$) gauge theories, the authors identified the insufficiency of current stabilization techniques and introduced a novel anisotropic kernel. 
This kernel enables complex Langevin simulations on discretized complex time paths, applied to SU($2$) Yang-Mills theory in $(3+1)$ dimensions. 
This advancement could pave the way for an ab initio real-time framework for QCD in and out of equilibrium. 
Real-time quantities, such as spectral functions and transport coefficients, are crucial for understanding the quark-gluon plasma. 
In Ref. [\refcite{Boguslavski:2023unu}], the authors achieved the first direct ab initio computation of unequal-time correlation functions in $(3+1)$-dimensional real-time Yang-Mills theory in thermal equilibrium. 
These unequal-time correlation functions, computed for the first time in non-Abelian lattice gauge theory, are essential for extracting real-time observables. 
This work lays the foundation for obtaining real-time quantities from lattice gauge theory using a first-principles real-time framework.

Non-perturbative studies of QCD at finite chemical potential remain among the most significant and notoriously challenging tasks due to the severe sign problem associated with well-established path integral Monte Carlo methods. 
The application of complex Langevin dynamics to finite-density lattice QCD has shown promising potential to circumvent the sign problem, reigniting interest in this approach, as demonstrated in Refs. [\refcite{Aarts:2008rr,Aarts:2008yw,deForcrand:2009zkb}].

As discussed earlier in this section, complex Langevin extends the gauge group from SU(3) to SL($3, \mathbb{C}$), where the non-compact nature of SL($3, \mathbb{C}$) can lead to runaway configurations, necessitating optimizations and stabilization techniques to ensure evolution remains close to the unitary manifold. 
Several works, including Refs. [\refcite{Aarts:2012yal,Aarts:2015yba,Felipe:2017wuf,Scherzer:2019weu,Scherzer:2020kiu}], have explored the validity of complex Langevin dynamics to address the sign problem in lattice QCD at non-zero baryon density, with a particular focus on the {\it overlap problem} and the {\it Silver Blaze problem}, as well as high-temperature and heavy dense regimes. 
For a comprehensive overview of the progress in complex Langevin simulations of the QCD phase diagram, see Refs. [\refcite{Aarts:2014fsa,Sexty:2014zya,Sinclair:2017zhn,Bloch:2018sof,Attanasio:2020spv}]. 
Complex Langevin method, stabilized with gauge cooling, was extended to full QCD at non-zero chemical potential in Ref. [\refcite{Sexty:2013ica}], with reliability compared to heavy quark QCD (HQCD) in the large quark mass limit. 
Ref. [\refcite{Fodor:2015doa}] compared complex Langevin simulations of finite-density QCD with reweighting from positive ensembles, discussing its applicability across the parameter space. 
In Ref. [\refcite{Nagata:2018mkb}], motivated by the successful application of complex Langevin methods at high temperatures and in the heavy dense limit, the large $\mu/T$ regime was investigated using four-flavor staggered fermions on a finite lattice. 
The study encountered a singular-drift problem, which was circumvented using the mass deformation technique initially developed in the context of matrix models \cite{Ito:2016efb}. 
In Ref. [\refcite{Sexty:2019vqx}], significant progress was made in studying QCD thermodynamics at finite chemical potential using four-flavor staggered quarks with Symanzik improvement on the bosonic sector. 
However, the study was limited to finite $16^3 \times 8$ lattices and restricted to high temperatures and larger-than-physical pion masses, $m_\pi \in [500,~700]~\mathrm{MeV}$. 
Similarly, in Ref. [\refcite{Scherzer:2020kiu}], a heavy pion mass of $\sim 1.3~\mathrm{GeV}$ was used with a two-flavor Wilson fermion action to compute the transition temperature at finite $\mu$ using third-order Binder cumulants. 
Building on these works, complex Langevin simulations were performed on lattices $8^3 \times 16$ and $16^3 \times 8$ with four flavors and a lattice spacing of $a^{-1} \approx 4.7~\mathrm{GeV}$, where the authors observed a plateau in the quark number dependence on chemical potential, suggesting connections to the Fermi surface and color superconductivity. 
Ref. [\refcite{Jaeger:2021wiv}] used complex Langevin simulations to explore the QCD phase diagram over a large range of chemical potentials and temperatures. 
In the simulations the authors used two flavors of dynamical Wilson fermions with a pion mass of approximately 480 MeV with a spatial volume of $24^3$. 
They found that at the lowest temperature the fermion density remains zero until $m_N/3$, in line with the expectations from the Silver Blaze phenomenon.

In Refs. [\refcite{Tsutsui:2018jva,Ito:2018jpo,Ito:2020mys,Namekawa:2021qtg}], the authors first identified the $T$ - $\mu_q$ validity region of complex Langevin as $\mu_q/T = 5.2~{-}~7.2$ on $8^3 \times 16$ lattices and $\mu_q/T = 1.6~{-}~9.6$ on $16^3 \times 32$ lattices using four-flavor staggered quarks. 
They later explored the flavor number dependence for $N_f = 2,~2+1,~3$ using Wilson fermions, mapping the validity region as an indispensable first step toward pragmatic calculations of QCD using the complex Langevin method. In Ref. [\refcite{Attanasio:2022mjd}], ab initio finite lattice results for two-flavor QCD thermodynamics were presented at finite $\mu$ for pion masses larger than physical values, with plans to incorporate strange quarks, improved actions, and continuum limits in future studies. 
The authors emphasize the significance of these investigations for understanding the hot QCD phase diagram and searching for the elusive critical endpoint.

Exploring the QCD phase diagram to the extremes of the color superconductivity phase in low-temperature and high-density QCD, Ref. [\refcite{Tsutsui:2021bog}] studied the diquark-antidiquark operator on an $8^3 \times 128$ lattice using four-flavor staggered fermions. 
The results exhibited violent fluctuations when the Fermi surface coincided with the energy levels of the quarks.

Initially, with correct implementations only available in limited regions of the QCD phase diagram, simpler models capturing aspects of QCD phenomenology were explored. 
For example, the relativistic Bose gas at finite chemical potential, which exhibits a Silver Blaze phenomenon and a sign problem similar to lattice QCD, was studied in Refs. [\refcite{Aarts:2008wh,Aarts:2009hn}]. 
Ref. [\refcite{Katz:2016azl}] provides a detailed comparison of various algorithms designed to address the sign problem in the O(3) non-linear sigma model in $(1+1)$ dimensions. 
Further insights into the complex Langevin method in lattice QCD have been gained from detailed studies of spatially reduced low-dimensional QCD models, specifically in $(0+1)$ dimensions (Refs. [\refcite{Aarts:2010gr,Pawlowski:2013pje,Fujii:2017oti}]) and $(1+1)$-dimensional reweighting trajectories (Refs. [\refcite{Bloch:2017sfg,Schmalzbauer:2016pbg,Bloch:2015coa,Makino:2015ooa}]). 
Thirring model was studied in Ref. [\refcite{Pawlowski:2013gag}] for $(2+1)$-dimensions at finite density, with comparisons in the heavy dense limit.  
Additionally, random matrix theory, which shares critical features with QCD such as spontaneous chiral symmetry breaking and the complex fermion determinant, has been employed with stabilization techniques near the chiral limit in the cold and dense regimes of QCD. 
Notable studies include Refs. [\refcite{Mollgaard:2013qra,Mollgaard:2014mga,Nagata:2015ijn,Ichihara:2016uld,Nagata:2016alq,Bloch:2016jwt,Bloch:2017sex}]. 
An excellent summary of the progress, current challenges, and potential solutions in lattice QCD is provided by Nagata in Ref. [\refcite{Nagata:2021ugx}] (translated into English by Hanada and Itou).

QCD matter under rotation is of particular interest in diverse systems such as rapidly rotating compact stars, quark-gluon plasma in heavy-ion collisions, and low-energy nuclear physics. 
Recently in Ref. [\refcite{Azuma:2023pzr}], the complex Langevin method was applied to a rotating system at thermal equilibrium with a finite angular momentum chemical potential, studying the U$(N)$ matrix quantum mechanics, where rotations affect the phase structure. 
Using the Langevin method, Ref. [\refcite{Azuma:2024bbi}] studied that Polyakov loop effective potentials in finite-temperature large-$N$ gauge theories obey a scaling relation, illustrated across various classes of Yang-Mills theories with lattice and Fourier discretization under varying imaginary angular velocities. 
Rotating systems are ubiquitous, and experimental realizations in condensed matter, such as ultracold atomic systems with spin-orbit coupling, have opened exciting new directions. 
However, non-perturbative characterizations require complex bosons, leading to a numerical sign problem. 
Refs. [\refcite{Hayata:2014kra,Yamamoto:2015ura,Berger:2018xwy}] demonstrated the implementation of interacting $(2+1)$-dimensional spin-orbit coupled bosons at finite angular momentum, revealing their thermodynamic properties. 
Ref. [\refcite{Attanasio:2019plf}] investigated the effects of artificial spin-orbit coupling on the density equation of state for a bosonic system with two pseudo-spins, with potential future studies on the interplay between spin-orbit coupling and rotation. 
In Ref. [\refcite{Rammelmuller:2018hnk}], the thermodynamics of a unitary Fermi gas over a wide range of temperatures and spin polarizations was studied using the complex Langevin method, focusing on the density equation of state, magnetization, and magnetic susceptibility. 
Recently, Ref. [\refcite{Keithley:2024gpf}] applied complex Langevin simulations based on the coherent-state, imaginary-time path integral representation to rotating Bose-Einstein condensates, addressing the inherent sign problem and paving the way for broader applications to rotating systems.

The quest to uncover the underlying topology of gauge theories, particularly in addressing the strong $\mathcal{CP}$ problem in the Standard Model, continues. 
Non-perturbative investigations of field theories with a purely imaginary topological $\theta$-term have utilized the complex Langevin method to tackle this issue. 
Relevant studies include Refs. [\refcite{Bongiovanni:2014rna,Bongiovanni:2015kia,Hirasawa:2020bnl}]. 
For instance, in Ref. [\refcite{Hirasawa:2020bnl}], the method was applied to a two-dimensional $U(1)$ theory with a topological $\theta$-term on a punctured torus in Ref. [\refcite{Matsumoto:2021zjf}]. 
Building on this, the complex Langevin method was extended to a four-dimensional SU(2) gauge theory with a topological $\theta$-term. 
In this context, the topology freezing problem was addressed using open boundary conditions. 
However, $\mathcal{CP}$ symmetry remained broken at $\theta = \pi$, restricting investigations related to 't Hooft's matching condition. 
To overcome this, the authors proposed a novel approach involving stout smearing, which restored the topological properties required for further study.

Traditionally, quantum theories require Hermitian Hamiltonians to ensure real eigenvalues and unitary evolution. 
However, $\mathcal{PT}$-symmetric theories, despite having positive spectra, are less studied due to the lack of a consistent probabilistic interpretation in Monte Carlo. 
In the early 2000s, even before the advent of correctness criteria, Refs. [\refcite{Bernard:2001wh,Bernard:2004st}] explored $\mathcal{PT}$-symmetric scalar theories by computing equal-time one-point and two-point Green's functions in zero and one dimension, providing insights into the probabilistic interpretation of path integrals and paving the way for future non-Hermitian complex Langevin studies. 
In Ref. [\refcite{Pehlevan:2007eq}], a connection was established between various solutions of the Schwinger-Dyson equations and stationary distributions of complex Langevin equations to study different phases of such theories. 
Building on these studies, incorporating fermions into the action and ensuring that the action remains invariant under the appropriate supersymmetric transformations, dynamical supersymmetry breaking in low-dimensional quantum field theories with complex actions was investigated in Refs. [\refcite{Joseph:2019sof,Joseph:2020gdh,Kumar:2022fas,Kumar:2023nya,Kumar:2023spc}]. 
These studies included superpotentials featuring $\mathcal{PT}$-symmetry.
 
Matrix models provide an essential framework for studying non-perturbative phenomena, including spontaneous symmetry breaking, but they often suffer from the sign problem. 
Complex Langevin has emerged as an ideal method to address these challenges. 
In Ref. [\refcite{Basu:2018dtm}], Gross-Witten-Wadia (GWW) phase transitions were observed in large-$N$ unitary matrix models through complex Langevin simulations. 
In matrix models, the determinant or Pfaffian arising from integrating out fermions becomes complex, and its phase plays a crucial role in determining the vacuum, but investigation suffers from the {\it singular-drift} problem. 
In Refs. [\refcite{Ito:2016efb,Ito:2016hlj}], the authors proposed addressing the above problem by deforming the action with a fermion bilinear term, with the original system recovered through extrapolations with respect to the deformation parameter. 
They demonstrated the effectiveness of this approach by applying it to a simple SO(4)-invariant matrix model, capturing spontaneous symmetry breaking. 
Exploring the deformation dependence on the vacuum, in Ref. [\refcite{Ito:2017wun}] they later considered three different types of deformation in a large-$N$ matrix model, which undergoes spontaneous symmetry breaking due to the phase of the fermion determinant, and compared these to ensure consistency. 
This work was further extended to investigate spontaneous rotational symmetry breaking in dimensionally reduced six-dimensional super Yang-Mills models with Euclidean signatures in Refs. [\refcite{Anagnostopoulos:2017gos,Anagnostopoulos:2019ptt}].

The above studies paved the way for extensive investigations into the spontaneous breaking of SO(10) rotational symmetry in the Euclidean IKKT/type IIB matrix model, as reported in Refs. [\refcite{Anagnostopoulos:2020xai,Anagnostopoulos:2020ebo,Anagnostopoulos:2020cwo}]. 
The authors concluded that an SO(3)-symmetric vacuum is consistent with studies using the Gaussian expansion method. 
These investigations were extended to Lorentzian signatures, revealing an expanding $(3+1)$-dimensional universe with exponential behavior at early times and power-law behavior at later times, as detailed in Refs. [\refcite{Nishimura:2019qal,Nishimura:2020blu,Hirasawa:2021xeh,Hirasawa:2023lpb,Hirasawa:2022qzg}]. 
Further studies in Refs. [\refcite{Hatakeyama:2022ybs,Hatakeyama:2021ake}] explored scenarios where the spacetime signature changes dynamically from Euclidean at early times to Lorentzian at late times. 
Recently, Ref. [\refcite{Nishimura:2022alt}]  proposed the addition of a Lorentz-invariant mass term, leading to exponential expansion consistent with a Lorentzian signature at late times. 
This study also observed the expansion of only one of nine spatial directions, corresponding to $(1+1)$-dimensional spacetime, which was explained in terms of the bosonic sector of the theory. 
Ref. [\refcite{Anagnostopoulos:2022dak}] provided an extensive review of progress in numerical studies of the IKKT matrix model using both complex Langevin and Monte Carlo methods. 
In Ref. [\refcite{Kumar:2022giw}], the dynamical breaking of SO(10) symmetry in the IKKT matrix model was revisited using a class of supersymmetry-preserving mass deformations, inspired by the BMN mass deformations \cite{Berenstein:2002jq}.

Recently, efforts have been made to integrate complex Langevin with machine learning optimization processes. 
In Ref. [\refcite{Alvestad:2023jgl}], it was proposed to systematically learn optimal kernels by leveraging insights from boundary terms through reinforcement learning, illustrating real-time dynamics in $(1+1)$-dimensional systems. 
Ref. [\refcite{Greensite:2014cxa}] reported comparisons between complex Langevin and mean-field methods applied to effective Polyakov line models. 
Additionally, there has been significant progress in comparing and potentially unifying complex Langevin and Lefschetz thimble methods, as discussed in Refs. [\refcite{Aarts:2013fpa,Aarts:2014nxa,Sexty:2014dxa,Tanizaki:2015rda,Hayata:2015lzj,Fukushima:2015qza,Tsutsui:2015tua,Nishimura:2017eiu,Bluecher:2018sgj,Nishimura:2017vav}].

\section{Complex Langevin simulations of zero-dimensional models} 
\label{sec:Complex_Langevin_simulations_of_zero-dimensional_models}

Supersymmetric quantum field theories in zero-dimensional spacetime serve as an elegant yet fundamental platform to explore spontaneous SUSY breaking and other essential quantum phenomena. 
In this realm, the system simplifies the study of probability distributions over single variables, with a non-positive definite weight reflecting the effects of SUSY. 
Since zero-dimensional theories lack spacetime extent, the field configurations evolve in Langevin time without the complications introduced by spacetime and its symmetries. 
As a result, the dynamics of field configurations and their equilibration can be understood without interference from more complex phenomena, making this framework an ideal testing ground for probing foundational concepts like spontaneous symmetry breaking (SSB).

The lack of spatial dimensions also eliminates many of the technical difficulties encountered in higher-dimensional SUSY theories, such as renormalization, making these systems easier to analyze numerically. 
Moreover, zero-dimensional SUSY models often provide valuable toy models for gaining intuition about quantum field theories in higher dimensions, particularly in regards to the behavior of their vacuum structures, correlation functions, and the nature of symmetry breaking. 
In this context, spontaneous SUSY breaking in zero dimensions serves as a conceptual bridge to understanding these phenomena in more complex settings.

\subsection{Bosonic models}

The class of Euclidean scalar quantum field theories described in Ref. [\refcite{Bender:1997ps}] is particularly intriguing due to its non-Hermitian Hamiltonian with $\mathcal{PT}$-symmetry. 
The Lagrangian for these theories has the form
\beq
\label{eqn:bos-bender}
{\cal L} = \hf (\partial_\mu \phi)^2 + \hf m^2 \phi^2 + W(\phi).
\eeq
The potential
\beq
W(\phi) = - \frac{g}{(2 + \delta)} (i \phi)^{(2 + \delta)},
\eeq
introduces the non-Hermitian nature via the imaginary unit $i$, yet retains $\mathcal{PT}$-symmetry. 
Here, $\phi$ is a scalar field with mass $m$, $g$ is the coupling parameter, and $\delta$ is a real parameter.

The non-Hermitian aspect arises because the potential involves $(i \phi)^{(2 + \delta)}$, where $\delta > -2$. 
Despite being non-Hermitian, there is significant evidence that these theories exhibit real and bounded energy spectra, which is one of the hallmarks of $\mathcal{PT}$-symmetric systems. 
In $\mathcal{PT}$-symmetric quantum mechanics, although the Hamiltonian is not Hermitian, it is symmetric under the combined action of parity ($\mathcal{P}$) and time reversal ($\mathcal{T}$), allowing for the possibility of a real spectrum.

These $\mathcal{PT}$-symmetric field theories provide an interesting window into non-traditional quantum field theories, particularly because they challenge the conventional requirement of Hermiticity for ensuring real energy eigenvalues. 
The fact that these theories can have real, stable spectra suggests that the requirement of Hermiticity might be relaxed in certain contexts, provided $\mathcal{PT}$-symmetry is preserved. 
This leads to a wider class of potentially physically meaningful models, especially when investigating phenomena that involve complex potentials, such as in quantum field theories with non-trivial topological terms or in systems with a chemical potential, where the action becomes complex.

The significance of these investigations extends beyond pure theory. 
They provide a framework for understanding non-Hermitian systems, not just in elementary particle physics, but in fields such as condensed matter physics and even optics, where $\mathcal{PT}$
symmetric systems have been experimentally realized.

In the context of zero-dimensional quantum field theories, the Lagrangian presented above reduces to
\beq
\label{eq:0d-bosonic}
{\cal L} = \hf m^2 \phi^2 + W(\phi),
\eeq 
with $W(\phi) = - \frac{g}{N} (i \phi)^N$ and $N = 2 + \delta$. 
This form allows for a direct computation of the partition function and correlation functions due to the absence of spacetime coordinates.

For the massless case, where $m = 0$, the Euclidean action reduces to the potential term
\beq
S = W(\phi) = - \frac{g}{N} (i \phi)^N.
\eeq
This simplification leads to a partition function of the form
\bea
Z &=& \frac{1}{2 \pi} \int_{-\infty}^{\infty} d\phi~ e^{-S} \nn \\
&=& \frac{1}{2 \pi} \int_{-\infty}^{\infty} d\phi~ \exp\left( \left[ \frac{g}{N} (i \phi)^N \right] \right).
\eea
This expression captures the non-Hermitian nature of the action, but the $\mathcal{PT}$-symmetry may help ensure that the theory has a real, positive partition function, as previously observed in such systems.

Similarly, the $k$-point correlation functions can be computed using
\beq
G_k = \langle \phi^k \rangle = \frac{1}{Z} \frac{1}{2 \pi} \int_{-\infty}^{\infty} d\phi ~\phi^k ~ \exp \left( \left[ \frac{g}{N} (i\phi)^N \right] \right).
\eeq
These correlation functions provide valuable insights into the field dynamics and structure of the zero-dimensional theory, especially regarding how the non-Hermitian potential influences the behavior of the field. 

\begin{figure*}[htp]
\centering
{\includegraphics[width=.6\textwidth]{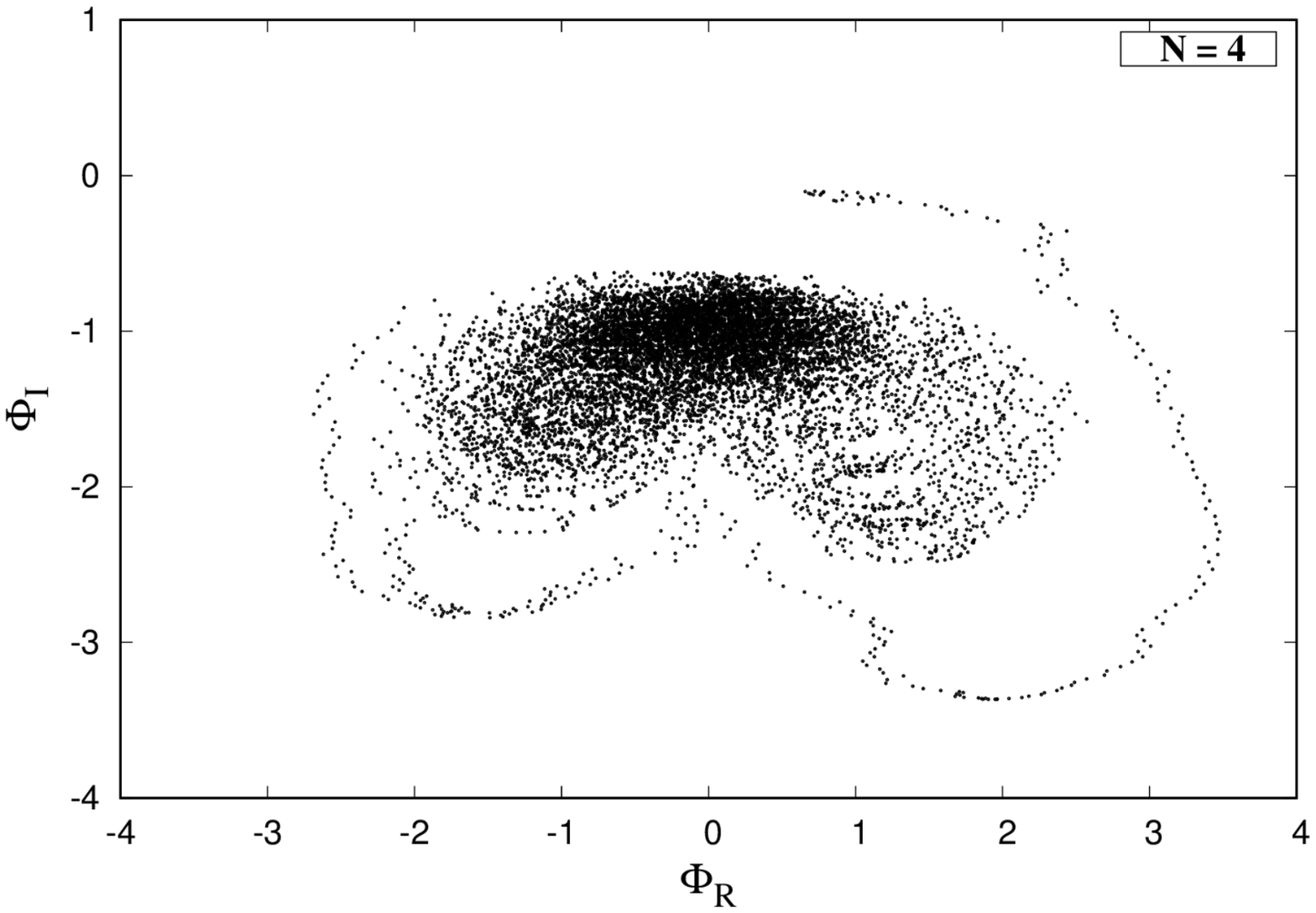}}
\caption{A scatter plot of the complexified field configurations in the $\phi_R - \phi_I$ plane for the zero-dimensional $ -(g/N) \left(i \phi \right)^N$ theory with coupling $g = 0.5$ and $N = 4$. The field trajectories during complex Langevin evolution are represented by black dots. After a large Langevin time, the field realizations form a cloud averaging around the point $(0.0,  - 1.163)$.}
\label{fig:n3-n4-clouds}
\end{figure*}

The one-point correlation function, $G_1$ takes the form \cite{Bender:1999ek}
\bea
G_1 = - \frac{i}{\sqrt{\pi}} \left(\frac{4 N}{g}\right)^{1/N}  { \Gamma\left(\frac{1}{N} + \hf\right) \cos \left(\frac{\pi}{N}\right) },
\eea
while the two-point correlation function, $G_2$, is given by
\bea
G_2 = \left(\frac{N}{g}\right)^{2/N} \frac{\Gamma\left(\frac{3}{N}\right)}{{\Gamma\left(\frac{1}{N}\right)}}{ \left[\sin^2\left(\frac{\pi}{N}\right) - 3 \cos^2\left(\frac{\pi}{N}\right)\right]}.
\eea
In a similar way, higher moments of $\phi$ can also be determined. 
Table \ref{tab:bosonic} provides a comparison between the results from complex Langevin simulations for $G_1$ and $G_2$ and their respective analytical expressions.
The analytical and numerical results match excellently and give confidence in using CLM.

\begin{table}[h]
\tbl{The correlation functions $G_1$ and $G_2$ obtained from complex Langevin simulations for zero-dimensional $- (g/N) (i \phi)^N$ theory for $N = 3, 4$. The parameters are: coupling constant $g = 0.5$, adaptive Langevin step size $\Delta \tau \leq 0.002$, thermalization steps $N_{\rm therm} = 10^6$, and generation steps $N_{\rm gen} = 10^7$. Measurements were taken with a gap of $1000$ steps. Numerically simulated values are compared with the exact results.}
{{\small \begin{tabular}{|l||	c |c |} 
	\hline
	$$ &   $~~~N=3$   &  $N=4~~~$  \\	
	\hline
	\hline
	$G_1^{\rm exact}$   & $0.0- i0.9185$   	 & $0.0 - i 1.1630$   	\\
	$G_1^{\rm cL}$  & $0.0112(121) - i 0.9183(41) $    & 	$0.0050(58) - i 1.1651(28)$		   \\
	\hline
	$G_2^{\rm exact}$   & $0.0 + i0.0$   	 & $-0.9560 + i 0.0$	    	\\
	$G_2^{\rm cL}$  & $-0.0001(16) - i0.0237(286)$  &	 $-0.9587 (45) -i 0.0122(158)$   \\
	\hline
	\end{tabular}} \label{tab:bosonic}}
\end{table}

\begin{figure*}[htp]
{\includegraphics[width=.49\textwidth,origin=c,angle=0]{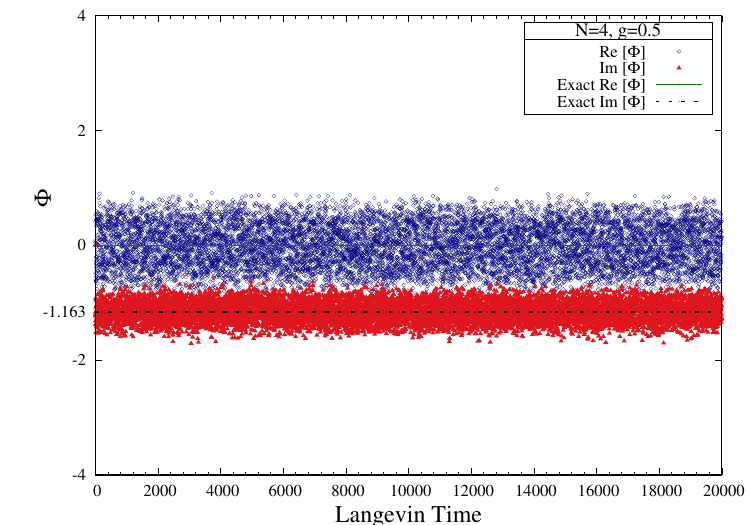}}
{\includegraphics[width=.49\textwidth,origin=c,angle=0]{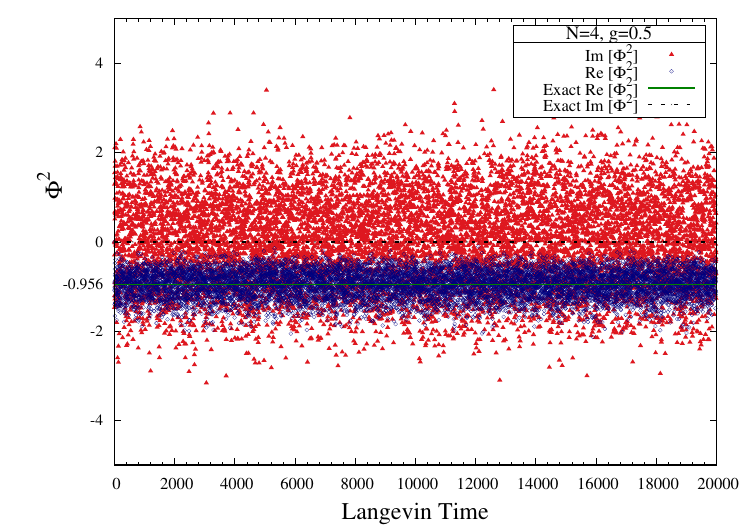}}
\caption{Langevin time history of one-point (Left) and two-point (Right) correlation functions. The model is $- (g/N) \phi^4$ theory with $g = 0.5$ and $N = 4$. Simulations were performed with adaptive Langevin step size $\Delta \tau \leq 0.002$, generation steps $N_{\rm gen} = 10^7$, and measurements were taken with a gap of $1000$ steps. The exact values are represented by solid and dashed lines.}
\label{fig:n4-history}
\end{figure*}

Figure \ref{fig:n3-n4-clouds} shows the complexified $\phi$ field configurations on the complex $\phi_R - \phi_I$ plane as it evolves in Langevin time. 
Figure \ref{fig:n4-history} shows the Langevin time history of $G_1$ and $G_2$ for the $N = 4$ case. 

\subsection{Exploring SUSY breaking}

The Lagrangian in Eq. \eqref{eqn:bos-bender} can be made supersymmetric by introducing the appropriate number of fermions. 
The two-dimensional supersymmetric Lagrangian takes the form \cite{Bender:1997ps}
\bea
{\cal L} &=& \hf \left(\partial_\mu \phi \right)^2 + \hf i \psib \slashed{\partial} \psi + \hf \psib W''(\phi) \psi + \hf \big [ W'(\phi) \big ]^2, \\
W(\phi) &=& - \frac{g}{(2 + \delta)} (i \phi)^{(2 + \delta)},
\eea
where $\psi$ and $\psib$ are Majorana fermions. 
This supersymmetric Lagrangian also breaks parity symmetry. 
We can ask if breaking of parity symmetry induces the breaking of supersymmetry in this model.

This question was addressed in Ref. [\refcite{Bender:1997ps}]. 
Through a perturbative expansion in $\delta$, the authors demonstrated that SUSY remains unbroken in this model. 
A non-perturbative exploration of SUSY breaking in this context, using the CLM, would be an interesting direction of study. 
Note that such a non-perturbative investigation via path integral Monte Carlo is not viable, as the action of this model can generally be complex.

Let us consider a zero-dimensional version of the supersymmetric model. 
In this case, we work with a general supersymmetric potential, $W(\phi)$.
The action is given by
\beq
S = \hf {\B}^2 + i \B W'(\phi) + \bar{\psi} W''(\phi) \psi,
\eeq
where $\phi$ is the bosonic field, $\psi$ and $\bar{\psi}$ are fermionic fields, and $\B$ is an auxiliary field. 
The prime denotes the derivative of the superpotential with respect to $\phi$.

In this theory, SUSY transforms fermionic fields into bosonic fields. 
There are two independent SUSY charges, $\Q$ and $\Qb$ in the theory. 
They correspond to $\cN = 2$ supersymmetry. 
We can obtain this action from dimensionally reducing one-dimensional SUSY QM with two supercharges.

The action is invariant under the following SUSY transformations.

For the $\Q$ charge
\beq
\label{eq:susy-transf-Q}
\Q \phi  = \psi, \quad \Q \psi = 0, \quad \Q \bar{\psi} = - i \B, \quad \Q \B = 0,
\eeq
and for the $\Qb$ charge
\beq
\label{eq:susy-transf-Qb}
\Qb \phi = - \bar{\psi}, \quad \Qb \bar{\psi} = 0, \quad \Qb \psi = - i \B, \quad \Qb \B = 0.
\eeq

The supercharges $\Q$ and $\Qb$ satisfy the following algebra:
\beq
\{ \Q, \Q \} = 0, \quad \{ \Qb, \Qb \} = 0, \quad \{ \Q, \Qb \} = 0.
\eeq

Additionally, the action can be written in a $\Q$- or $\Q \Qb$-exact form, specifically
\bea
S &=& \Q \psib \left( \frac{i}{2} \B - W' \right) \nn \\
&=& \Q \Qb \left( \hf \psib \psi + W(\phi) \right),
\eea
which demonstrates that the action is invariant under both SUSY charges
\beq
\Q S = 0, \quad \Qb S = 0.
\eeq

The auxiliary field $\B$ has been introduced for the off-shell completion of the SUSY algebra. 
We can be eliminated it with the help of the equation of motion:
\beq
\B = - i W'(\phi).
\eeq

The partition function for this model is
\bea
Z &=& \frac{1}{2 \pi} \int d\B d\phi d\psi d\psib \ e^{-S} \nn \\
&=& \frac{1}{2 \pi} \int d\B d\phi d\psi d\psib \exp \left[ - \left( \hf {\B}^2 + i \B W'(\phi) + \psib W''(\phi) \psi \right) \right].
\eea

After completing the square and integrating out $\B$ we can simplify the partition function to
\beq
Z =  \frac{1}{\sqrt{2 \pi}} \int d\phi d\psi d\psib \ \exp \left[ - \left( \hf  {W'(\phi)}^2 + \psib W''(\phi) \psi \right) \right].
\eeq

Finally, integrating out the fermions yields
\beq
Z = - \frac{1}{\sqrt{2 \pi}} \int d\phi \ W''(\phi) \ \exp \left[ -\hf {W'(\phi)}^2 \right].
\eeq

If SUSY is broken, the partition function vanishes. 
In such cases, the expectation values of observables, when normalized by the partition function, may become ill-defined.

The expectation value of the auxiliary field plays a key role in investigating SUSY breaking. 
It can be calculated as:
\bea
\langle \B \rangle &=& \frac{1}{Z} \frac{1}{2 \pi} \int d\B d\phi d\psi d\psib ~ \B ~ e^{-S} \nn \\
&=& - \frac{1}{Z} \frac{i}{\sqrt{2 \pi}} \int_{-\infty}^{\infty} d\phi \ \frac{\partial}{\partial \phi} \ \exp \left[ -\hf {W'(\phi)}^2 \right].
\eea
For SUSY broken case, the normalized expectation value of $\B$ becomes indeterminate, leading to a form like $0/0$, signaling issues in determining the value directly.

To address this, we can introduce an external field, which will allow us to detect spontaneous SUSY breaking by lifting the potential ambiguity. 
This approach is similar to detecting the spontaneous breaking of other symmetries: an external field is applied to resolve ground-state degeneracy, and once the symmetry-breaking behavior is identified, the external field is set to zero in the thermodynamic limit.

For example, in the Ising model, a magnetic field is introduced as the external field to detect spontaneous magnetization, with the spin operator acting as the order parameter. 
In the same spirit, we can detect SUSY breaking by introducing an external field in this model. 

One method to introduce such an external field is by modifying the boundary conditions for the fermions, replacing periodic or anti-periodic boundary conditions with twisted boundary conditions. 
This external field approach offers a way to probe spontaneous SUSY breaking.

\subsubsection{Theory on one-site lattice}

We can interpret the zero-dimensional theory as the one coming from the dimensional reduction of a one-dimensional SUSY QM model. 
The one-dimensional action is integrated over a compactified time circle of circumference \(\beta\) in Euclidean space.
It has the form
\beq
S  =  \int_0^\beta d\tau \left[ \ \hf {\B}^2 + i\B \left(\dot{\phi} + W'\right) + \psib \left( \dot{\psi} + W'' \psi \right) \ \right],
\eeq
where $\dot{\phi}$ and $\dot{\psi}$ denote derivatives with respect to the Euclidean time $\tau \in [0,~\beta]$. 
The above action is invariant under $Q$ but not under $\Qb$.

To discretize the theory on a one-dimensional lattice with $N_\tau$ sites, we use finite differences for the derivatives. 
The lattice action then becomes
\bea
S &=& \sum_{n=0}^{N_\tau - 1} \ \Bigg[\ \hf {\B}^2(n) +  i \B(n) \Big( \phi (n+1) - \phi (n) + W'(\phi (n)) \Big) \nn \\
&&~~~~~~~~~~~~~ + \psib(n) \ \Big( \psi(n+1) - \psi(n)+ W''(\phi (n)) \ \psi(n) \Big) \ \Bigg],
\eea
where $n$ represents the lattice site. 
Note that the fields and coupling parameters have been rescaled so that the action is expressed in dimensionless variables. 
This lattice action preserves one of the supercharges, $\Q$. The $\Qb$ SUSY is not maintained when $N_\tau \geq 2$.

For the simplest case with one lattice point, i.e., $N_\tau = 1$, the action simplifies to
\bea
S &=& \hf {\B}^2(0) +  i \B(0)  \Big( \phi (1) - \phi (0) + W'(\phi (0))  \Big) \nn \\ 
&&~~~~~~~~~~~~~ + \psib(0)  \Big(  \psi(1) - \psi(0)+ W''(\phi (0)) \ \psi(0) \Big),
\eea
where $\phi(1)$ and $\psi(1)$ are determined by the boundary conditions. For periodic boundary conditions, we have
\bea
	\phi(1)  =  \phi(0),&&~~
	\psi(1) = \psi(0), \nn \\
	\psib(1) = \psib(0),&&~~
	\B(1) =  \B(0).
\eea
The action reduces to
\beq
S = \hf {\B}^2 + i \B W' + \psib W'' \psi.
\eeq

Thus, we see that the action of the zero-dimensional supersymmetric model with $\cN = 2$ SUSY is equivalent to the dimensional reduction of a SUSY quantum mechanics model with periodic boundary conditions.

\subsubsection{Boundary conditions with a twist}

Instead of using periodic boundary conditions, we introduce twisted boundary conditions for the fermions, to address the indefinite form of the expectation values encountered previously. 
Twisted boundary conditions have been explored in the context of supersymmetric models by Kuroki and Sugino in Refs. [\refcite{Kuroki:2009yg,Kuroki:2010au}]. 
Under twisted boundary conditions, the field configurations satisfy:
\bea
	\phi(1)  =  \phi(0),&&~~
	\psi(1)  = e^{i\alpha} \psi(0), \\
	\psib(1)  = e^{- i\alpha} \psib(0),&&~~
	\B(1)  =  \B(0).
\eea

The action takes the form:
\beq
S_\alpha =  \hf {\B}^2 + i \B W' + \psib \Big( e^{i \alpha} - 1 + W'' \Big) \psi.
\eeq

The introduction of the twist parameter $\alpha$ results in a soft breaking of SUSY, as evidenced by
\beq
Q S_\alpha = -i \Qb S_\alpha = \psib \left( e^{i \alpha} - 1 \right) \psi,
\eeq 
and SUSY is recovered in the limit $\alpha \to 0$.

The partition function for this twisted model is given by
\bea
\label{eq:Z-twist}
Z_\alpha &=& \frac{1}{2 \pi} \int d\B d\phi d\psi d\psib \ e^{ -S_\alpha } \nn \\
&=& - \frac{1}{\sqrt{2 \pi}} \int d\phi ~ \Big( e^{i \alpha} - 1 + W'' \Big)  \exp \left[ - \hf W'^2 \right].
\eea

The expectation value of the auxiliary field $\B$ with twisted boundary conditions is given by
\bea
\langle \B \rangle_\alpha &=&  \frac{1}{Z_\alpha} \frac{1}{2 \pi} \int d\B d\phi d\psi d\bar{\psi} \ \B \ e^{ -S_\alpha }  \nn \\
&=&  \frac{1}{Z_\alpha} \frac{i}{\sqrt{2 \pi}} \int d\phi \ W' \Big( e^{i \alpha} - 1 + W'' \Big) \exp \left[ - \hf W'^2 \right].
\eea
This expression is now well-defined, as the twist parameter $\alpha$ serves as a regularization parameter, resolving the indefinite form $\langle \B \rangle = 0/0$ encountered with periodic boundary conditions. 
In the limit $\alpha \to 0$, a vanishing expectation value of $\langle \B \rangle_\alpha$ would suggest that SUSY is preserved, while a non-zero value indicates SUSY breaking.

The effective action with twisted boundary conditions is given by
\beq
S_\alpha^{~\text{eff}} = \hf  W'^2 -  \text{ln} \left[ e^{i \alpha} - 1 + W'' \right],
\eeq
and the gradient of this effective action, which is required for the complex Langevin method, is
\bea
\label{eq:clm-drift}
\frac{\partial S_\alpha^{~\text{eff}}}{\partial \phi} &=& \frac{\partial}{\partial \phi}  \left( \hf  W'^2 - \ln \left[ e^{i \alpha} - 1 + W'' \right] \right) \nn \\
&=& W' W'' - \frac{W'''}{\Big ( e^{i \alpha} - 1 + W''\Big)}.
\eea

We can explore various quantum mechanics models depending on the choice of the potential. 
Examples include models with real or complex double-well potentials, general polynomial-type potentials, and scarf-type potentials. 
In the following section, we will focus on models featuring $\mathcal{PT}$-symmetric potentials.

\subsection{$\mathcal{PT}$-symmetric models}

Let us consider the superpotential
\beq
W(\phi) = - \frac{g}{(2 + \delta)} (i \phi)^{(2 + \delta)}.
\eeq
This is the same as the one we looked at earlier, for the case of bosonic models.

The twisted partition function has the form
\bea
Z_\alpha &=& -\frac{1}{\sqrt{2 \pi}} \int_{-\infty}^\infty d\phi \ \Big( e^{i \alpha} - 1 + W'' \Big) \exp\left[ - \hf W'^2 \right]  \nn \\
&=& -\frac{1}{\sqrt{2 \pi}} \int_{-\infty}^\infty d\phi \ \Big( e^{i \alpha} - 1 + g (1 + \delta) (i \phi)^{\delta} \Big) \exp \left[ \hf g^2 (i \phi)^{2(1+\delta)} \right] .
\eea

The auxiliary field $\B$ has the following expectation value
\bea
\langle \B \rangle_\alpha &=&- \frac{1}{Z_\alpha} \frac{1}{\sqrt{2 \pi}} \int_{-\infty}^\infty d\phi \ (-iW') \Big( e^{i \alpha} - 1 + W'' \Big) \exp\left[ - \hf W'^2 \right]  \nn \\
&=& \frac{1}{Z_\alpha} \frac{1}{\sqrt{2 \pi}} \int_{-\infty}^\infty d\phi \  g  (i \phi)^{1 + \delta} \Big( e^{i \alpha} - 1 + g (1 + \delta) (i\phi)^\delta \Big) \nn \\
&& ~~~ \times \exp \left[ \hf g^2 (i\phi)^{2(1+\delta)} \right] .~~
\eea

Let us analyze this model for various integer values of $\delta$ and investigate whether SUSY is broken or preserved. 

For the case $\delta = 0$, we can easily perform analytical evaluations. 
The partition function is
\bea
Z_\alpha [\delta = 0] &=& -\frac{1}{\sqrt{2 \pi}} \int_{-\infty}^\infty d\phi \ \Big( e^{i \alpha} - 1 + g \Big) \ \exp \left[ - \hf g^2 \phi^2 \right]  \nn \\
&=& -\frac{1}{\sqrt{2 \pi}} \Big( e^{i \alpha} - 1 + g \Big) \sqrt{\frac{2 \pi}{g^2}}.
\eea

Once we turn the external field off ($\alpha \to 0$) we obtain a non-zero value for the partition function:
\beq
Z_{\alpha=0} [\delta = 0] = -\frac{1}{\sqrt{2 \pi}} g \sqrt{\frac{2 \pi}{g^2}} = -1.
\eeq
This implies that SUSY is preserved in the system.

We also have
\bea
\langle \B \rangle_\alpha   [\delta = 0] &=&  \frac{1}{Z_\alpha} \frac{1}{\sqrt{2 \pi}} \int_{-\infty}^\infty d\phi \ ig \phi \Big( e^{i \alpha} - 1 + g \Big) ~ \exp \left[ -\hf g^2 \phi^{2} \right] \nn \\
&=&-\frac{ ig \int_{-\infty}^\infty d\phi \ \phi \Big( e^{i \alpha} - 1 + g \Big) \exp \left[ - \hf g^2 \phi^2 \right] }{ \Big( e^{i \alpha} - 1 + g \Big) \sqrt{\frac{2 \pi}{g^2}}} \nn \\
&=& -\frac{i g \int_{-\infty}^\infty d\phi \ \phi \ \exp \left[ - \hf g^2 \phi^2 \right]}{ \sqrt{\frac{2 \pi}{g^2}}} = 0,
\eea
implying that SUSY is preserved in the theory when $\delta = 0$.

When $\delta = 2$, we have the twisted partition function:
\bea
Z_\alpha [\delta = 2] &=&- \frac{1}{\sqrt{2 \pi}} \int_{-\infty}^\infty d\phi \ \Big( e^{i \alpha} - 1 - 3g  \phi^{2} \Big) \exp\left[ -\hf g^2 \phi^{6} \right] \nn \\
&=& -\frac{\Big( e^{i \alpha} - 1\Big)}{\sqrt{2 \pi}} \int_{-\infty}^\infty d\phi \ \exp\left[ -\hf g^2  \phi^{6} \right]  \nn \\
&& ~~~ ~~~ +  \frac{3g}{\sqrt{2 \pi}} \int_{-\infty}^\infty d\phi \ \phi^{2} \exp\left[ -\hf g^2  \phi^{6} \right].
\eea

Once the external field is turned off, $\alpha \to 0$, we get a non-zero partition function
\beq
Z_{\alpha = 0} [ \delta = 2 ] =  \frac{3g}{\sqrt{2 \pi}} \int_{-\infty}^\infty d\phi \ \phi^{2} \exp\left[ -\hf g^2  \phi^{6} \right] = 1.
\eeq
Again, suggesting that SUSY is preserved in this system. 
The expectation value of the $\B$ field has the form
\beq
\langle \B \rangle_\alpha [\delta = 2] =   - \frac{ig}{Z_\alpha\sqrt{2 \pi}} \int_{-\infty}^\infty d\phi \ \phi^{3} \left( e^{i \alpha} - 1 - 3g \phi^2 \right) \exp\left[ -\hf g^2 \phi^{6} \right] = 0.
\eeq
This value also confirms that SUSY is preserved when $\delta = 2$. 
We can perform similar calculations for the case $\delta = 4$ and show that SUSY is preserved in the theory.

We can simulate this model with the $\delta$-potential using complex Langevin dynamics. 
The drift term coming from the $\delta$-potential is
\bea
\frac{\partial S_\alpha^{~\text{ eff}}}{\partial \phi} &=& \frac{\partial}{\partial \phi}  \left[ \hf  W'^2 - \ln \left( e^{i \alpha} - 1 + W'' \right) \right] \nn \\
&=& W' W'' - \frac{W'''}{\Big ( e^{i \alpha} - 1 + W''  \Big)} \nn \\
&=& - i g^2 (1+ \delta) (i \phi)^{2\delta +1} - \frac{i g \delta (1+\delta)  (i \phi)^{\delta - 1}}{\Big(   e^{i \alpha} - 1 + g (1+\delta) (i \phi)^{\delta}  \Big)}.
\eea

The results from complex Langevin simulations are tabulated in Table \ref{tab:delta_1p0_3p0} and \ref{tab:delta_2p0_4p0}. 
They clearly show that the expectation value of the auxiliary field, $\langle \B \rangle_\alpha$, goes to zero in the limit $\alpha \to 0$. 
Thus, we can conclude that SUSY is unbroken in the model with a $\delta$-potential for values of $\delta = 1, 2, 3, 4$.

\begin{table}[h]
\tbl{The expectation values for the auxiliary field obtained using complex Langevin simulations for models with superpotential $W'(\phi) = -ig (i \phi)^{(1+\delta)}$, with coupling $g = 0.5$ and $\delta = 1$ and $3$.}
{\begin{tabular}{| c | c | c | c |} 
		\hline
		$ ~~~~~\delta~~~~~$  & $~~~~ \alpha ~~~~$ &  $ \langle \B \rangle |_{\alpha} $  & $~~~~$ SUSY $~~~~$ \\
		\hline
		\hline
		\multirow{7}{*}{$1.0$}
		&0.4	& $-0.2498 (224)   - i 0.2109 (487) $ & \multirow{7}{*}{Preserved}\\ \cline{2-3}
		&0.5	& $-0.2580 (202)   - i 0.2998 (450) $ & \\ \cline{2-3}
		&0.6	& $-0.2617 (186)   - i 0.3504 (420)$ & \\ \cline{2-3}
		&0.7	& $-0.2726 (172)   - i 0.3719 (403) $ & \\ \cline{2-3}
		&0.8	& $-0.2858 (160)   - i 0.3998 (391) $& \\  \cline{2-3}
		&0.9	& $-0.3113 (149)   - i 0.3978 (391) $ & \\  \cline{2-3}	
		& $\alpha \rightarrow 0$ &	 $ - 0.2433(2213) + i 0.0742(5080)  $  & \\
		\hline

		\multirow{7}{*}{$3.0$}
		&0.3	& $ 0.0567 (32) + i 0.4452 (566) $ &  \multirow{7}{*}{Preserved} \\ \cline{2-3}
		&0.4	& $ 0.0738 (32) + i 0.4544 (538) $ & \\ \cline{2-3}
		&0.5	& $ 0.0870 (34) + i 0.4387 (475) $ &   \\  \cline{2-3}
		&0.6	& $ 0.0961 (43) + i 0.4284 (416) $ & \\  \cline{2-3}
		&0.7	& $ 0.1034 (53) + i 0.3946 (441) $ & \\  \cline{2-3}
		&0.8	& $ 0.1027 (64) + i 0.3539 (398) $ & \\  \cline{2-3}	
		& $\alpha \rightarrow 0$ &	 $ 0.0054(311) + i 0.3625(4025)  $ & \\
		\hline
	\end{tabular} \label{tab:delta_1p0_3p0}}
\end{table}

\begin{table}[]
\tbl{The expectation values $\langle \B \rangle_\alpha$ obtained using complex Langevin simulations for the models with superpotential $W'(\phi) = -ig (i \phi)^{(1+\delta)}$, with coupling $g = 0.5$ and $\delta = 2$ and $4$.}
{\begin{tabular}{| c | c | c | c |} 
		\hline
		$ ~~~~~~\delta ~~~~~~$  & $~~~~ \alpha ~~~~$ &  $ ~~~~~~~~~~~~~~~~~~~\langle \B \rangle |_{\alpha}~~~~~~~~~~~~~~~~ $  & $~~~~$ SUSY $~~~~$ \\
		\hline
		\hline
		\multirow{7}{*}{$2.0$}
		&0.05	& $ 0.0014 (36)  - i 0.0609 (1416) $ & \multirow{7}{*}{Preserved}\\ \cline{2-3}
		&0.1	& $ 0.0102 (50)  - i 0.1986 (1101) $ & \\ \cline{2-3}
		&0.2	& $ 0.0079 (80)  - i 0.0679 (1004) $ & \\ \cline{2-3}
		&0.4	& $ 0.0134 (96)  - i 0.0627 (701) $ & \\ \cline{2-3}
		&0.6	& $ 0.0079 (120) - i 0.0208 (655)$ & \\  \cline{2-3}
		&0.8	& $-0.0068 (126) + i 0.0294 (595) $ & \\  \cline{2-3}	
		& $\alpha \rightarrow 0$ &	 $ 0.0019(84) - i 0.1423(1932)$  & \\
		\hline
		
		\multirow{7}{*}{$4.0$}
		&0.05	& $-0.0005 (20) - i 0.0155 (1257) $ & \multirow{7}{*}{Preserved}\\ \cline{2-3}
		&0.1	& $-0.0017 (37) - i 0.0435 (1043) $ & \\ \cline{2-3}
		&0.2	& $ 0.0059 (48) + i 0.0787 (817) $ & \\ \cline{2-3}
		&0.4	& $ 0.0016 (64) + i 0.0108 (648) $ & \\ \cline{2-3}
		&0.6	& $ 0.0132 (70) + i 0.0761 (526)$ & \\  \cline{2-3}
		&0.8	& $ 0.0063 (68) + i 0.0258 (418) $ & \\  \cline{2-3}	
		& $\alpha \rightarrow 0$ &	 $ -0.0018(48) - i 0.0092(1712)$  & \\
		\hline
	\end{tabular} \label{tab:delta_2p0_4p0}}
\end{table}

\subsection{How reliable are complex Langevin simulations?}

In this section, we show how we can justify the simulation methods by introducing two recent approaches for validating their accuracy. 
First method is based on the Fokker-Planck equation as a criterion for correctness, and the second exploits the nature of the decay of the probability distribution of the magnitude of the drift term. 

\subsubsection{Justification using Fokker-Planck equation}

In this section, we examine the validity of the simulations by applying criteria based on the behavior of holomorphic observables and the Langevin operator. 

The evolution of holomorphic observables $\cO[\phi, \tau]$ in the complex Langevin method is governed by the equation:
\beq
\frac{\partial \cO[\phi, \tau]}{\partial \tau} = \widetilde{L} \cO[\phi, \tau], 
\eeq
where $\widetilde{L}$ is the Langevin operator defined as:
\beq
\widetilde{L} = \left[ \frac{\partial}{\partial \phi}  - \frac{\partial}{\partial \phi} S[\phi] \right] \ \frac{\partial}{\partial \phi}.
\eeq

At equilibrium, assuming that it has been reached, the observables become independent of $\tau$, which implies:
\beq
C_\cO \equiv \langle \widetilde{L} \cO[\phi] \rangle = 0.
\eeq
This condition serves as a criterion for the correctness of the complex Langevin method. 
The efficacy of this criterion has been studied in various models, as detailed in Refs. [\refcite{Aarts:2009uq,Aarts:2011ax,Aarts:2013uza}].

\begin{figure*}[htp]
	
	{\includegraphics[width=.49\textwidth,origin=c,angle=0]{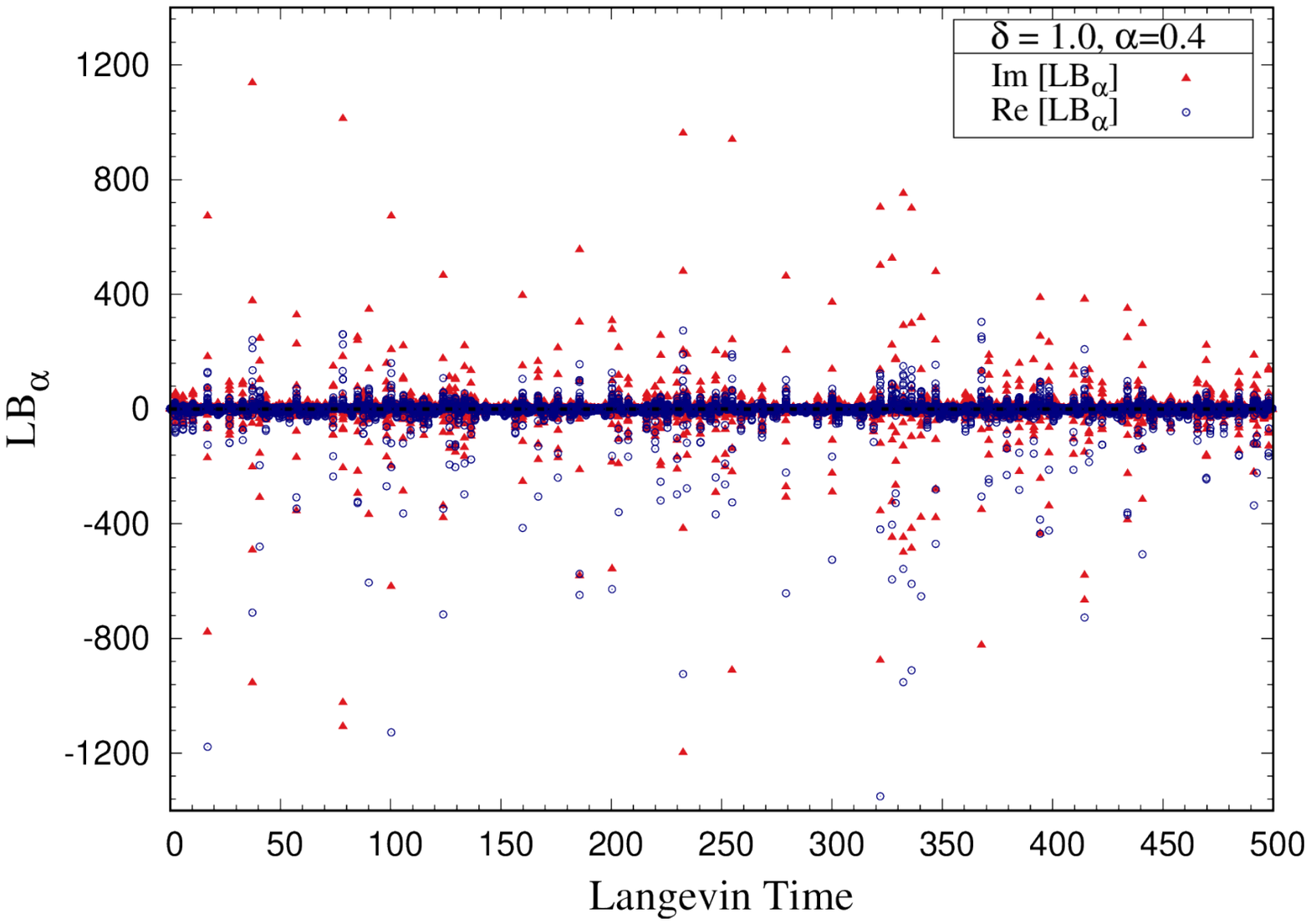}}
	{\includegraphics[width=.49\textwidth,origin=c,angle=0]{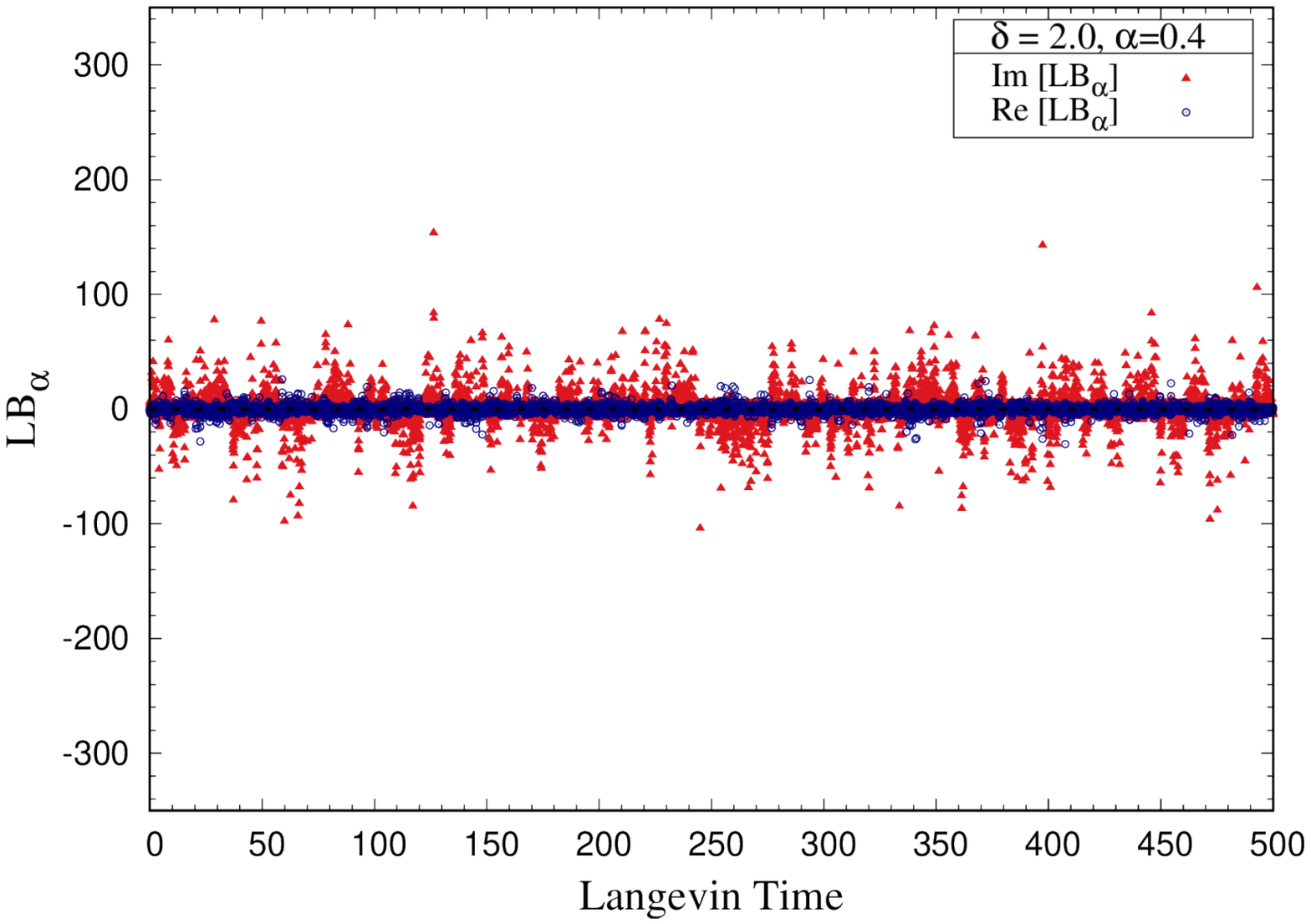}}
	
	{\includegraphics[width=.49\textwidth,origin=c,angle=0]{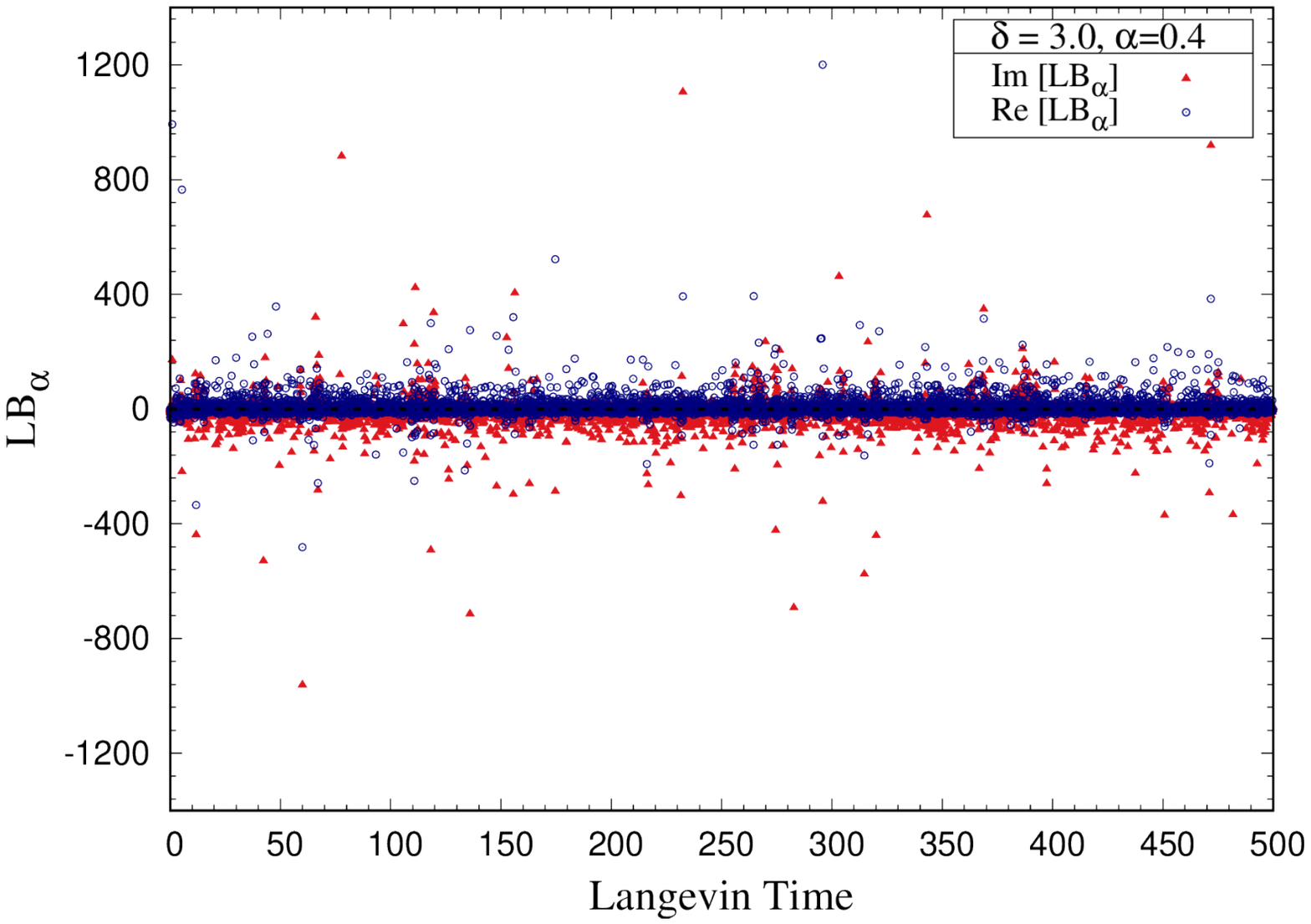}}
	{\includegraphics[width=.49\textwidth,origin=c,angle=0]{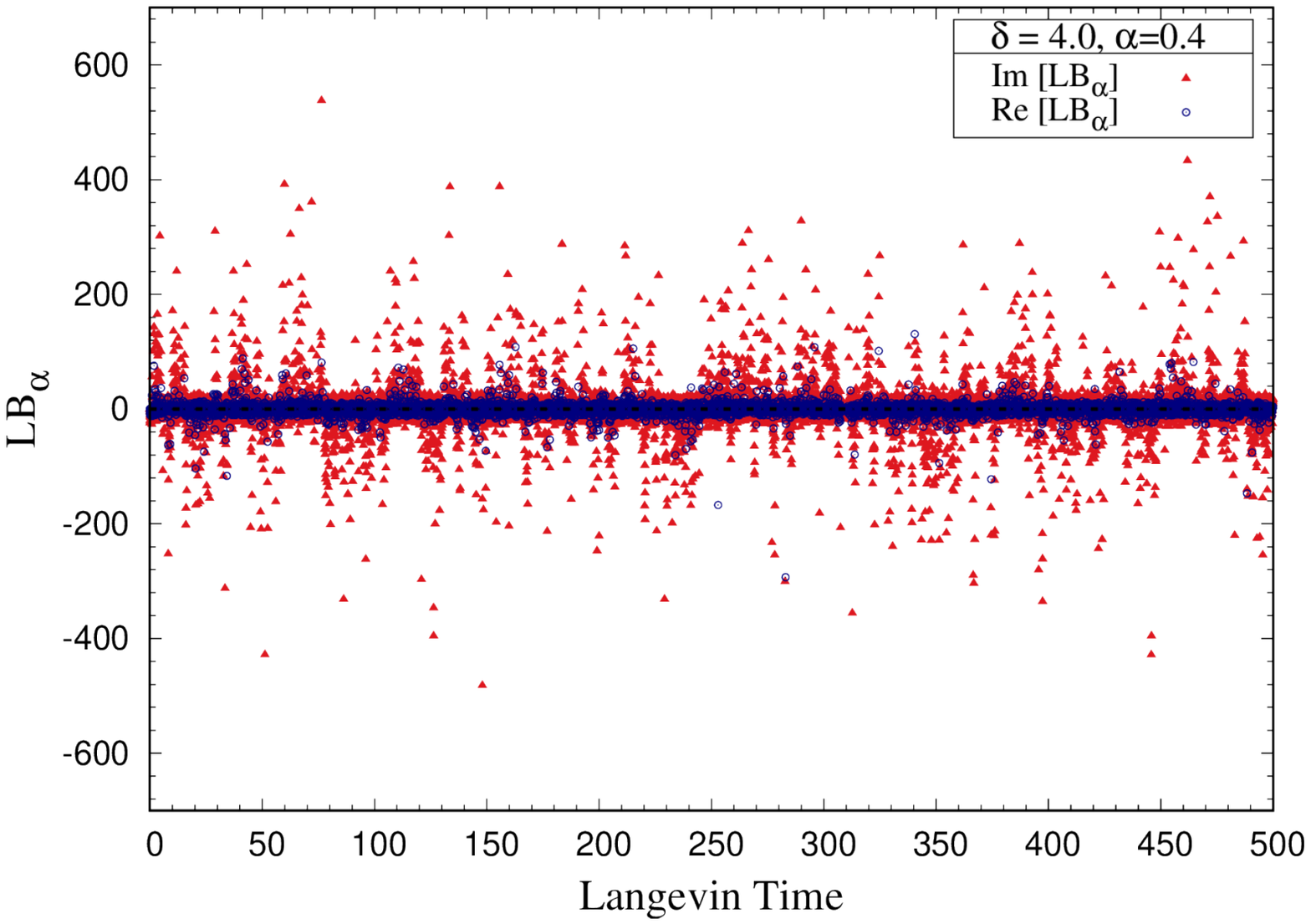}}
	
	\caption{The Langevin time history of $\widetilde{L} \B$ for regularization parameter, $\alpha = 0.4$. The simulations were performed for superpotential $W'(\phi) = -ig (i \phi)^{(1 + \delta)}$ with coupling $g = 0.5$. The delta values are: $\delta = 1$ (Top-Left), $\delta = 2$ (Top-Right), $\delta = 3$ (Bottom-Left) and $\delta = 4$ (Bottom-Right). The simulations were performed using adaptive Langevin step size $\Delta \tau \leq 5\times10^{-5}$, generation steps $N_{\rm gen} = 10^7$, and measurements were taken with a gap of $500$ steps. The exact result at equilibrium distribution is $\widetilde{L}\B = 0$.}
\label{fig:LO_delta}
	
\end{figure*}

For the observable $\cO$ corresponding to the auxiliary field $\B$, the Langevin operator $\widetilde{L}$ acts as follows:
\bea
\widetilde{L} \cO = \widetilde{L} \B =  -i W''' + i W' {W''}^2 - \frac{i W'' W'''}{\Big ( e^{i \alpha} - 1 + W''\Big)}.
\eea
Figure \ref{fig:LO_delta} shows the Langevin history of the correctness criterion, $\widetilde{L} \B$, with a regularization parameter $\alpha = 0.4$ for the superpotential $W' = -ig (i \phi)^{1 + \delta}$. 
The simulation data fluctuates around the expected value, suggesting the correctness of the method.

Tables \ref{tab:delta_LO_1p0_3p0} and \ref{tab:delta_LO_2p0_4p0} present the simulated values of $\langle \widetilde{L} \B \rangle_\alpha$ for the superpotential $W'(\phi) = -ig (i \phi)^{(1+\delta)}$ with coupling parameter $g = 0.5$ and various regularization parameters $\alpha$.

\begin{table}[]
\tbl{The simulated values of $\widetilde{L} \B_\alpha$ for models with superpotential $W'(\phi) = -ig (i \phi)^{(1+\delta)}$and coupling $g = 0.5$. The values of delta are: $\delta = 1, 3$.}
{\begin{tabular}{| c | c | c |} 
		\hline
		$~~~~~~\delta~~~~~~$  & $~~~~~~~\alpha~~~~~~~$  & $~~~~~~~~~~~~~~~~~~~~~~~\langle \widetilde{L} \B \rangle |_\alpha~~~~~~~~~~~~~~~~~~~~~$  \tabularnewline
		\hline 
		\hline 
		\multirow{4}{*}{1.0}
		& 0.4  & $-0.6263 (3592)    + i 0.0042 (3062) $  \tabularnewline
		\cline{2-3} 
		& 0.5  & $-0.1442 (2127)    + i 0.0202 (1752)$  \tabularnewline
		\cline{2-3}
		& 0.6  & $-0.0239 (1517)    + i 0.0400 (1375)$  \tabularnewline
		\cline{2-3} 
		& 0.7  & $~~0.0198 (1192)    + i 0.0387 (1171)$  \tabularnewline
		\cline{2-3} 
		& 0.8  & $-0.0107 (1169)    + i 0.0494 (988)$  \tabularnewline
		\cline{2-3} 
		& 0.9  & $-0.0401 (990)     + i 0.0104 (915)$  \tabularnewline
		\cline{2-3} 
		& $\alpha \rightarrow 0$ 
		& $ -1.2716 (2.421) - i 0.1173 (2.122) $  \tabularnewline
		\hline 
		\multirow{4}{*}{3.0}
		& 0.3  & $ 0.1846 (5176)    + i 0.1366 (3738) $  \tabularnewline
		\cline{2-3} 
		& 0.4  & $-0.3282 (1845)    + i 0.0443 (3164)$  \tabularnewline
		\cline{2-3}
		& 0.5  & $-0.2215 (1856)    + i 0.1869 (2377)$  \tabularnewline
		\cline{2-3} 
		& 0.6  & $-0.2046 (1456)    + i 0.2870 (1969)$  \tabularnewline
		\cline{2-3} 
		& 0.7  & $ 0.0022 (1476)    + i 0.2841 (2076)$  \tabularnewline
		\cline{2-3} 
		& 0.8  & $-0.0483 (1412)    + i 0.1976 (1960)$  \tabularnewline
		\cline{2-3} 
		& $\alpha \rightarrow 0$
		& $-0.3031 (2.181) - i 0.2210 (2.335)  $   \tabularnewline
		\hline 	
	\end{tabular} \label{tab:delta_LO_1p0_3p0}}
\end{table}

\begin{table}[]
\tbl{The simulated values of $\widetilde{L} \B_\alpha$ for models with superpotential $W'(\phi) = -ig (i \phi)^{(1+\delta)}$and coupling $g = 0.5$. The values of delta are: $\delta = 2, 4$.}
{\begin{tabular}{| c | c | c |}
		\hline 
		$~~~~~~\delta~~~~~~$  & $~~~~~~~\alpha~~~~~~~$  & $~~~~~~~~~~~~~~~~~~~~~~~\langle \widetilde{L} \B \rangle |_\alpha~~~~~~~~~~~~~~~~~~~~~$  \tabularnewline
		\hline 
		\hline 
		\multirow{4}{*}{2.0} 
		& 0.05 & $ 0.0036 (49)  - i 0.1572 (1315)$  \tabularnewline
		\cline{2-3}  
		& 0.1 	& $ 0.0082 (94)  - i 0.2145 (1273)$  \tabularnewline
		\cline{2-3}  
		& 0.2  & $ 0.0113 (156) - i 0.1480 (1359)$  \tabularnewline
		\cline{2-3} 
		& 0.4  & $ 0.0066 (246) - i 0.1409 (1300)$  \tabularnewline
		\cline{2-3} 
		& 0.6  & $-0.0014 (312) - i 0.1029 (1280)$  \tabularnewline
		\cline{2-3} 
		&  0.8 & $-0.0023 (348) - i 0.1132 (1245)$  \tabularnewline
		\cline{2-3} 
		& $\alpha \rightarrow 0$ 
		& $  0.0034 (142) - i 0.1906 (2223) $  \tabularnewline
		\hline 
		\multirow{4}{*}{4.0}
		& 0.05 & $-0.0086 (127)  + i 0.3919 (2944)$  \tabularnewline
		\cline{2-3}
		& 0.1	& $-0.0292 (202)  + i 0.3050 (2945)$  \tabularnewline
		\cline{2-3}  
		& 0.2  & $-0.0127 (310)  + i 0.5222 (2910)$  \tabularnewline
		\cline{2-3} 
		& 0.4  & $ 0.0295 (503)  + i 0.4377 (2889)$  \tabularnewline
		\cline{2-3} 
		& 0.6  & $ 0.0497 (595)  + i 0.3674 (2690)$  \tabularnewline
		\cline{2-3} 
		& 0.8  & $-0.0781 (1796) + i 0.1504 (3194)$  \tabularnewline
		\cline{2-3} 
		& $\alpha \rightarrow 0$
		& $  -0.0171 (361) + i 0.3794 (5019) $   \tabularnewline
		\hline 
	\end{tabular} \label{tab:delta_LO_2p0_4p0}}
\end{table}

\subsubsection{Drift term decay}

Another method for validating the complex Langevin dynamics, as discussed in Refs. [\refcite{Nagata:2016vkn,Nagata:2018net}], involves examining the probability distribution $P(u)$ of the magnitude of the drift term $u$ at large values. 
The drift term magnitude is defined as:
\beq
u \equiv \left| \frac{\partial S}{\partial \phi} \right|.
\eeq

According to Refs. [\refcite{Nagata:2016vkn,Nagata:2018net}], the correctness of the complex Langevin method is indicated if the probability of observing large drift magnitudes is exponentially suppressed. 
This suppression ensures that large fluctuations do not dominate and the method remains valid.

\begin{figure*}[htp]
	
	{\includegraphics[width=.49\textwidth,origin=c,angle=0]{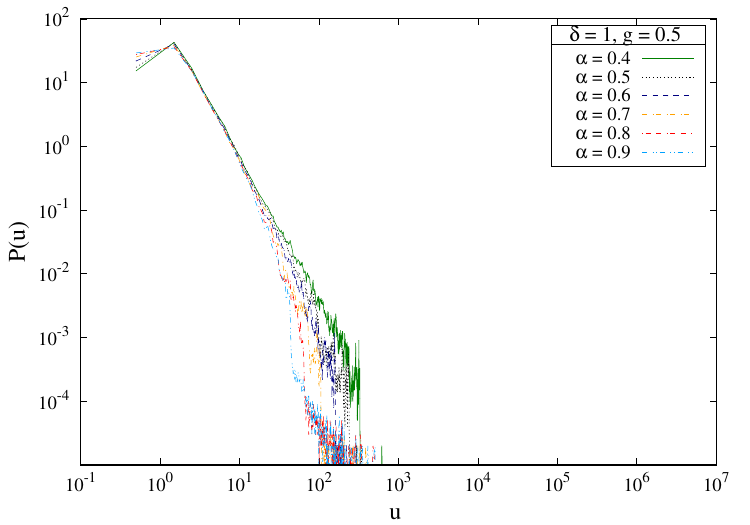}}
	{\includegraphics[width=.49\textwidth,origin=c,angle=0]{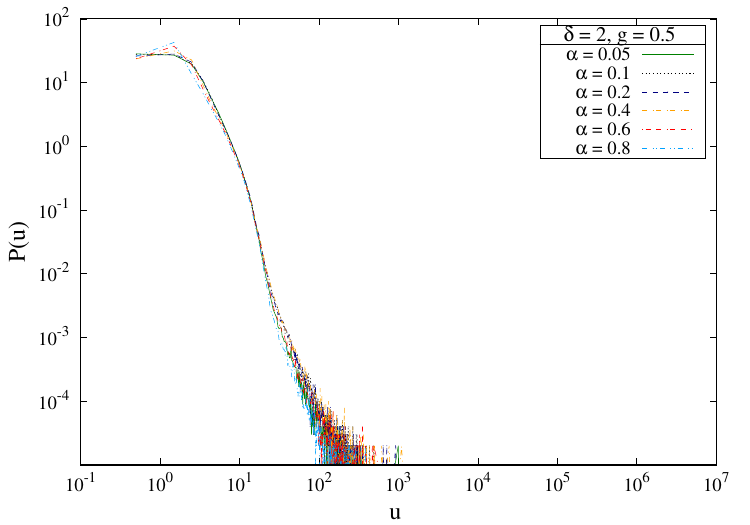}}
	
	{\includegraphics[width=.49\textwidth,origin=c,angle=0]{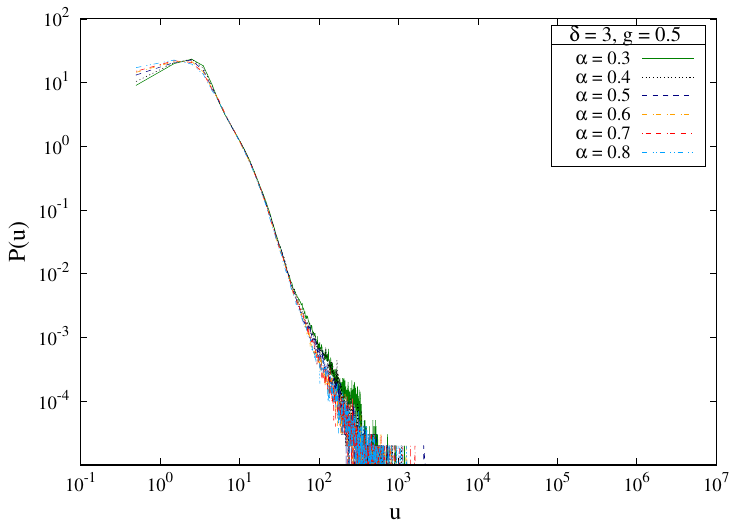}}
	{\includegraphics[width=.49\textwidth,origin=c,angle=0]{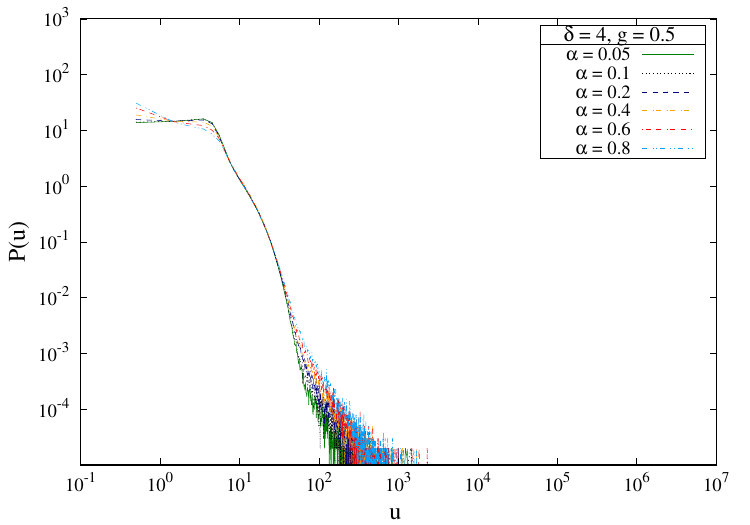}}
	
	\caption{The probability distribution $P(u)$ of the drift term magnitude $u$ for the model with a superpotential $W'(\phi) = -ig (i \phi)^{(1+\delta)}$, with coupling $g = 0.5$, and generation steps $N_{\rm gen} = 10^7$, on a log-log plot. The results are for $\delta = 1$ (Top-Left), $\delta = 2$ (Top-Right), $\delta = 3$ (Bottom-Left), and $\delta = 4$ (Bottom-Right). The Langevin step takes adaptive values $\Delta \tau \leq 5 \times 10^{-5}$.}
	\label{fig:drift_delta}
	
\end{figure*}

In some of the models described in this section, however, we observe that the probability distribution $P(u)$ falls off as a power law with respect to $u$, rather than exponentially. 
Despite this, the simulations show excellent agreement with the corresponding analytical results wherever applicable. 
Figure \ref{fig:drift_delta} illustrates the probability distribution $P(u)$ of the drift term's magnitude $u$ for the superpotential $W'(\phi) = -ig (i \phi)^{(1+\delta)}$. 
On a log-log plot, the distribution exhibits a power-law decay for large $u$ values.

\section{Complex Langevin method for SUSY quantum mechanics} 
\label{sec:Complex_Langevin_method_for_SUSY_quantum_mechanics}

In the early developments of supersymmetry, quantum mechanics models became a crucial testing ground for examining the complexities of dynamical SUSY breaking. 
These models provided a controlled setting to explore the mechanisms behind SUSY breakdown, particularly highlighting the non-perturbative factors that are essential in field theories. 
A landmark contribution came in 1981 from Witten, who introduced the concept of the \textit{Witten index}. 
This topological index became a key tool in investigating dynamical SUSY breaking.

Over the years, many researchers have explored non-perturbative mechanisms that drive SUSY breaking in quantum mechanics. 
These studies have significantly enhanced our understanding of how SUSY can be spontaneously broken, even in relatively simple systems, contributing to a broader comprehension of the phenomenon. 
As research progressed, it became evident that supersymmetric quantum mechanics was more than just a tool for testing field theory techniques -- it evolved into a fascinating and independent area of study with its own unique phenomena and implications.

In this section, we delve into non-perturbative SUSY breaking in lattice-regularized supersymmetric quantum mechanics with complex superpotentials, using complex Langevin simulations to aid our investigation.

\subsection{Supersymmetric quantum mechanics}
\label{sec:Supersymmetric_quantum_mechanics}

Let us consider the action $S[\phi, \psi, \psib]$ for a supersymmetric quantum mechanics with two supercharges.  
The superpotential is denoted by $W(\phi)$. 
The degrees of freedom of the model consist of a scalar field $\phi$ and two fermions, $\psi$ and $\psib$. 
The action is formulated over a compactified Euclidean time circle with circumference $\beta$. 
Its explicit form is given by
\bea
\label{cont-action-1d}
S[\phi, \psi, \psib] &=& \int_0^\beta d \tau
\Bigg[ \hf \B(\tau)^2 + i \B \left( \frac{\partial}{\partial{\tau}}{\phi(\tau)} + \frac{\partial }{\partial{\phi}} W(\phi(\tau)) \right) \nn \\ 
&& \hspace{2cm} + {\psib(\tau)} \left(\frac{\partial}{\partial{\tau}} + \frac{\partial^2 }{\partial{\phi}^2} {W} (\phi(\tau)) \right) \psi(\tau) \Bigg].
\eea
Here, the derivatives with respect to $\phi$ and $\tau$ are denoted by a prime and a dot, respectively, and $\B$ is an auxiliary field.

The action remains invariant under the following SUSY transformations
\beq
\label{Q-transf}
\Q \phi = \psi,~~~ \Q \psi = 0,~~~ \Q \psib =- i \B,~~~ \Q \B = 0,
\eeq
and
\beq
\label{Qb-transf}
\Qb \phi = - \psib,~~~ \Qb \bar{\psi} = 0,~~~ \Qb \psi = - i \B + 2 \dot{\phi},~~~ \Qb \B = 2 i \dot{\psib}.
\eeq
Here, $\Qb$ and $\Q$ represent the two supercharges. 
They obey the algebra
\beq
\label{eq:algebra}
\{ \Q, \Qb \} = 2 \partial_\tau, \quad \{ \Q, \Q \} = 0, \quad \{ \Qb, \Qb \} = 0.
\eeq

Additionally, the action can be expressed in both $\Q$-exact and $\Q \Qb$-exact forms
\bea
\label{QQb-cont-exact}
S &=& \Q \int_0^\beta d\tau ~ \psib \left[ \frac{i}{2} \B - \left(\frac{\partial \phi}{ \partial \tau} + W^{'}(\phi)  \right) \right] \nn \\
&=& \Q \Qb  \int_0^\beta d\tau \left( \hf \psib \psi + W (\phi) \right).
\eea
This formulation highlights the structure of the action in terms of the supercharges and the role of the superpotential $W(\phi)$ in defining the dynamics of the model.

In the path integral formalism, we can write down the partition function as
\beq
\label{eqn:cont-pf}
Z \equiv \int \cD \B \cD \phi \cD \psi  \cD \psib ~e^{-S[\phi, \psi, \psib]}.
\eeq
It is assumed that periodic temporal boundary conditions are imposed on the fields.

In a system where SUSY is unbroken, we consider the Hamiltonian $H$, which corresponds to the Lagrangian in Eq. \eqref{cont-action-1d}, and has energy levels $E_n$ with $n = 0, 1, 2, \cdots$, where the ground state energy is $E_0 = 0$. 
The bosonic and fermionic excited states form a SUSY multiplet, given by
\beq
\label{eqn:susy_multiplet_unbroken}
| b_{n+1} \rangle = \frac{1}{\sqrt{2 E_{n+1}}} \Qb |f_n \rangle, ~~~| f_n \rangle = \frac{1}{\sqrt{2 E_{n+1}}} \Q | b_{n+1} \rangle,
\eeq
which satisfies the SUSY algebra from Eq. \eqref{eq:algebra}. 
The state $| b_0 \rangle$ represents the ground state of the system. 
Let us assume that $| b_n \rangle$ and $| f_n \rangle$ have fermion number charges $F = 0$ and $F = 1$, respectively. 
Then, the partition function $Z$, under periodic boundary conditions for both bosonic and fermionic fields, becomes equivalent to the Witten index $\Delta_W(\beta)$\cite{Witten:1982df}.

We can express the partition function as
\bea
\label{eqn:cont-Witten_index_unbroken}
Z  &=& \Delta_W \left(\beta \right) = \Tr \left[ (-1)^F e^{- \beta H} \right] \nn \\
&=&  \langle b_0 | b_0 \rangle + \sum_{n = 0}^\infty \left[ \left(\langle b_{n+1} | b_{n+1} \rangle - \langle f_n | f_n \rangle  \right) e^{- \beta E_{n+1}} \right].
\eea
From above, we can see that the partition function does not vanish due to the presence of a normalizable ground state. 
Consequently, the normalized expectation values of observables in the theory are well-defined.

The auxiliary field has the following normalized expectation value
\beq
\label{eqn:aux_field_exp_val}
\langle \B \rangle = \frac{1}{\Delta_W \left(\beta \right)} \left[\langle b_0 | \B | b_0 \rangle + \sum_{n = 0}^\infty \left(\langle b_{n+1} | \B | b_{n+1} \rangle - \langle f_n | \B |f_n \rangle \right) e^{- \beta E_{n+1}} \right].
\eeq
We can use this as an order parameter to detect SUSY breaking \cite{Kuroki:2009yg}. 
It is important to note that the unpaired state mentioned in Eqs. \eqref{eqn:cont-Witten_index_unbroken} and \eqref{eqn:aux_field_exp_val} need not necessarily be a bosonic state or a unique state. 

The auxiliary field $\B$ was introduced to complete the off-shell SUSY algebra. 
From the transformation properties of $\B$ under the $\Q$-supersymmetry in Eq. \eqref{Q-transf}, and the fact that the ground state is annihilated by the supercharges, it follows that in a system with unbroken SUSY, the normalized expectation value of the auxiliary field vanishes. 
However, in the case of SUSY breaking, the situation is more complex.

When SUSY is spontaneously broken, the Hamiltonian $H$ has a positive ground state energy, i.e., $0 < E_0 < E_1 < E_2 < \cdots$, and the SUSY multiplets are now defined as
\beq
\label{eqn:susy_multiplet}
| b_n \rangle = \frac{1}{\sqrt{2 E_n}} \Qb |f_n \rangle, ~~~| f_n \rangle = \frac{1}{\sqrt{2 E_n}} \Q | b_n \rangle,
\eeq
satisfying the algebra in Eq. \eqref{eq:algebra}. 
Unlike the unbroken case, when SUSY is broken, the supersymmetric partition function
\beq
\label{eqn:cont-Witten_index}
Z  = \Delta_W = \Tr \left[ (-1)^F e^{- \beta H} \right] = \sum_{n = 0}^\infty \left[ \left(\langle b_n | b_n \rangle - \langle f_n | f_n \rangle \right) e^{- \beta E_n} \right]
\eeq
vanishes as a result of bosonic and fermionic states cancelling each other. 
Consequently, the normalized expectation values of observables, such as the auxiliary field, become ill-defined.

The normalized expectation value of the auxiliary field can be written as
\beq
\label{eqn:cont-B-expectation}
\langle \B \rangle = \frac{\sum_{n = 0}^\infty \left[ \left(\langle b_n | \B | b_n \rangle - \langle f_n | \B |f_n \rangle \right) e^{- \beta E_n} \right]}{\Delta_W \left(\beta \right)}.
\eeq

However, in a SUSY-broken system, the numerator vanishes due to $\Q$-supersymmetry, since $\langle b_n | \B | b_n \rangle = \langle f_n | \B | f_n \rangle$. 
This results in a $0/0$ indeterminate form, making the normalized expectation of the auxiliary field ill-defined.

Kuroki and Sugino introduced a regulator \cite{Kuroki:2009yg}, that breaks supersymmetry explicitly, but resolves the vacuum degeneracy by selecting a single vacuum state where SUSY is broken. 
This regulator, denoted as $\alpha$ is implemented by imposing twisted boundary conditions (TBC) on the fermions. 
Specifically, the boundary conditions for fermions are modified as follows:
\beq
\psi (\tau + \beta) = e^{i \alpha} \psi(\tau) \quad {\rm and} \quad \psib(\tau +\beta) = e^{- i \alpha} \psib(\tau).
\eeq
Kuroki and Sugino showed that for non-zero $\alpha$, the partition function does not vanish, and thus, the normalized expectation value of $\B$ becomes well-defined. 
It is easy to see that when $\alpha \to 0$, the periodic boundary conditions (PBC) are recovered, and SUSY is restored. 
Hence, $\alpha$ acts as a regularization parameter, resolving the indeterminate form in Eq. \eqref{eqn:cont-B-expectation}. 
The behavior of the auxiliary field expectation value as $\alpha \to 0$ provides insight into SUSY breaking: a vanishing value indicates unbroken SUSY, while a non-zero value signals SUSY breaking. 
The twist parameter $\alpha$ will be incorporated in the lattice regularized theory discussed in Sec. \ref{sec:SUSY_models_on_a_lattice}.

The auxiliary field $\B$ can be integrated out using its equation of motion
\beq
\B = - i \left( \frac{\partial \phi}{ \partial \tau} + W^{'}(\phi) \right),
\eeq
leading to the on-shell form of the action:
\beq
\label{eqn:act-cont}
S = \int_0^\beta d\tau
\left[ \hf \left({\frac{\partial \phi}{ \partial \tau}} + W^{'}(\phi) \right)^2 + {\psib} \left( \frac{\partial }{ \partial \tau} + W''(\phi) \right) \psi \right].
\eeq
By applying the Leibniz rule and removing the total derivative term $\left( \frac{\partial \phi}{\partial \tau} W'(\phi) \right)$, the action simplifies to
\beq
S = \int_0^\beta d\tau
\left[ \hf { \left(\frac{\partial \phi }{ \partial \tau } \right) }^2 + \hf \left[ W'(\phi) \right]^2 + {\psib} \left( \frac{\partial}{\partial \tau} + W''(\phi) \right) \psi \right].
\eeq

This total derivative term, though omitted in the continuum theory, becomes crucial when discretizing the theory on a lattice. 
Its inclusion is necessary to ensure the exactness of $\Q$-supersymmetry on the lattice. 
Therefore, for the lattice analysis, we will use Eq. \eqref{eqn:act-cont} as the target theory in the continuum.

To verify the $\Q$- and $\Q \Qb$-exact forms of the continuum action, given in Eq. \eqref{QQb-cont-exact}, we can use the on-shell action, obtained by integrating out $\B$:
\beq
i \B = \frac{\partial \phi}{\partial \tau} + W'(\phi).
\eeq
This results in the action
\bea
S &=& - \Q \int_0^\beta d{\tau} ~ \hf \psib \left( \frac{\partial \phi}{\partial \tau} + W^{'}(\phi) \right) \nn \\
&=& \Q \Qb \int_0^\beta d\tau \left( \hf \psib \psi + W (\phi) \right).
\eea

\subsection{SUSY models on a lattice}
\label{sec:SUSY_models_on_a_lattice}

Let us discretize the action given in Eq. \eqref{eqn:act-cont} on a lattice with extent in one direction. We define the lattice as $\Lambda$, consisting of $N_\tau$ equally spaced sites, and with lattice spacing $a$. 
The total physical extent of the lattice is $\beta = N_\tau a$, and the integral over time is replaced by a Riemann sum, while derivatives are replaced by discrete difference operators.

The continuum action is discretized as follows:

The time integral is replaced by a sum over lattice points: 
\beq
\int_0^\beta d\tau \rightarrow a \sum_{i=1}^{N_\tau}.
\eeq

The continuum derivative is replaced by a symmetric difference operator on the lattice. 
For a scalar function $f$ defined on the lattice sites, the symmetric difference operator is
\beq
\nabla^S_{ij} = \frac{1}{2} \left( \nabla^+_{ij} + \nabla^-_{ij} \right),
\eeq
where the indices $i, j$ represent lattice sites, and $\nabla^+_{ij}$ and $\nabla^-_{ij}$, respectively, are the forward and backward difference operators. 
They act on $f_j$ in the following way
\beq
\nabla^+_{ij} f_j = \frac{1}{a} (f_{i+1} - f_i) \quad {\rm and} \quad \nabla^-_{ij} f_j = \frac{1}{a} (f_i - f_{i-1}).
\eeq

However, this symmetric discretization leads to the fermion doubling problem, where extra unphysical fermionic states appear. 
To avoid this issue, we apply the \textit{Wilson discretization}. 
The Wilson term adds a correction to the difference operator to suppress the extra modes responsible for the doubling problem. 
The modified difference operator, including the Wilson term, is
\beq
\nabla^W_{ij}(r) = \nabla^S_{ij} - \frac{r a}{2} \square_{ij},
\eeq
where $\square_{ij} = \nabla^+_{ik} \nabla^-_{kj}$ is the lattice Laplacian, and $r$ is the Wilson parameter, typically chosen from the interval $r \in [-1, 1] \setminus \{0\}$.

For one-dimensional systems, a choice of $r = \pm 1$ simplifies the Wilson difference operator to
\beq
\nabla^W_{ij}(\pm 1) = \nabla^{\mp}_{ij},
\eeq
indicating that using either the forward or backward difference operator alone can resolve the fermion doubling problem in this case.

This prescription maintains lattice supersymmetry by adding appropriate improvement terms to account for the discretization of continuum surface integrals, ensuring that the lattice theory preserves the symmetries of the continuum theory \cite{Bergner:2007pu}. 
Therefore, by carefully choosing the lattice difference operators and adding necessary improvements, the lattice regularization can maintain manifest supersymmetry and avoid the fermion doubling issue.

In this lattice regularization approach, we follow the symmetric derivative prescription with a Wilson mass matrix as suggested in Ref. [\refcite{Catterall:2000rv}]. 
The discretized action in Eq. \eqref{eqn:lat-reg-action} is constructed to ensure that the fermion doubling problem is handled effectively while maintaining supersymmetry on the lattice. 

The lattice action $\mcS$ is given by
\beq
\label{eqn:lat-reg-action}
\mcS = a \sum_{i = 0}^{N_\tau - 1} \left[ \hf \left( \sum_{j = 0}^{N_\tau - 1} \nabla^S_{ij} \phi_j + \Omega'_i  \right)^2 + \psib_i \sum_{j = 0}^{N_\tau - 1} \left( \nabla^S_{ij} + \Omega''_{ij} \right) \psi_j \right],
\eeq
where $\nabla^S_{ij}$ is the symmetric lattice derivative operator, $\Omega'_i$ is defined as 
\beq
\Omega'_i \equiv \sum_{j = 0}^{N_\tau - 1} K_{ij} \phi_j + W'_i, 
\eeq
where $K_{ij}$ is the Wilson mass matrix, and $\Omega''_{ij} = K_{ij} + W''_{ij} \delta_{ij}$ is the derivative of $\Omega'_i$.

The matrix $K_{ij}$ is constructed to mitigate the fermion doubling problem. It is defined as
\beq
K_{ij} \equiv m \delta_{ij} - \frac{r a}{2} \square_{ij},
\eeq
where $m$ is the mass parameter.

To simplify the analysis and obtain a dimensionless form of the action, we introduce the following rescalings:
\beq
\label{eqn:rescale}
\widetilde{\phi} = a^{-1/2} \phi, \quad \widetilde{\nabla}^S = a \nabla^S, \quad \widetilde{\Omega}' = \sqrt{a} \Omega', \quad \widetilde{\Omega}'' = a \Omega''.
\eeq

After rescaling, the action becomes dimensionless:
\beq
\label{eqn:lat-dimless-action}
\widetilde{\mcS} = \sum_{i = 0}^{N_\tau - 1} \left[ \hf \left( \sum_{j = 0}^{N_\tau - 1} \widetilde{\nabla}^S_{ij} \widetilde{\phi}_j + \widetilde{\Omega}'_i \right)^2 + \psib_i \sum_{j = 0}^{N_\tau - 1} \left( \widetilde{\nabla}^S_{ij} + \widetilde{\Omega}''_{ij} \right) \psi_j\right].
\eeq

This dimensionless action is more convenient for numerical simulations and analytic calculations, as it eliminates the dependence on the lattice spacing $a$. 
The form of the action maintains supersymmetry on the lattice by carefully including the Wilson term and ensuring that the derivative operators are appropriately defined.

\subsubsection{Lattice discretization}

From now on, we will remove the tilde symbol on lattice variables and it is understood that they are dimensionless. 
To accommodate the Wilson mass terms in the lattice formulation, the SUSY transformations for the fields are modified accordingly. 
The transformations for a given lattice site $k$ are:
\beq
\label{eqn:lat-Wilson-Q}
\Q \phi_k = \psi_k, \quad \Q \psib_k = - N_k, \quad \Q \psi_k = 0, 
\eeq
and
\beq
\label{eqn:lat-Wilson-Qbar}
\Qb \phi_k = - \psib_k, \quad \Qb \psi_k = \overline{N}_k, \quad \Qb \hspace{0.05cm} \psib_k = 0,
\eeq
where the variables $N_k$ and $\overline{N}_k$ are defined as:
\beq
N_k = \nabla^S{\phi}_k + \Omega'_k \quad {\rm and} \quad \overline{N}_k = \nabla^S{\phi}_k - \Omega'_k.
\eeq
These expressions introduce the Wilson mass terms via $\Omega'_k$, which account for the fermion doubling issue in the lattice theory.

The supercharges satisfy the following algebra:
\beq
\label{eqn:lat-Q-Qb-algebra}
\{ \Q, \Q \} = 0, \quad \{ \Qb, \Qb \} = 0, \quad{\rm and }\quad \{ \Q, \Qb \} = 2 \nabla^S.
\eeq
This algebra holds on the lattice, but it is important to note that the discretized derivative $\nabla^S$ modifies the relationship between the supercharges compared to the continuum case.

The lattice-regularized action is given by:
\beq
\label{eqn:lat-action}
\mcS = \sum_{i = 0}^{N_\tau - 1} \left[ \hf \left( \sum_{j = 0}^{N_\tau - 1} \nabla^S_{ij} \phi_j + \Omega'_i \right)^2 + \psib_i \sum_{j = 0}^{N_\tau - 1} \left( \nabla^S_{ij} + \Omega''_{ij} \right) \psi_j \right].
\eeq

However, a critical aspect of lattice regularization is the breakdown of exact SUSY due to the failure of the Leibniz rule for lattice derivatives. 
This means that although the action preserves the $\Q$-supercharge (i.e., $\Q \mcS = 0$), the $\Qb$-supercharge is broken for $N_\tau \geq 2$. 
Therefore, the full lattice theory only retains $\Q$-invariance, leading to:
\beq
\Q \mcS = 0 \quad \text{but} \quad \Qb \mcS \neq 0.
\eeq

This partial breaking of supersymmetry is a known issue in lattice formulations of SUSY theories. 
Although the $\Q$-supercharge remains intact, maintaining full supersymmetry on the lattice generally requires more intricate constructions, such as adding improvement terms to recover the $\Qb$-invariance, or employing other specialized lattice formulations like twisted or topological lattices.

The $\Qb$ supersymmetry is broken for a finite lattice size $N_\tau$ because it is impossible to define a corresponding $\Qb$-invariant transformation on the lattice variables that maintains the algebra $\{ \Q, \Qb \} = 2 \nabla^S$ \cite{Kanamori:2007yx}. 
Nevertheless, $\Q$-exactness is crucial and sufficient to eliminate any SUSY-breaking counter-terms and mitigate lattice artifacts. 
For further discussion on this, see Refs. [\refcite{Catterall:2000rv,Catterall:2003wd,Giedt:2004vb,Kuroki:2009yg}].

As discussed in Section \ref{sec:Supersymmetric_quantum_mechanics} for the case of  the theory in the continuum, when SUSY is broken, the partition function vanishes. 
Consequently, the expectation values of observables, normalized by the partition function, can become ill-defined. 
To tame this problem in the lattice regularized theory, we employ periodic boundary conditions for bosons and twisted boundary conditions for fermions \cite{Kuroki:2009yg, Kuroki:2010au}.

Once we introduce the twist, we have
\beq
\phi_{N_\tau} = \phi_0, \quad \psi_{N_\tau} = e^{i \alpha} \psi_0, \quad \psib_{N_\tau} = e^{-i \alpha} \psib_0.
\eeq 
When SUSY is dynamically broken in the theory, when $\alpha = 0$, the Witten index vanishes. 
This leads to the fermion determinant changing its sign depending on the bosonic field configurations. 
As a result, we have a sign problem in models with dynamical supersymmetry breaking.

Upon using the twist, the partition function, given in Eq. \eqref{eqn:cont-pf}, becomes
\bea
\label{eqn:lat-pf}
Z_\alpha = \left( \frac{1}{\sqrt{2 \pi}} \right)^{N_\tau}  \int \left( \prod_{k = 0}^{N_\tau - 1} d\phi_k d\psi_k d\psib_k \right) e^{ - \mcS_\alpha },
\eea
where $\mcS_\alpha$ is the lattice regularized action respecting the twisted boundary conditions.

Explicitly, $\mcS_\alpha$ is given by
\bea
\label{eqn:lat-action-r1}
\mcS_\alpha &=& \sum_{i = 0}^{N_\tau - 1} \hf \bigg( \phi_i - \phi_{i-1} + m \phi_i + W'_i \bigg)^2 \nn \\
&& ~~~ ~~~ + \sum_{i = 0}^{N_\tau - 1} \psib_i \bigg( \psi_i - \psi_{i-1} + \left( m + W''_{ii} \right) \psi_i \bigg).
\eea

We can incorporate the mass term into the superpotential $W$ and introduce a new potential $\Xi$ in the following way
\beq
\label{eqn:lat-anho-pot}
\Xi \equiv \frac{1}{2} m \phi^2 + W.
\eeq

With this redefinition, the action, under twisted boundary conditions has the form
\bea
\label{eqn:lat-action-r1-Xi}
\mcS_\alpha &=& \sum_{i = 0}^{N_\tau - 1} \frac{1}{2} \left( \sum_{j = 0}^{N_\tau - 1} \nabla^{-}_{ij} \phi_j + \Xi'_i \right)^2 + \sum_{i = 0}^{N_\tau - 1} \psib_i \left( \sum_{j = 0}^{N_\tau - 1} \nabla^{-}_{ij} + \Xi^{''}_{ij} \right) \psi_j.  
\eea
Here, the expressions for $N_i$ and $\overline{N}_i$ are updated as
\bea
N_i &=& \sum_{j = 0}^{N_\tau - 1} \nabla^S_{ij} \phi_j + \Omega'_i = \sum_{j = 0}^{N_\tau - 1} \nabla^{-}_{ij} \phi_j + \Xi'_i, \\
\overline{N}_i &=& \sum_{j = 0}^{N_\tau - 1} \nabla^S_{ij} \phi_j - \Omega'_i = \sum_{j = 0}^{N_\tau - 1} \nabla^+_{ij} \phi_j - \Xi'_i.
\eea

We can integrate out the fermions. Then, the fermionic contribution to the partition function, as given in Eq. \eqref{eqn:lat-pf}, becomes
\bea
\label{eqn:z-alpha-F}
Z_\alpha^F &=& \prod_{k = 0}^{N_\tau - 1} \left(1 + \Xi^{''}_{kk} \right) - e^{i \alpha}.
\eea
It represents the determinant of the twisted Wilson fermion matrix $\mathcal{W}_{\alpha}^F$:
\beq
Z_\alpha^F = \det \left[\mathcal{W}_\alpha^F \right].
\eeq
For periodic boundary conditions ($\alpha = 0$), this matches the expression found in Ref. [\refcite{Catterall:2000rv}].

The full partition function is then given by
\beq
\label{eqn:lat-pf-bos}
Z_\alpha = \left( \frac{1}{\sqrt{2 \pi}} \right)^{N_\tau} \int \left( \prod_{k = 0}^{N_\tau - 1} d\phi_k \right) \exp \left[ - \mcS^{\text{eff}}_\alpha \right],
\eeq
where
\bea
\label{eqn:lat-eff-action}
\mcS_{\alpha}^{\text{eff}} &=& \mcS^{B} - \ln \left( \det \left[ \mathcal{W}_\alpha^F \right] \right) \nn \\ 
&=& \sum_{k = 0}^{N_\tau - 1} \frac{1}{2} \left( \phi_k - \phi_{k-1} + \Xi'_k \right)^2 - \ln \left( \prod_{k = 0}^{N_\tau - 1} \left(1 + \Xi^{''}_{kk} \right) - e^{i \alpha} \right).
\eea

For an observable $\mathcal{O}$, its expectation value is computed as
\bea
\label{eqn:lat-exp-obs}
\langle \mathcal{O} \rangle &=& \lim_{\alpha \to 0} \langle \mathcal{O} \rangle_\alpha \nn \\
&=& \lim_{\alpha \to 0} \frac{1}{Z_\alpha} \left( \frac{1}{\sqrt{2 \pi}} \right)^{N_\tau} \int \left( \prod_{k = 0}^{N_\tau - 1} d\phi_k \right) \mathcal{O} \exp \left[ - \mcS_{\alpha}^{\text{eff}} \right].
\eea

When we use the complex Langevin method, the gradient of the action must be computed to update the field configurations. 
The drift term, being the negative gradient of the action, includes the fermion determinant in the denominator, which may have zeroes in the complexified space. 
This can lead to the singular-drift problem, where the dynamical variables approach singularities in the drift term. 
A recent study has shown that such problems are not limited to logarithmic singularities but can arise generally where the stochastic process involves a singular-drift term \cite{Nishimura:2015pba}.

\subsubsection{Observables: correlation functions}

Correlation functions are fundamental to lattice field theory, providing essential insights into the statistical and dynamical properties of quantum systems. 
They are crucial for understanding mass and energy spectra in quantum field theories, which are central to fields such as nuclear, particle, and condensed matter physics.

In lattice QFT, we use high-dimensional lattice-regulated path integrals to compute correlation functions. We define the correlation functions for bosonic and fermionic fields at site $k$:
\begin{equation}
G^B_\alpha (k) \equiv \langle \phi_0 \phi_k \rangle_\alpha,
\end{equation}
and
\begin{equation}
G^F_\alpha(k) \equiv \langle \psib_0 \psi_k \rangle_\alpha.
\end{equation}

These functions provide valuable information about the interactions and dynamics within the theory.

For the fermionic correlation function, we have
\bea
\langle \psib_0 \psi_k \rangle_\alpha &=& 
 \frac{1}{Z_\alpha} \left( \frac{1}{\sqrt{2 \pi}} \right)^{N_\tau} \int \left(\prod_{t = 0}^{N_\tau - 1} d\phi_t \right) \nn \\
&& \times \underbrace{\left\{ \int \left(\prod_{t = 0}^{N_\tau - 1} d\psi_t d\psib_t \right) \psib_0 \psi_k \exp \left[- \sum_{t = 0}^{N_\tau - 1} \psib_t \left[ \left( 1 + \Xi^{''}_{tt} \right) \psi_t - \psi_{t - 1} \right] \right] \right\}}_{{\langle \psib_0 \psi_k \rangle}^F} \nn \\
&& \times \exp\left[ -\sum_{t = 0}^{N_\tau - 1} \frac{1}{2} \left( \phi_t - \phi_{t - 1} + \Xi'_t \right)^2 \right] \\
&=& \frac{1}{Z_\alpha} \left( \frac{1}{\sqrt{2 \pi}} \right)^{N_\tau} \int \left( \prod_{i = 0}^{N_\tau - 1} d\phi_i \right) \underbrace{ \left( - \frac{{ \langle \psib_0 \psi_k \rangle^F }}{ \det \left[ \mathcal{W}_\alpha^F \right] } \right) }_{\left[ \psib_0 \psi_k \right]^L_\alpha} \exp \left[ - \mcS_\alpha^{\text{eff}} \right],
\eea
where \(\langle \psib_0 \psi_k \rangle^F\) is computed as
\beq
\langle \psib_0 \psi_k \rangle^F = - \prod_{t = k + 1}^{N_\tau - 1} \left[1 + \Xi^{''}_{tt}\right]. 
\eeq
Comparing with Eq. \eqref{eqn:lat-exp-obs}, we can define the Langevin observable that corresponds to the fermionic correlator as
\beq
\left[ \psib_0 \psi_k \right]^L_\alpha = - \frac{ \prod_{i = k + 1}^{N_\tau - 1} \left[1 + \Xi^{''}_{ii} \right] }{ \prod_{i = 0}^{N_\tau - 1} \left[ 1 + \Xi^{''}_{ii} \right] - e^{i \alpha}}.
\eeq

The computation of the bosonic correlation function is straightforward compared to the fermionic case. 
In this scenario, the Langevin observable directly corresponds to the bosonic correlation function itself.

The Langevin observable is simply the product $\phi_0 \phi_k$ for the $k$-th lattice site. 
The bosonic correlation function $\langle \phi_0 \phi_k \rangle_\alpha$ is given by:
\bea
\langle \phi_0 \phi_k \rangle_\alpha &=& \frac{1}{Z_\alpha} \left( \frac{1}{\sqrt{2 \pi}} \right)^{N_\tau} \int \left( \prod_{i = 0}^{N_\tau - 1} d\phi_i \right) \phi_0 \phi_k ~ \exp \left[ - \mcS_\alpha^{\text{eff}} \right].
\eea

\subsubsection{Observables: Ward identities}

Ward identities are fundamental tools in lattice field theory for analyzing symmetry properties, including the study of SUSY breaking. 
They emerge from the invariance principles of quantum field theories and are crucial for understanding how lattice formulations of SUSY theories behave, particularly regarding broken symmetries.

In lattice field theories, the introduction of a spacetime lattice as a regulator breaks continuous symmetries, such as Poincar\'e invariance and SUSY. 
For instance, a non-zero gluino mass in supersymmetric Yang-Mills theory introduces additional soft SUSY breaking. 
Consequently, lattice-adapted SUSY Ward identities become essential for examining the critical parameters where the broken symmetries might be restored \cite{Ali:2017nug}.

For the supersymmetric variations of the fields given by Eqs. \eqref{eqn:lat-Wilson-Q} and \eqref{eqn:lat-Wilson-Qbar}, demanding the invariance of the lattice action yields a set of Ward identities relating the fermionic and bosonic correlators. 

We can introduce source terms $J$, $\chi$, and $\overline{\chi}$, to the partition function. 
Then, 
\bea
\label{eqn:lat-pf-source}
Z_\alpha \left( J, \chi, \overline{\chi} \right) &=& \left( \frac{1}{\sqrt{2 \pi}} \right)^{N_\tau} \int \left( \prod_{k = 0}^{N_\tau - 1} d\phi_k d\psi_k d\psib_k \right) \nn \\ 
&& \times \exp \left[ - \mcS_\alpha + \sum_{k = 0}^{N_\tau - 1} \left( J_k \phi_k + \chi_k \psib_k + \overline{\chi}_k \psi_k \right) \right].
\eea

The invariance of the partition function under the $\Q$-transformations implies that
\bea
\Q Z_\alpha \left( J, \chi, \overline{\chi} \right) &=& \left( \frac{1}{\sqrt{2 \pi}} \right)^{N_\tau} \int \left( \prod_{t = 0}^{N_\tau - 1} d\phi_t d\psi_t d\psib_t \right) \nn \\
&& \times \exp \left[ - \mcS_\alpha + \sum_{t = 0}^{N_\tau - 1} \left( J_t \phi_t + \theta_t \psib_t + \overline{\theta}_t \psi_t \right) \right] \nn \\
&& \times \left( - \Q \mcS_\alpha + \sum_{t = 0}^{N_\tau - 1} \left( J_t \Q \phi_t + \theta_t \Q \psib_t \right) \right) \nn \\
&=& 0.
\eea

Furthermore, the derivative of the partition function with respect to the source terms $J_j$ and $\chi_i$ yields a set of supersymmetric Ward identities
\beq
\langle \psib_i \psi_j \rangle + \langle N_i \phi_j \rangle = 0.
\eeq

To investigate spontaneous SUSY breaking, we can use the following Ward identity
\beq
\label{eqn:lat-ward-iden}
W_1 : 
\begin{array}{l}
	\langle \psib_0 \psi_k \rangle + \langle N_0 \phi_k \rangle = 0.
\end{array}
\eeq

\subsection{Complex Langevin method for SUSY quantum mechanics}

In simulations of quantum mechanics models, several key observables are crucial for understanding dynamical SUSY breaking and verifying lattice SUSY. 
Here is a summary of these observables and their significance:

\begin{itemize}
\item[1.] {\it Expectation Value of the Auxiliary Field:}
\beq
\B_\alpha = -i \left( \nabla^S_{ij} \phi_j + \Omega'_i \right) = -i \left( \nabla^{-}_{ij} \phi_j + \Xi'_i \right).
\eeq
The expectation value of the auxiliary field $\langle \mathcal{B} \rangle$ serves as an order parameter for SUSY breaking. 
Studies have shown that a non-zero value indicates SUSY breaking, while a vanishing value suggests SUSY is preserved \cite{Kuroki:2009yg, Kuroki:2010au, Joseph:2019sof}. 
However, in some cases, a vanishing value might be accidental. 
To confirm SUSY breaking, higher powers of $\mathcal{B}$ may be analyzed in addition to other observables.

\item[2.] {\it Bosonic Action:} The expectation value of the bosonic action, $\mcS^B_\alpha$, provides insight into the degrees of freedom on the lattice. 
For exact lattice SUSY, this value is independent of interaction couplings \cite{Catterall:2001fr}. 
In supersymmetric quantum mechanics, we expect
\beq
\langle \mcS^B \rangle = \lim_{\alpha \to 0} \langle \mathcal{S}^B_\alpha \rangle
   \begin{cases}
       \neq \frac{N_\tau}{2} & \text{SUSY broken} \\
       = \frac{N_\tau}{2} & \text{SUSY preserved}.
   \end{cases}
\eeq

\item[3.] {\it Mass Gaps:} The mass gaps can be extracted using fits to the correlation functions, such as $\cosh \big[ m a( t - \frac{N_\tau}{2}) \big]$ or exponential fits over specific time slices \cite{Catterall:2001fr, Giedt:2004vb}. 
The equality of fermionic and bosonic mass gaps provides additional information about SUSY breaking.

\item[4.] {\it Ward Identity:} The Ward identity $W_1$, given by
\beq
   W_1 : 
   \begin{array}{l}
       \langle \psib_0 \psi_k \rangle + \langle N_0 \phi_k \rangle = 0,
   \end{array}
\eeq
is crucial for confirming exact lattice SUSY. For SUSY-preserved theories, this identity should hold
\beq
\lim_{\alpha \to 0} W_1:
   \begin{cases}
       -\langle \psib_0 \psi_k \rangle_\alpha = \langle N_0 \phi_k \rangle_\alpha & \text{SUSY preserved} \\
       -\langle \psib_0 \psi_k \rangle_\alpha \neq \langle N_0 \phi_k \rangle_\alpha & \text{SUSY broken}.
   \end{cases}
\eeq
\end{itemize}

In the limit $\alpha \to 0$, these observables help determine whether the system retains exact lattice SUSY. 
For models with preserved SUSY, the partition function is well-defined, allowing for meaningful numerical investigations of normalized expectation values. 
However, for models with spontaneously broken SUSY, the partition function vanishes, leading to ill-defined normalized expectation values.

\subsubsection{Theories exhibiting $\mathcal{PT}$-symmetry}

Quantum mechanics and quantum field theory are traditionally framed using Hermitian Hamiltonians and Lagrangians. 
Recently, however, there has been growing interest in extending these theories to non-Hermitian contexts, especially those with $\mathcal{PT}$-symmetry, which can have real spectra. 
These non-Hermitian theories have found applications in various fields, such as optoelectronics and phase transitions. 
Furthermore, it has been demonstrated that concepts from Hermitian quantum field theory, like spontaneous symmetry breaking and the Higgs mechanism, can be adapted to $\mathcal{PT}$-symmetric non-Hermitian theories. 
For instance, Ref. [\refcite{Alexandre:2020wki}] explores $\mathcal{PT}$-symmetric $\cN = 1$ supersymmetric quantum field theories in $3+1$ dimensions, revealing that these models, despite their explicit supersymmetry, introduce a novel non-Hermitian channel for soft SUSY breaking.

By imposing $\mathcal{PT}$-symmetric boundary conditions on the functional-integral formulation of the four-dimensional $- \lambda \phi^4$ theory, it is possible to achieve a spectrum that is bounded below. 
This interaction results in a quantum field theory that has a real and bounded spectrum, in addition to exhibiting perturbative renormalizability and asymptotic freedom. It also has potential utility in describing the Higgs sector of the Standard Model.

Consider the following potential:
\beq
\label{eqn:lat-pt-symm-pot}
\Xi(\phi) = - \frac{g}{(2 + \delta)} \left(i \phi \right)^{\left(2 + \delta \right)},
\eeq
where $\Xi^{'}(\phi) = - i g ~ (i \phi)^{(1 + \delta)}$ and $\delta$ is a continuous parameter. 
The supersymmetric Lagrangian for this $\mathcal{PT}$-symmetric theory breaks parity symmetry, raising the question of whether this parity breaking also leads to SUSY breaking. 
Ref. [\refcite{Bender:1997ps}] addressed this for a two-dimensional model and found that SUSY remains unbroken despite perturbative expansions in $\delta$. 
Using the complex Langevin method, we explore the presence or absence of non-perturbative SUSY breaking in one-dimensional analogs of these models. 
Path integral Monte Carlo methods are inadequate here due to the generally complex nature of the action in these models\footnote{Ref. [\refcite{Dhindsa:2020ovr}] showed, using Monte Carlo simulations, that $\mathcal{PT}$ invariance is intact in SUSY QM models with $\delta = 0, 2, 4$. Also see Refs. [\refcite{Kadoh:2015zza,Kadoh:2018ivg,Kadoh:2018ele,Kadoh:2019bir}] for other related work on SUSY QM on a lattice.}.

{\it Even $\delta$ case:} When $\delta = 0$, the model reduces to the supersymmetric harmonic oscillator.

Table \ref{tab:d2-4bSb} presents simulation results for $\delta = 2$ and $4$ taken from Ref. [\refcite{Joseph:2020gdh}]. Simulations, with physical parameter $g_{\rm phys} = 0.5$, were conducted on lattices with $N_\tau = 4, 8, 12$, and with twist $\alpha = 0$. 
The expectation value of the auxiliary field $\langle \B \rangle$ vanishes, and the expectation value of the bosonic action $\langle \mcS^B \rangle$ is consistent with $\hf N_\tau$ within errors.
Additionally, $\langle \mcS^B \rangle$ is independent of the coupling $g$. 
These results suggest that SUSY is preserved in these models.

\begin{table}[]
\tbl{The observables $\B_\alpha$ and $\mcS^B_{\alpha}$ for the $\mathcal{PT}$-symmetric SUSY QM models with $\delta = 2, 4$.}
{\begin{tabular}{| c | l | r | c | c |} 
				\hline
				$\Xi'(\phi)$&$N_\tau$ &  $a = N_\tau^{-1}$ &    $~\langle \B_{\alpha} \rangle$   &  $~\langle \mcS^B_{\alpha} \rangle$ \\   
				\hline
				\hline
				&${4}$	&${0.25}$	&
				$0.0000(0) 	+ i0.0005(282)$ & $2.0130(102)+ i0.0000(0)$ \\	
				$\delta = 2$
				&${8}$ &${0.125}$	
				& $0.0000(0) 	+ i0.0128(750)$ & $4.0326(157)+ i0.0000(0)$ \\ 	
				&${12}$	&${0.0833}$ &
				$0.0000(0) 	- i0.0071(263)$ & $6.0354(58)+ i0.0000(0)$\\
				\hline
				&${4}$	&${0.25}$	&
				$0.0000(0) 	+ i0.0167(679)$ & $1.9975(47)+ i0.0000(0)$ \\		
				$\delta = 4$
				&${8}$ &${0.125}$	
				& $0.0000(0) 	+ i0.0142(567)$ & $4.0058(54)+ i0.0000(0)$ \\	
				&${12}$	&${0.0833}$ &
				$0.0000(0) 	- i0.0309(1022)$ & $6.0018(74)+ i0.0000(0)$ \\ 
				\hline
			\end{tabular} 
			\label{tab:d2-4bSb}
			}
\end{table}

\begin{figure*}[tbp]
	\centering	
	\includegraphics[width=.6\textwidth,origin=c,angle=0]{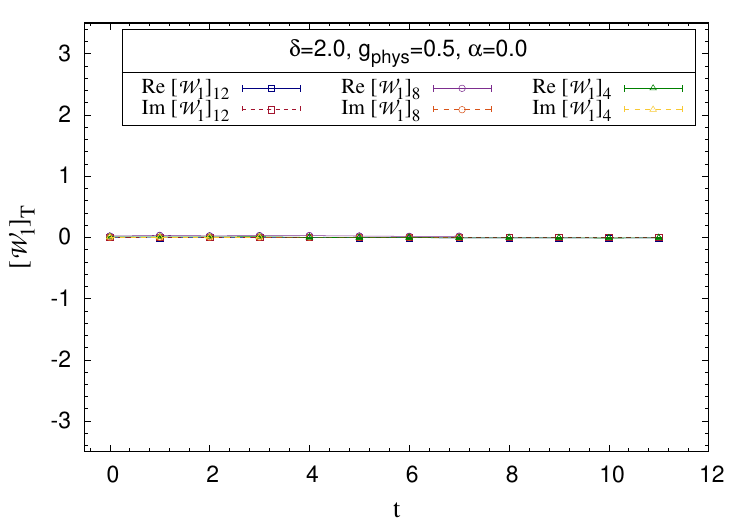}\\
	\includegraphics[width=.6\textwidth,origin=c,angle=0]{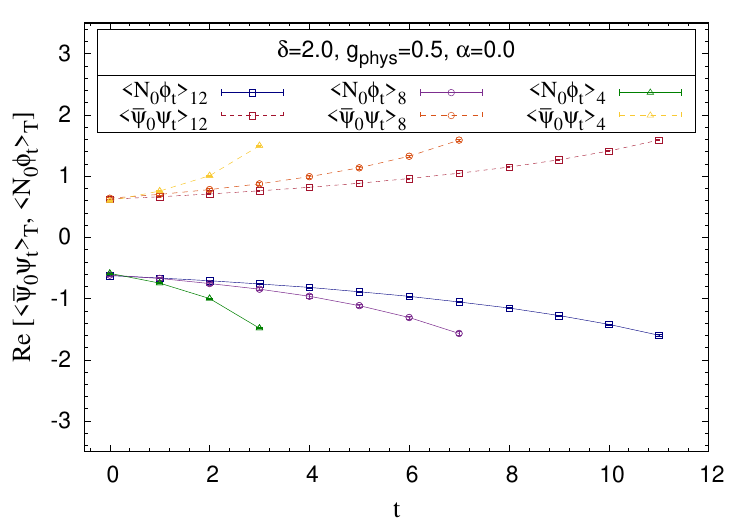}
	\includegraphics[width=.6\textwidth,origin=c,angle=0]{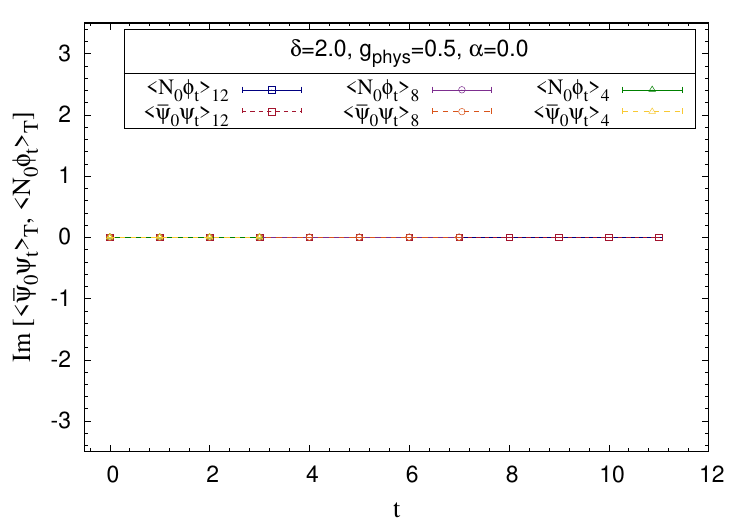}
	
	\caption {The $\mathcal{PT}$-symmetric SUSY QM with $\delta = 2$. Top panel shows the full Ward identity, the middle and bottom panels show, respectively, the real and imaginary parts of bosonic and fermionic contributions to Ward identity. The lattices used are $N_\tau = 4, 8$, and $12$.}
	\label{fig:lat-susy-d2-T8-B-Sb}	
\end{figure*}

\begin{figure*}[tbp]
	\centering	
	\includegraphics[width=.6\textwidth,origin=c,angle=0]{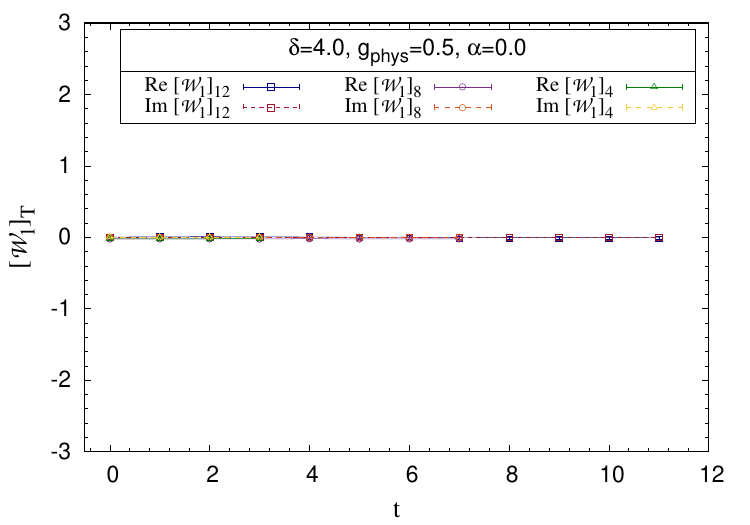}\\
	\includegraphics[width=.6\textwidth,origin=c,angle=0]{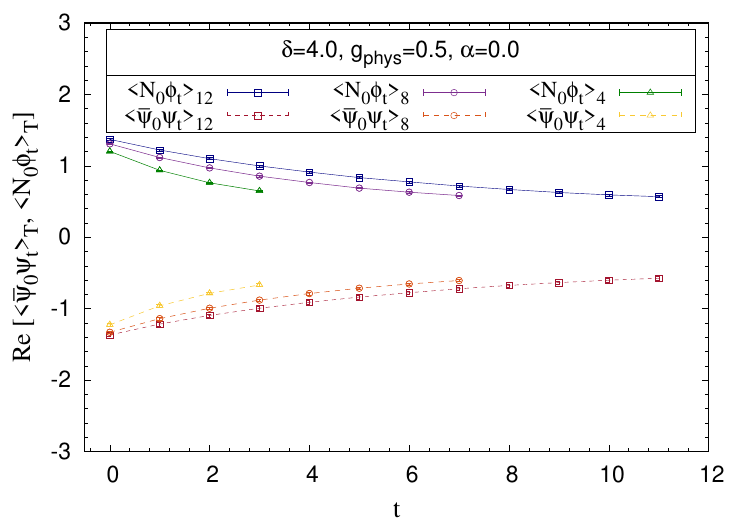}\\
	\includegraphics[width=.6\textwidth,origin=c,angle=0]{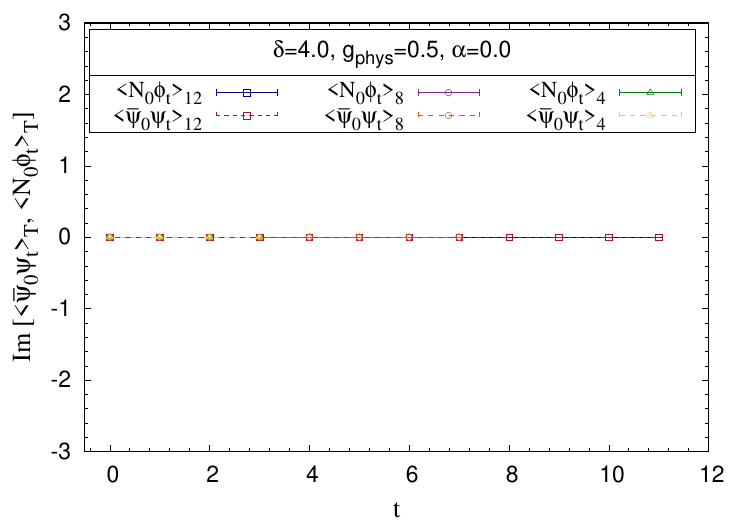}
	
	\caption {The $\mathcal{PT}$-symmetric SUSY QM with $\delta = 4$. Top panel shows the full Ward identity, the middle and bottom panels show, respectively, the real and imaginary parts of bosonic and fermionic contributions to Ward identity. The lattices used are $N_\tau = 4, 8$, and $12$.}
	\label{fig:lat-susy-d4-T8-B-Sb}	
\end{figure*}

Figures \ref{fig:lat-susy-d2-T8-B-Sb} and \ref{fig:lat-susy-d4-T8-B-Sb} show the Ward identities for $\delta = 2$ and $\delta = 4$, respectively, on an $N_\tau = 8$ lattice. 
The top panels show the complete Ward identity $W_1$, while the middle and bottom panels display the real and imaginary parts of the bosonic and fermionic contributions to this identity.

The simulations indicate that the bosonic and fermionic contributions effectively cancel each other within statistical uncertainties. 
This confirms that the Ward identities are satisfied, suggesting that SUSY is preserved in these models with $\mathcal{PT}$-symmetric $\delta$-even potentials.

{\it Odd $\delta$ case:} To address the singular-drift problem in these models, one can introduce a deformation parameter, $\mu_{\rm phys}$, and extract results in the limit as this parameter approaches zero.

In Ref. [\refcite{Joseph:2020gdh}] simulations were conducted for various non-zero values of $\mu_{\rm phys}$, and the $\delta = 1$ model is recovered by taking the limit $\mu_{\rm phys} \to 0$. 
The results suggest that when $\mu_{\rm phys}$ exceeds a certain threshold, the correctness criteria for the simulations are satisfied, and the probability of absolute drift decreases exponentially. 
Only the parameter space where complex Langevin method is reliable is considered and the results are analyzed in the limit $\mu_{\rm phys} \to 0$. 

\begin{figure*}[tbp]
	\centering
	\includegraphics[width=.48\textwidth,origin=c,angle=0]{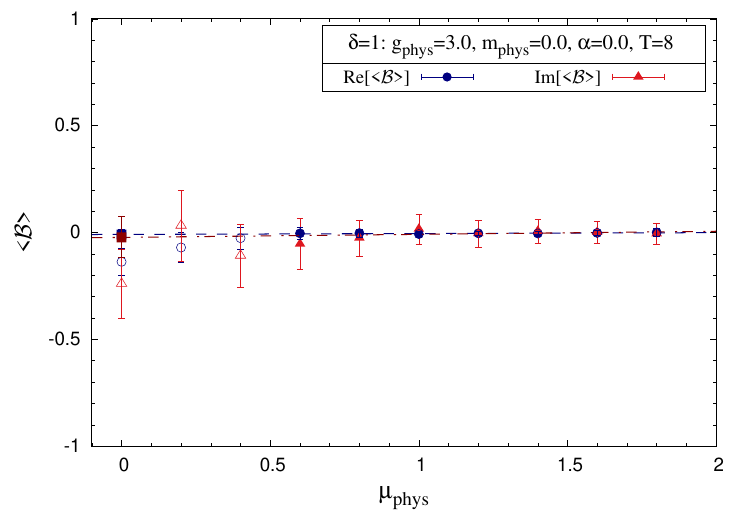} \includegraphics[width=.48\textwidth,origin=c,angle=0]{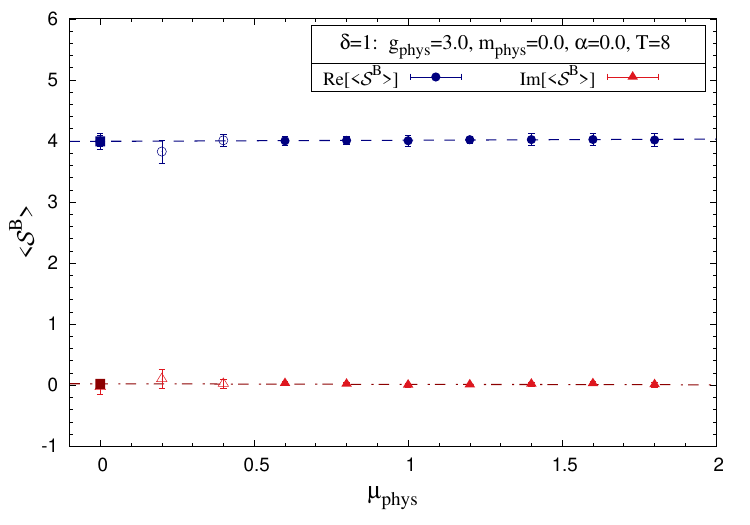}
	
	\caption{The $\mathcal{PT}$-symmetric SUSY QM with $\delta = 1$. Left panel shows the expectation values of $\B_\alpha$ and the right panel shows $\mathcal{S}^B_\alpha$, for various $\mu_{\rm phys}$ and $\lim \mu_{\rm phys} \to 0$ on a $N_\tau = 8$ lattice. The simulations were performed for $g_{\rm phys} = 3$, $m_{\rm phys} = 0$, and $\alpha = 0$. }
	\label{fig:d1bSb}	
\end{figure*}

Figure \ref{fig:d1bSb} shows the expectation values $\langle \B \rangle$ (left) and $\langle \mcS^B \rangle$ (right) for a lattice with $N_\tau = 8$ across different values of $\mu_{\rm phys}$. 
The filled data points show the results where complex Langevin simulations are deemed trustworthy, while unfilled data points indicate regions where the correctness criteria are not met. 
The lines show linear fits for the reliable parameter space, and solid squares denote the values of the respective observables in the limit $\mu_{\rm phys} \to 0$.

The simulation results demonstrate that $\langle \B \rangle$ vanishes as $\mu_{\rm phys} \to 0$, and the expectation value of the bosonic action, $\langle \mcS^B \rangle = \frac{1}{2} N_\tau$, is independent of the physical parameters in this limit. 
Thus, SUSY is preserved in the $\delta = 1$ model.

For the $\delta = 3$ model, inspired by techniques from Ref. [\refcite{Ito:2016efb}] (that have been successfully applied in Refs. [\refcite{Anagnostopoulos:2017gos,Anagnostopoulos:2020xai}]), we can introduce a fermionic deformation term in the action to manage the singular-drift problem. 
The fermionic action then becomes
\beq
\label{eqn:df}
\mcS^F = \sum_{i = 0}^{N_\tau - 1} \psib_i  \bigg(\sum_{j = 0}^{N_\tau - 1}  \nabla^{-}_{ij} + d_f + \Xi^{''}_{ij} \bigg) \psi_j,
\eeq
where $d_f$ is the deformation parameter. 
The values for $d_f$ are selected such that they ensure the correctness of the complex Langevin method. 
The $\delta = 3$ model is recovered in the limit as $d_f \to 0$. 
The simulations indicate that the correctness criteria are met above a certain $d_f$ value, with the probability of absolute drift decreasing exponentially.

\begin{figure*}[tbp]
	\centering	
	\includegraphics[width=.48\textwidth,origin=c,angle=0]{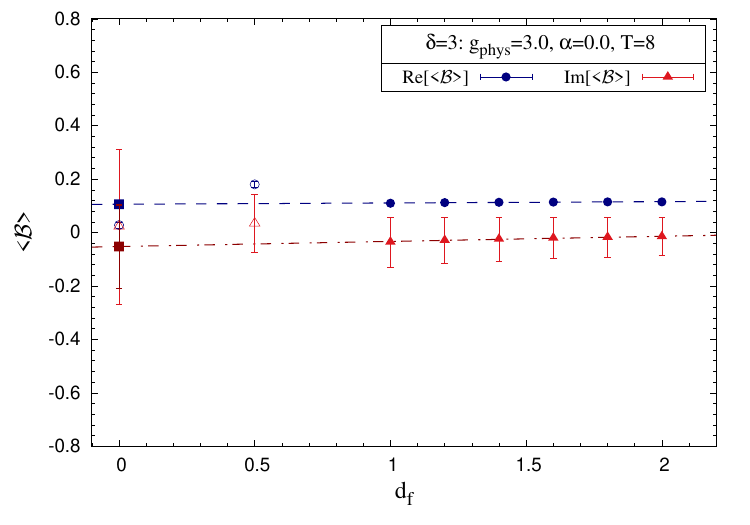}	
	\includegraphics[width=.48\textwidth,origin=c,angle=0]{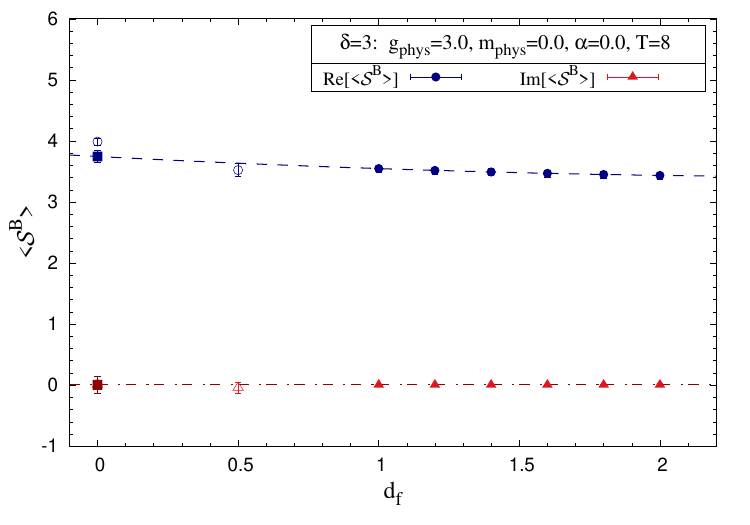}
	
	\caption{The $\mathcal{PT}$-symmetric SUSY QM with $\delta = 3$. The left and right panels show the expectation values of $\B_\alpha$ and $\mathcal{S}^B_\alpha$, respectively, against mass deformation $d_f$ parameter on a $N_\tau = 8$ lattice. The simulations were performed for the parameters $g_{\rm phys} = 3$, $m_{\rm phys} = 0$, and $\alpha = 0.0$. Extrapolations to the $d_f \to 0$ limit are represented using the dashed curves.}
	\label{fig:d3bSb}	
\end{figure*}

Figure \ref{fig:d3bSb} shows the expectation values $\langle \B \rangle$ (left) and $\langle \mcS^B \rangle$ (right) on a $N_\tau = 8$ lattice for various $d_f$ values. 
The filled data points (red triangles for the imaginary part and blue circles for the real part) correspond to parameter ranges where the complex Langevin method is reliable. 
Conversely, unfilled data points represent regions where the correctness criteria are not met. 
Dashed curves indicate linear fits for $\langle \B \rangle$ data and quadratic fits for $\langle \mcS^B \rangle$ data. Solid squares denote the values of the observables in the limit $d_f \to 0$.

The simulation results suggest that $\langle \B \rangle$ vanishes as $d_f \to 0$, and $\langle \mcS^B \rangle$ approaches $\frac{1}{2} N_\tau$, within error bars, in this limit. 
Additionally, the expectation value of $\mcS^B$ becomes independent of the physical parameters used in the model. 
These results suggest that SUSY is intact in the $\delta = 3$ model.

\subsection{Checking reliability of simulations in SUSY QM models}

To ensure the reliability of the simulations, here also, we can employ the recently proposed methods - (i) the Langevin operator criterion and (ii) tracking the decay of the probability distribution of the drift term's magnitude.

\subsubsection{Action of Langevin operator on observables}

For monitoring the correctness of our simulations, we can consider the evolution of observables $\cO_i[\phi, \theta]$ at the $i$-th site, given by:
\beq
\frac{\partial \cO_i[\phi, \theta]}{\partial \theta} = \widetilde{L}_i \cO_i[\phi, \theta], 
\eeq
where $\widetilde{L}_i$ is the Langevin operator for the $i$-th site, defined as:
\beq
\widetilde{L}_i = \left(\frac{\partial}{\partial \phi_i}  - \frac{\partial \mcS^{\rm eff}[\phi]}{\partial \phi_i}  \right) \ \frac{\partial}{\partial \phi_i}.
\eeq

Once equilibrium is achieved, we can eliminate the $\theta$ dependence from the observables. 
At this point, the condition $C_{\cO_i} \equiv \langle \widetilde{L}_i \cO_i [\phi] \rangle = 0$ serves as a criterion for the correctness of the simulations.

Considering $\B_i$ as the observable at the $i$-th site, we have:
\beq
\widetilde{L}_i \B_i =  -i \Xi^{'''}_{iii} + i \Xi^{''}_{ii} \frac{\partial \mcS^{\rm eff}}{\partial \phi_i}.
\eeq

The observable $\widetilde{L} \B$ respects the translational symmetry on the lattice. 
Thus, we can monitor its averaged value across all lattice sites.

Table \ref{tab:lat-susy-pt-symm-d1-4-LB} presents the expectation values of $\widetilde{L} \B$ for $\mathcal{PT}$-symmetric SUSY QM with $\delta = 2$ and 4.

\begin{table}[]
\tbl{The observable $\widetilde{L}\B_\alpha$ for the $\mathcal{PT}$-symmetric potentials given in Eq. \eqref{eqn:lat-pt-symm-pot} for the cases $\delta = 2, 4$. The parameters used in simulations are $\beta = 1$, $g_{\rm phys} = 0.5$, and $\alpha = 0$.}
{\begin{tabular}{| c | l | r | c |}  
				\hline
				$\Xi'(\phi)$&$N_\tau$ &  $a = N_\tau^{-1}$   &  $\langle \widetilde{L}\B_{\alpha} \rangle$  \\   
				\hline
				\hline
				&${4}$&${0.25}$  & $-0.0000(0) - i0.0104(72)$ \\
				$\delta={2}$&${8}$&${0.125}$ & $-0.0000(0) + i0.0006(59)$ \\
				&${12}$&${0.0833}$  & $-0.0000(0) - i0.0104(72)$ \\
				\hline
				&${4}$&${0.25}$ & $0.0000(0) + i0.0403(244)$ \\
				$\delta={4}$&${8}$&${0.125}$  & $0.0000(0) + i0.0027(91)$ \\
				&${12}$&${0.0833}$  & $0.0000(0) - i0.0098(64)$ \\ 
				\hline
			\end{tabular}
			\label{tab:lat-susy-pt-symm-d1-4-LB}
			}
\end{table}

\begin{figure}[tbp]
	\centering
	\includegraphics[width=.48\textwidth,origin=c,angle=0]{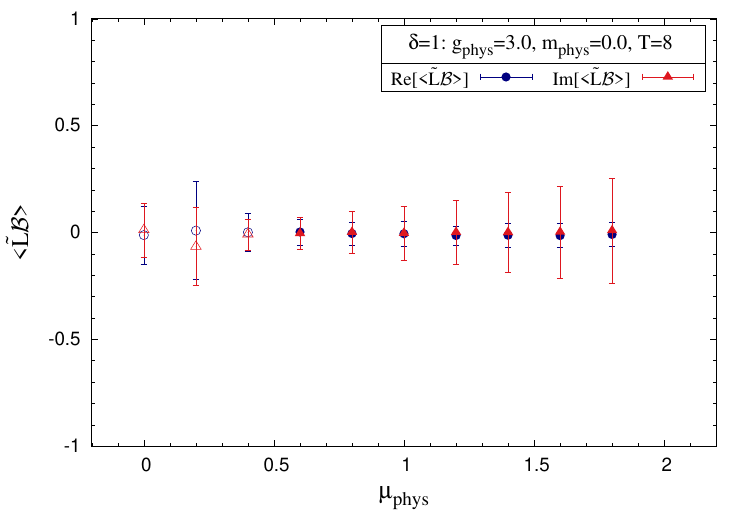} $~$ \includegraphics[width=.48\textwidth,origin=c,angle=0]{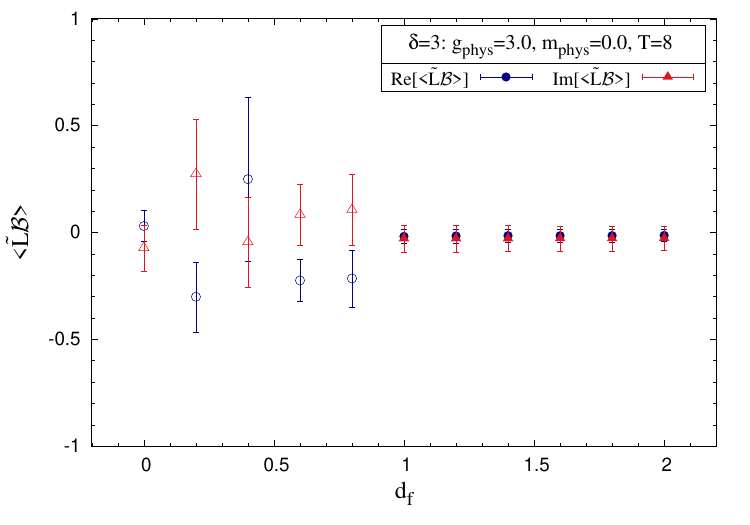}	
	
	\caption{The observable $\widetilde{L} \B_\alpha$ for the $\mathcal{PT}$ symmetric SUSY QM on a $N_\tau = 8$ lattice. Left panel shows the case $\delta = 1$ and right panel is for the case $\delta = 3$. The simulations parameters are $g_{\rm phys} = 3$, $m_{\rm phys} = 0$, and $\alpha = 0.0$.}
	\label{fig:lbd1_fig:lbd3}
\end{figure}

Figure \ref{fig:lbd1_fig:lbd3} (left) shows the expectation values of $\widetilde{L} \B$ for the $\delta = 1$ model at various $\mu_{\rm phys}$ values. 
The filled data points indicate simulation results that meet the decay-of-the-drift-term criterion, while the unfilled data points represent results that do not.

Figure \ref{fig:lbd1_fig:lbd3} (right), shows $\widetilde{L} \B$ for the $\delta = 3$ model across different values of the deformation parameter $d_f$. 
Similarly, the filled data points correspond to simulation results that satisfy the decay-of-the-drift-term criterion, whereas the unfilled points do not.

\subsection{Decay of the drift terms}

As indicated earlier, for the complex Langevin method to be reliable, the probability distribution of the drift term magnitude should decay exponentially or faster as for large values of the magnitude. 
For the models under consideration, we can define the magnitude of the mean drift in the following way
\beq
u \equiv \sqrt{\frac{1}{N_\tau} \sum_{i = 0}^{N_\tau - 1} \left| \frac{\partial \mcS^{\text{eff}}}{\partial \phi_i}\right|^2}. 
\eeq

To address the singular drift problem appropriate deformation parameters have been introduced. 
The final results are then obtained after extrapolating to the limit where these deformation parameters vanish.

\begin{figure}[htbp]
	\centering
	\includegraphics[width=.48\textwidth,origin=c,angle=0]{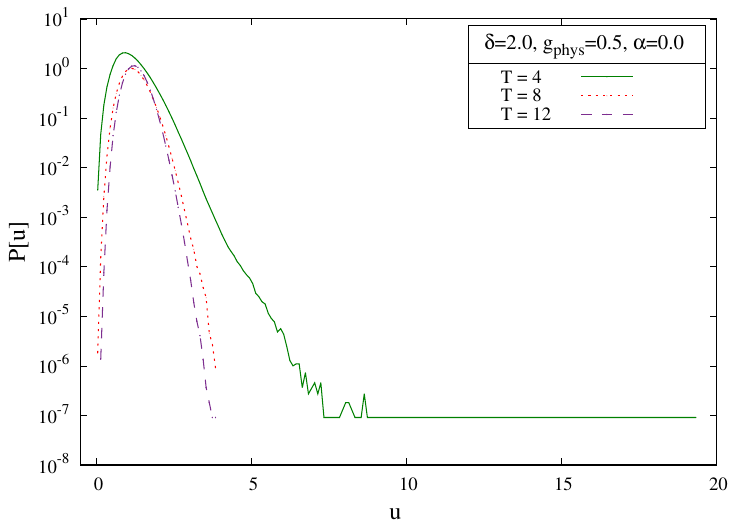}	$~$ \includegraphics[width=.48\textwidth,origin=c,angle=0]{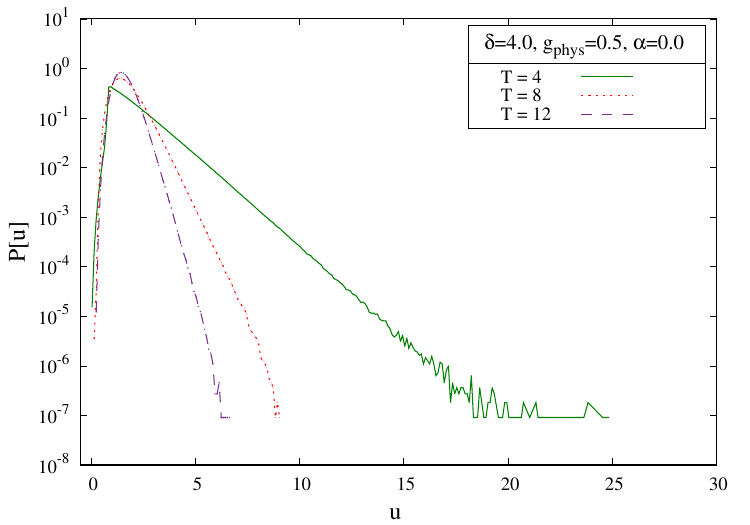}	
	\caption{The decay of the drift terms for the $\mathcal{PT}$ symmetric SUSY QM with $\delta =  2$ (left) and $\delta =  4$ (right). The simulation parameters are $g_{\rm phys} = 0.5$ and $\alpha = 0$.}
	\label{fig:pdSd2-4}
\end{figure}

The decay of the drift term for $\mathcal{PT}$ symmetric models with $\delta = 2$ and $\delta = 4$ are shown in Fig. \ref{fig:pdSd2-4}.
When the twist parameter $\alpha = 0$, the drift terms decay exponentially or faster, confirming that the simulations in this parameter regime are reliable.

\begin{figure}[tbp]
	\centering	
	\includegraphics[width=.48\textwidth,origin=c,angle=0]{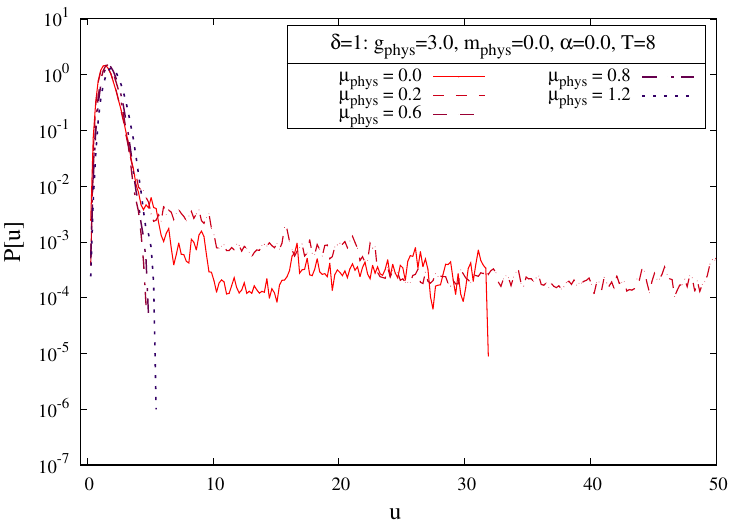} $~$ \includegraphics[width=.48\textwidth,origin=c,angle=0]{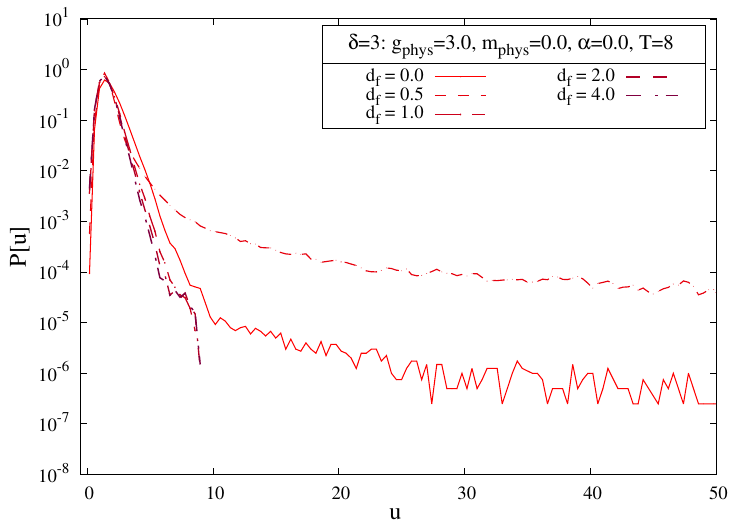}
	\caption{The decay of the drift terms for the $\mathcal{PT}$ symmetric SUSY QM with odd $\delta$ on $N_\tau = 8$ lattice. The simulation parameters were $g_{\rm phys} = 3$, $m_{\rm phys} = 0$, and $\alpha = 0.0$. The left panel shows the $\delta = 1$ case. Simulations were performed for various $\mu_{\rm phys}$ values. The right panel shows the $\delta = 3$ case. Simulations were performed for several fermionic mass deformation parameter values $d_f$.}
	\label{fig:pdSd1_fig:pdSd3}
\end{figure}

Figure \ref{fig:pdSd1_fig:pdSd3} (left) shows the drift term decay for the $\mathcal{PT}$ symmetric model with $\delta = 1$ for several values of $\mu_{\rm phys}$. 
The results show decay of the drift terms at exponentially or faster rate when $\mu_{\rm phys} \geq 0.6$, indicating that the simulations are reliable in this parameter range. 
In Fig. \ref{fig:pdSd1_fig:pdSd3} (right), the decay of the drift terms for different values of the fermionic mass deformation parameter $d_f$ are shown in the $\delta = 3$ case of the $\mathcal{PT}$ symmetric model. 
When $d_f > 1.0$, the drift terms decay exponentially or faster, suggesting that the simulations are trustworthy in this regime of parameters.

\section{Complex Langevin simulations of two-dimensional models} 
\label{sec:Complex_Langevin_simulations_of_two-dimensional_models}

In this section, we explore the application of the complex Langevin method to field theories in two dimensions. 
We will focus on a two-dimensional model with minimal supersymmetry, specifically the $\mathcal{N} = 1$ Wess-Zumino model. 
Lattice regularized version of this model is examined in scenarios where the superpotential is represented by a double-well potential.

\subsection{Scalar field theories}

In a two-dimensional Euclidean scalar field theory, the Lagrangian density is given by
\beq
\mathcal{L}_E = \hf \partial_\mu \phi  \partial_\mu \phi +\hf m^2 \phi^2 + W(\phi),
\eeq
where $\phi$ represents a dimensionless scalar field, $m$ is the mass parameter, and $W(\phi)$ denotes the interaction potential.

The corresponding Euclidean action is then expressed as:
\beq
S_E = \int d^2 x~\mathcal{L}_E.
\eeq 

We want to simulate this model using the complex Langevin method. 
First, we discretize it on a two-dimensional toroidal lattice with the following setup: The temporal and spatial extents are $\beta_t = \beta_x = La$, where $L$ is the number of lattice sites in each direction and $a$ is the lattice spacing. 
The continuous integral $\int d^2 x$ is replaced by the discrete sum $a^2 \sum_x$.

For the discretized version of the Lagrangian, the Laplacian operator can be expressed as:
\beq
\left(\partial_\mu \phi \right)^2 = - \phi \partial_\mu^2 \phi = -\frac{1}{a^2} \left[ \phi_{x}\phi_{x+\mu} + \phi_{x}\phi_{x-\mu} - 2 \phi^2_x \right],
\eeq
where $\phi_{x \pm \mu}$ denotes the field values at neighboring sites in the $\pm \mu$-th direction.

The complex Langevin update for the field configuration $\phi_x$ at lattice site $x$, at Langevin time $\theta$, with step size $\epsilon$, is given by:
\beq
\phi_{x, \theta + \epsilon} = \phi_{x, \theta} + \epsilon v_{x, \theta} + \eta_{x, \theta} \sqrt{\epsilon}, 
\eeq
where the drift term is $v_{x, \theta} = - \frac{\partial S_E}{\partial \phi_{x, \theta}}$ and $\eta_{x, \theta}$ is a real Gaussian noise term.

\subsubsection{Model with a $\phi^4$ term}

As a straightforward example, we consider the potential $W(\phi) = \lambda \phi^4$. 
Classically, this model exhibits discrete $\mathbb{Z}_2$ symmetry, where $\phi \to -\phi$. 
There is a possibility that in quantum theory, this symmetry is dynamically broken. 
We can use the expectation value of the scalar field, $\langle \phi \rangle$, as an order parameter to monitor this symmetry breaking. 
If $\langle \phi \rangle = 0$, the theory is in the symmetric phase; if $\langle \phi \rangle \neq 0$, the theory is in the symmetry-broken phase.

We use this model to test our Langevin analysis, employing a lattice regularization with dimensionless parameters $m_0^2 = m^2 a^2$ and $\lambda_0 = \lambda a^2$. 
Additionally, we introduce new parameters $\kappa$ and $\tilde{\lambda}$ (see Ref. [\refcite{De:2005ny}]) defined by:
\bea
m_0^2 &\to& \frac{1-2\tilde{\lambda}}{\kappa} -4,\\
\lambda_0 &\to& 6\frac{\tilde{\lambda}}{\kappa^2}, ~~ {\rm and} \\ 
\phi &\to& \sqrt{2\kappa} \Phi.
\eea

The corresponding lattice action is:
\beq
S = - 2 \kappa \sum_x \sum_\mu \Phi_x \Phi_{x + \mu} + \sum_x \Phi_x^2 + \tilde{\lambda} \sum_x \left( \Phi_x^2 - 1 \right)^2.
\eeq

\begin{figure}
	\centering
	\includegraphics[width=.6\textwidth]{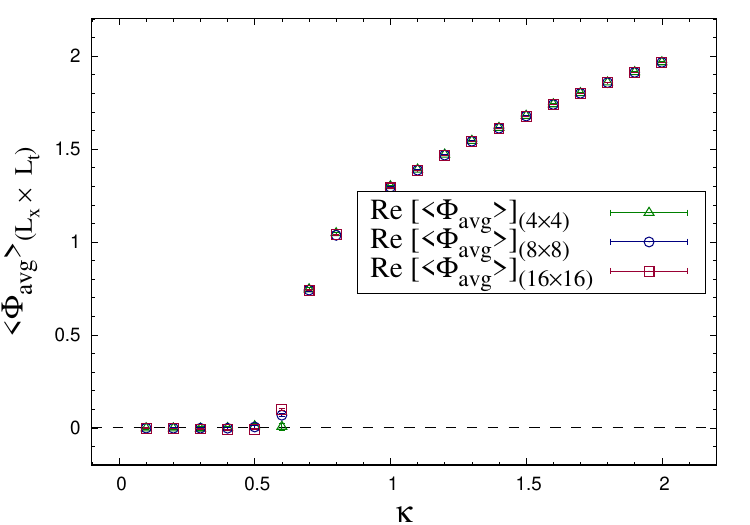}
	\includegraphics[width=.6\textwidth]{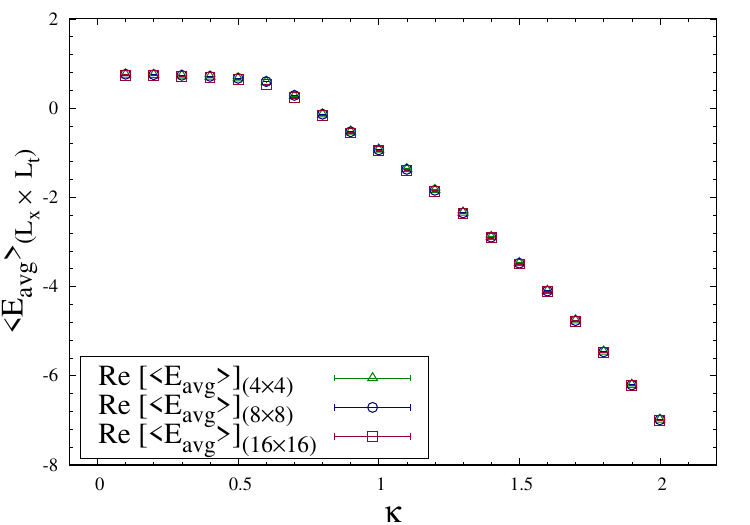}
	\includegraphics[width=.6\textwidth]{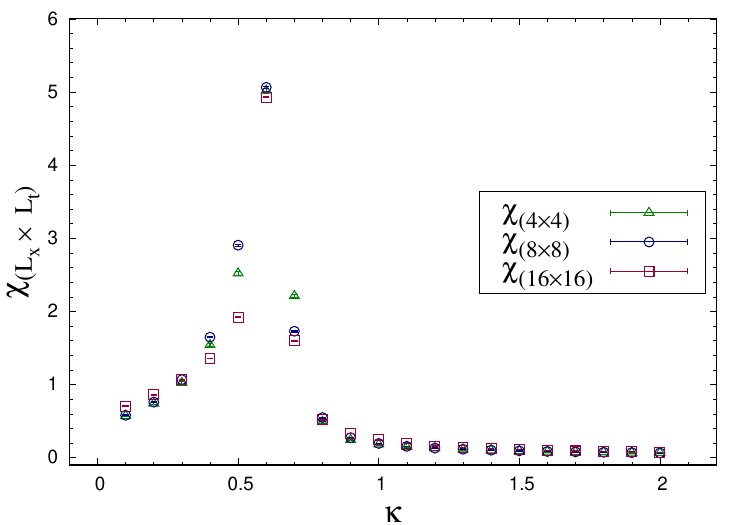}
	\caption{The two-dimensional model with $\phi^4$ potential. Expectation values of the order parameter $\Phi_{\text{avg}}$, energy $E$, and susceptibility $\chi$ against $\kappa$ are shown on the top, middle, and bottom panels, respectively. The coupling used is $\tilde{\lambda} = 0.5$.}
	\label{fig:phi4}
\end{figure}

In our simulations, we track the following observables as a function of $\kappa$: the average field $\langle \Phi_{\text{avg}} \rangle$, which is the order parameter, the energy $E$, and the susceptibility $\chi$. 
Figure \ref{fig:phi4} shows the simulation results for different lattice sizes with a fixed $\tilde{\lambda} = 0.5$. 
They suggest a phase transition around $\kappa = 0.6$. 
For $\kappa \ge 0.6$, $\langle \Phi_{\text{avg}} \rangle \neq 0$, indicating a $\mathbb{Z}_2$ symmetry-broken phase.

\subsubsection{Model with a $\mathcal{PT}$-symmetric term}

Let us look into the two-dimensional $\mathcal{PT}$-invariant scalar field theory with the potential $W(\phi) = -\lambda (i\phi)^{(2 + \delta)}$. 
Here, $\lambda$ has dimensions of $m^2$ and $\delta$ is a real parameter. 
For these models, the spectrum is real and bounded below when $\delta > 0$ and the mass parameter is non-zero. 
The theoretical reasoning behind this positivity can be illustrated with a specific example.

Consider the case where $\delta = 1$. The Lagrangian for this theory is given by:
\beq
\mathcal{L}_E = \frac{1}{2} \left(\partial_\mu \phi\right)^2 + \frac{1}{2} m^2 \phi^2 + i \lambda \phi^3.
\eeq
In a conventional real $\lambda \phi^3$ theory, Green's functions are expressed as a formal power series in $\lambda^2$ within the weak coupling expansion. 
Although this series is real, it does not alternate in sign, making it non-Borel summable. The lack of summability indicates that the spectrum is not bounded below.

However, when the coupling $\lambda$ is replaced by $i \lambda$, the theory becomes $\mathcal{PT}$-symmetric. 
In this case, the power series remains real and alternating in sign, leading to a summable perturbation series. 
This suggests that the theory indeed has a real and positive spectrum \cite{Bender:1997ps,Bender:1998gh,Milton:2003av}.

For the $\delta = 1$ model, the lattice action is given by:
\beq
S = -\sum_x \sum_\mu \phi_x \phi_{x + \mu} + \left( 2 + \frac{m_0^2}{2} \right) \sum_x \phi_x^2 + i \lambda_0 \sum_x \phi_x^3,
\eeq
where $m_0$ and $\lambda_0$ are dimensionless parameters for mass and coupling, respectively.

\begin{figure}
	\centering
	\includegraphics[width=.7\textwidth]{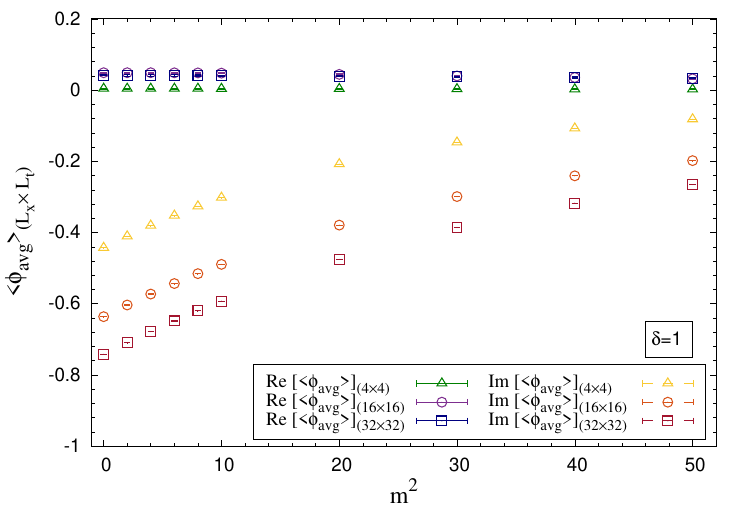}
 
	\includegraphics[width=.7\textwidth]{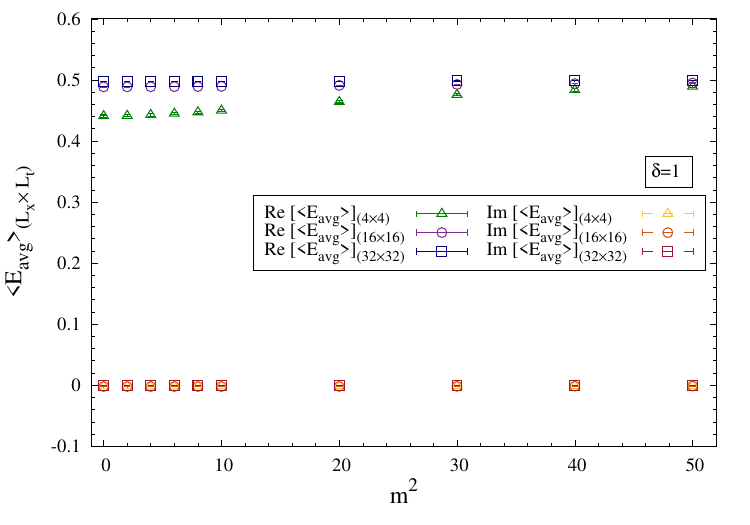}

	\caption{The two-dimensional bosonic $\mathcal{PT}$-symmetric model with $\delta = 1$ potential. Top panel shows the expectation values of the real and imaginary parts of the order parameter $\phi$ against physical mass parameter $m^2$. The bottom panel shows the energy $E$ against physical mass parameter $m^2$. Both the plots are for different lattice extents and at coupling $\lambda = 10.0$.}
	\label{fig:pt1}
\end{figure}

Figure \ref{fig:pt1} shows the simulation results for the bosonic $\mathcal{PT}$-symmetric theory with potential $\delta = 1$.

On the top panel, the expectation values of the real and imaginary parts of the average field $\phi$ versus the physical mass $m^2$ is shown for various lattice extents and with the coupling $\lambda = 10.0$. 
On the bottom panel, the ground state energy $E$ versus $m^2$ is shown for the same lattice extents and coupling. 

\begin{figure}
	\centering

	\includegraphics[width=.7\textwidth]{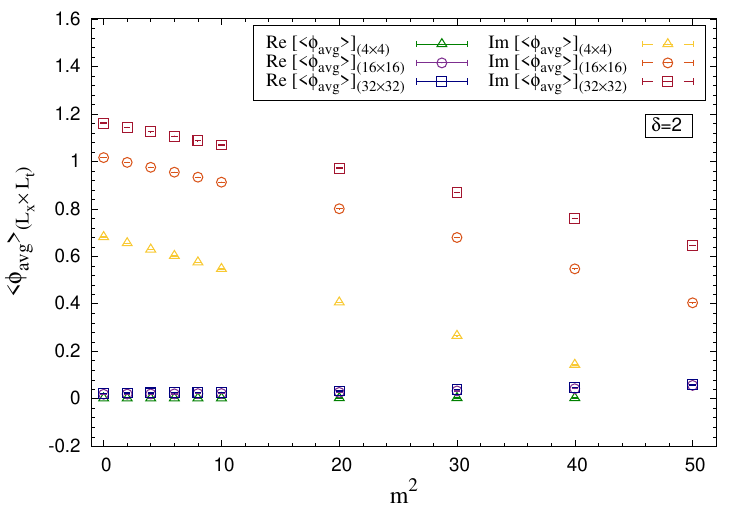}
 
	\includegraphics[width=.7\textwidth]{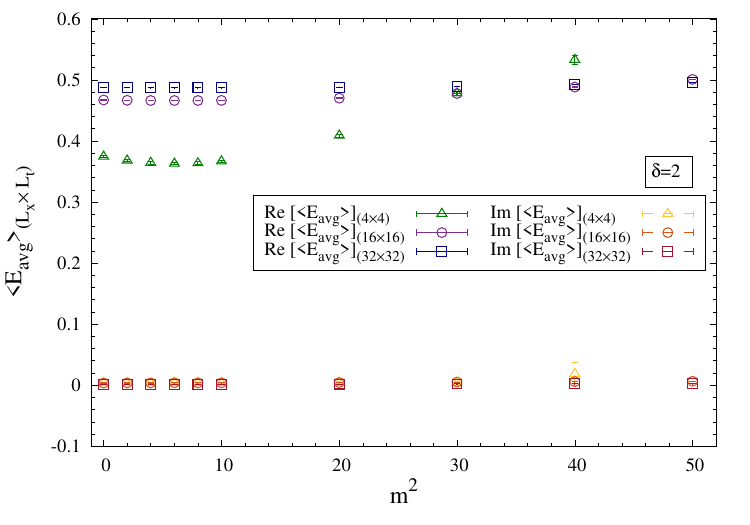}
	
	\caption{The two-dimensional bosonic $\mathcal{PT}$-symmetric model with $\delta = 2$ potential. Top panel shows the expectation values of the real and imaginary parts of the order parameter $\phi$ against physical mass parameter $m^2$. The bottom panel shows the energy $E$ against physical mass parameter $m^2$. Both the plots are for different lattice extents and at coupling $\lambda = 10.0$.}
	\label{fig:pt2}
\end{figure}

Figure \ref{fig:pt2} shows the simulation results for the same model but with $\delta = 2$.

The simulation results indicate that $\langle \phi_{\rm avg} \rangle \neq 0$, signifying that parity is broken for both $\delta = 1$ and $\delta = 2$. 
Furthermore, the expectation value of the energy is real and positive, with Re [$\langle E_{\rm avg} \rangle] > 0$ and Im [$\langle E_{\rm avg} \rangle] = 0$, confirming a real and bounded below spectrum for these interactions. 
These findings align with the analytical predictions (see Ref. [\refcite{Bender:1998gh}]).

\subsection{$\mathcal{N} = 1$ Wess-Zumino model}

In this section, we explore a supersymmetric extension of the model discussed earlier. 
(The zero- and one-dimensional cousins of this model were studied in Refs. [\refcite{Joseph:2019sof,Joseph:2020gdh}] and discussed in the previous sections.) 
We introduce fermions to the framework and examine the two-dimensional $\mathcal{N} = 1$ Wess-Zumino model, which is the simplest example of a supersymmetric quantum field theory. 

The model comprises a scalar field $\phi$ and a two-component Majorana spinor $\psi$. 
The Euclidean action for this model is given by:
\beq
S_E = \int d^2 x ~\hf \left[ \left(\partial_\mu \phi\right)^2 + \bar{\psi} \mathcal{M} \psi + W^2\left( \phi \right) \right],
\eeq
where $\mathcal{M} = \gamma^\mu \partial_\mu + W'(\phi)$ represents the fermion matrix and $W(\phi)$ is the superpotential. 

The action is invariant under supersymmetry transformations:
\beq
\delta \phi = \bar{\epsilon} \psi, 
\eeq
\beq
\delta \psi = \left[ \gamma^\mu \partial_\mu \phi - W(\phi) \right] \epsilon,
\eeq
\beq
\delta \bar{\psi} = 0.
\eeq

In this context, the Majorana spinor satisfies the relation:
\beq
\bar{\psi} = \psi^T \mathcal{C},
\eeq
where $\mathcal{C}$ is the charge conjugation matrix in Euclidean space:
\beq
\mathcal{C} = \begin{pmatrix}
0 & -1 \\
1 & 0
\end{pmatrix}.
\eeq

To perform a thorough, non-perturbative analysis of the model, we place it on a symmetric toroidal lattice using a lattice formulation developed by Golterman and Petcher \cite{Golterman:1988ta}. 
This approach separates the Euclidean continuum action into distinct bosonic and fermionic components, denoted $S_b$ and $S_f$, respectively.

The total lattice action is given by:
\beq
S = S_b + S_f,
\label{eqn:n1wz-Slat}
\eeq
where
\beq
S_b = \frac{1}{2} \left( - \phi_r \Box^2_{rr'} \phi_{r'} + W^2_r \right),
\eeq
\beq
S_f = \ln \left[ \text{Pf}(\mathcal{M}) \right] = - \frac{1}{2} \text{tr} \left[ \ln \mathcal{M} \right].
\eeq

Here, $\phi_r$ represents the bosonic field at lattice site $r$, and $W_r$ is the value of the potential at that site. 
The fermion matrix $\mathcal{M}$ is defined as:
\beq
\mathcal{M} \equiv \mathcal{M}^{\alpha \beta}_{r r'} = \gamma^{\mu}_{\alpha \beta} \mathcal{D}^{\mu}_{r r'} + \delta_{\alpha \beta} W'_{rr'},
\eeq
where $\gamma^\mu_{\alpha \beta}$ are the Dirac gamma matrices in Euclidean space, and $W'_{rr'}$ is the derivative of the potential at lattice sites $r$ and $r'$. 
The Pfaffian of $\mathcal{M}$, denoted $\text{Pf}(\mathcal{M})$, is used in the fermionic part of the action.

The symmetric difference operators used are defined as:
\beq
\mathcal{D}^\mu_{rr'} \equiv \frac{1}{2} \left[ \delta_{r + e_\mu, r'} - \delta_{r - e_\mu, r'} \right],
\eeq
\beq
\Box^n_{rr'} \equiv \frac{1}{2} \sum_\mu \left[ \delta_{r + ne_\mu, r'} + \delta_{r - ne_\mu, r'} - 2 \delta_{rr'} \right].
\eeq

These operators help in discretizing the differential operators for the bosonic and fermionic parts of the action on the lattice.

To study the theory with various superpotentials using the complex Langevin method, we apply the Euler discretized Langevin equation for the field configuration at lattice site $s$ at Langevin time $\theta$. 
The drift term for the complex Langevin update is given by:
\bea
v_{s, \theta} &=& - \frac{\partial S}{\partial \phi_{s, \theta}} \nn \\
&=& \Box^2_{s r'} \phi_{r', \theta} - W_{r'} W'_{r' s} + \left( \frac{\partial \mathcal{M}}{\partial \phi_s} \right)^{\alpha \beta}_{rr'} \left( \mathcal{M}^{-1} \right)^{\beta \alpha}_{r'r}.
\eea

Here, $\Box^2_{s r'}$ represents the lattice Laplacian operator acting on the field, $W_{r'}$ is the potential, $W'_{r' s}$ is the derivative of the potential, and $\mathcal{M}^{-1}$ denotes the inverse of the fermion matrix. 

As discussed before, to ensure the reliability of the complex Langevin simulations we analyze the distribution $P(u)$ of the magnitude $u$ of the drift term. 
At a given Langevin time $\theta$, the magnitude of the drift term is defined as:
\beq
u \equiv u_\theta = \sqrt{ \frac{1}{L^2}  \sum_s \Big | v_{s,\theta} \Big |^2}.
\eeq

By examining the decay of the distribution $P(u)$, we can verify the reliability of our simulations and ensure that the results are consistent with theoretical expectations.

Let us consider a double-well superpotential \cite{Catterall:2003ae,Baumgartner:2011jw,Wozar:2011gu} given by:
\beq
W(\phi) = \lambda \phi^2 - \frac{m^2}{4\lambda}, \quad \text{with} \quad \lambda \neq 0.
\eeq
(Note that we have used a real potential for simplicity. It is possible to consider simulations of a more general and complex superpotential.)

This potential has two classical vacua at $\phi = \pm \frac{m}{2\lambda}$. 
In the lattice formulation, we use dimensionless couplings $\lambda_0$ and $m_0$, related to the continuum parameters by $\lambda_0 = \lambda a$ and $m_0 = ma$. 
The potential and its derivative on the lattice are given by:
\beq
W_r = \lambda_0 \phi_r^2 - \frac{m_0^2}{4\lambda_0} - \frac{1}{2} \Box^1_{rr'} \phi_r,
\eeq
\beq
W'_{r r'} \equiv \frac{\partial P_r}{\partial \phi_{r'}} = 2 \lambda_0 \phi_r \delta_{rr'} - \Box^1_{r r'},
\eeq
where $\Box^1_{r r'}$ is the Wilson mass operator, which eliminates the fermion doubling problem at finite lattice spacing but does not vanish in the continuum limit. 
Due to the introduction of the Wilson term, the lattice action loses parity invariance, making the two vacuum states distinct. 
As a result, field configurations tend to reside near one of the classical vacua. 
For large values of $m_0^2 / \lambda_0$, the $\mathbb{Z}_2$ symmetry is spontaneously broken, and the field $\phi$ settles into a definite ground state.

\begin{figure}[t]
	\centering
	\includegraphics[width=.8\textwidth,origin=c,angle=0]{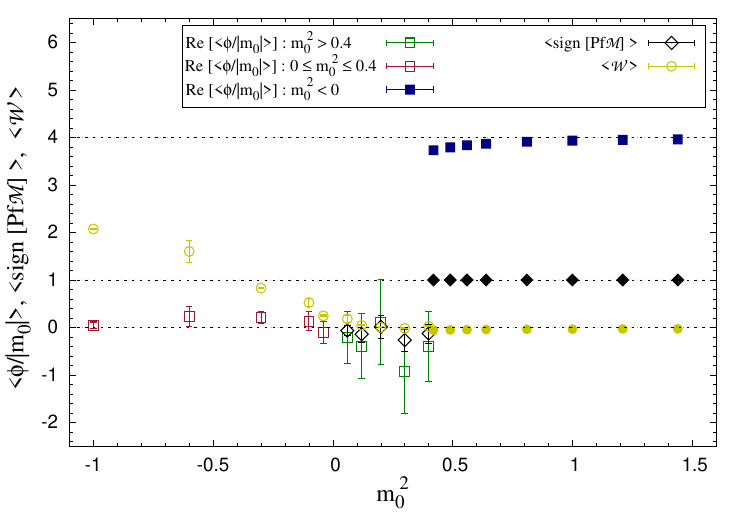}	
	
        \caption{Complex Langevin simulation results for $\mathcal{N} = 1$ Wess-Zumino model with a double-well superpotential. The observables are (on the $y$-axis) the field $\langle \phi \rangle$, the Pfaffian sign $\langle {\rm sign} \left[ {\rm Pf} \mathcal{M}\right] \rangle$, and the Ward identity $\langle \mathcal{W} \rangle$.  The lattice mass $m_0^2$ is on the $x$-axis. A fixed lattice extent, $L = 4$ and lattice coupling, $\lambda_0 = 0.125$ are used in the simulations.}
	\label{fig:wz_quad}
\end{figure}

In Fig. \ref{fig:wz_quad} we show the simulation results for the field $\langle \phi \rangle$, the sign of the Pfaffian $\langle {\rm sign} \left[ {\rm Pf} \mathcal{M}\right] \rangle$, and the Ward identity $\langle \mathcal{W} \rangle$ against lattice mass $m_0^2$. 
Simulations were performed for a fixed lattice extent $L = 4$ and lattice coupling $\lambda_0 = 0.125$.

For $m_0^2 = +1$, the field configurations (blue squares) are confined with small fluctuations around one of the classical vacua, $\phi = \frac{m_0}{2\lambda_0}$, indicating that the theory is in the $\mathbb{Z}_2$ broken phase. 
At $m_0^2 = +0.16$, we observe the tunneling behavior: field configurations (green triangles) exhibit large fluctuations, oscillating between the two classical vacua at $\phi = \pm \frac{m_0}{2\lambda_0}$. 
This behavior is further illustrated by the change in the Pfaffian sign, shown as black diamonds. 
For $m_0^2 = -1$, the field configurations, shown as red circles, undergo small fluctuations around a single vacuum state. 
This behavior is consistent with the system having an $\mathbb{Z}_2$ symmetry.

\begin{figure}[t]
	\centering
	\includegraphics[width=.7\textwidth,origin=c,angle=0]{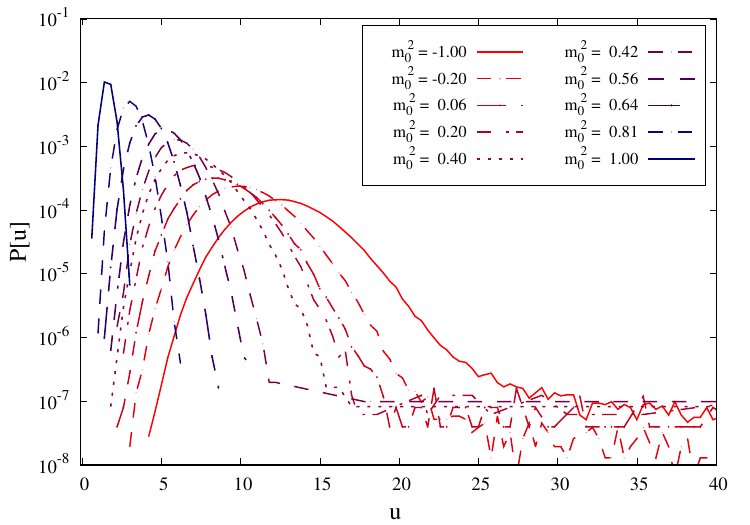}	
	
        \caption{The decay of the absolute values of the drift terms for various lattice mass $m_0^2$ are shown for the $\mathcal{N} = 1$ Wess-Zumino model with a double-well superpotential. The simulations were done for a fixed lattice extent $L = 4$ and lattice coupling $\lambda_0 = 0.125$.}
	\label{fig:wz_quad_drift}
\end{figure}

For the case of the continuum theory (infinite volume), for large $m^2/\lambda$, the scalar field selects a single, unique ground state, signaling a broken $\mathbb{Z}_2$ symmetry and unbroken supersymmetry. 
The simulations show that for $m_0^2 \geq m_{0, c}^2$, where $m_{0, c}^2$ is a critical value, the scalar field (blue squares) adopts the ground state $\phi = \frac{m_0}{2\lambda_0}$, and the sign of the Pfaffian (black diamonds) approaches $+1$. 

As $m_0^2$ decreases below this critical value, tunneling effects become apparent, with the scalar field (green squares) fluctuating between the two vacua and approaching zero, $\langle \phi \rangle \sim 0$. 
This behavior is directly linked to the Pfaffian flipping its sign, reflected in $\langle \text{sign}~\text{Pf} \mathcal{M} \rangle \sim 0$. 
These observations suggest a restoration of $\mathbb{Z}_2$ symmetry and potential dynamical SUSY breaking.

The Ward identity (yellow circles) provides further support for these findings. 
As $m_0^2$ decreases, the Ward identity $\langle \mathcal{W} \rangle$ no longer vanishes, indicating a transition from an unbroken SUSY phase to a broken SUSY phase. 
When $m_0^2 < 0$, one can see a $\mathbb{Z}_2$ symmetric phase with the scalar field (shown in red squares) around zero, $\langle \phi \rangle \sim 0$, and broken SUSY with $\langle \mathcal{W} \rangle \neq 0$.

The drift term decay is shown in Fig. \ref{fig:wz_quad_drift} - there is an exponential or faster decay for $m_0^2 > 0.42$, while for $m_0^2 \leq 0.42$, the drift term exhibits a power-law behavior. 
This power-law behavior could be indicative of the singular-drift problem, where the drift term's behavior changes due to critical phenomena or other underlying issues in the simulations.

\section{Complex Langevin analysis of the IKKT model} 
\label{sec:Complex_Langevin_analysis_of_the_IKKT_model}

Non-perturbative investigations of ten-dimensional superstring theories are crucial for understanding the emergence of spacetime. 
We live in a four-dimensional spacetime. 
An explanation for how six extra dimensions are compactified is essential for making these theories phenomenologically viable. 
Matrix models serve as powerful tools for exploring the non-perturbative aspects of superstrings. 
One such model is the IKKT (type IIB) matrix model, introduced in 1996 as a constructive definition of type IIB superstring theory in ten dimensions \cite{Ishibashi:1996xs}. 
This model, in the Schild gauge \cite{Green:1983wt}, provides a matrix regularization of the type IIB superstring action. 

Formally, one can obtain the IKKT matrix model as the zero-volume limit of ten-dimensional $\mathcal{N} = 1$ super Yang-Mills theory with SU($N$) gauge group.
In the large-$N$ limit, this equivalence between the IKKT matrix model and type IIB superstring theory is expected.
The model accommodates ten-dimensional extended $\mathcal{N} = 2$ SUSY, ensuring the inclusion of gravity. In this context, the $N \times N$ bosonic matrices represent the gravitational degrees of freedom, with their eigenvalues corresponding to spacetime points. 
Thus, spacetime is dynamically generated from the matrix degrees of freedom. 
In the large-$N$ limit, one anticipates the emergence of a smooth spacetime manifold from these eigenvalues. 
As a result of the compactification of the extra dimensions, the distribution of the eigenvalues should collapse to a lower-dimensional manifold. 
In Euclidean signature, this process requires spontaneous breaking of the ten-dimensional rotational symmetry of the model.

In this section, we explore the possibility of spontaneous symmetry breaking of SO$(10)$ symmetry in the IKKT matrix model with Euclidean signature. 
The model faces a severe sign problem due to the inherently complex Pfaffian obtained after integrating out fermions. 
The Pfaffian phase is crucial for determining the correct vacuum state of the model. 
Complex Langevin simulations of the Euclidean IKKT matrix model presents challenges, including the singular-drift problem. 
To address this, we propose introducing mass deformations to the IKKT model, in a supersymmetry preserving manner. 
Complex Langevin simulations can then be used to explore the nature of SSB in this model, at finite values of the mass deformation parameters, with a zero-mass extrapolation applied to recover the target IKKT matrix model.

\subsection{A brief overview of the IKKT model}

The dimensional reduction of ten-dimensional maximally supersymmetric Yang-Mills theory with gauge group SU($N$) to a point yields the IKKT matrix model. 
The action of the IKKT matrix model is given by:
\bea
S_{\rm IKKT} &=&  S_\text{b} + S_\text{f}, \\
S_\text{b}  &=&  -\frac{1}{4g^2} \text{tr} \Big( [X_\mu, X_\nu][X^\mu, X^\nu] \Big), \\
S_\text{f}   &=& -\frac{1}{2g^2} \text{tr} \Big( \psib \Gamma^\mu[X_\mu,\psi] \Big).
\label{Actikkt}
\eea

In the above, $g$ is the Yang-Mills coupling in $(0 + 0)$ dimensions, $X_\mu$ represents a ten-dimensional vector with SO($9,1$) symmetry, and $\psi$ is a sixteen-component spinor in the Majorana-Weyl representation of Spin($9,1$).

The matrices for the bosonic sector $X_\mu$ (where $\mu$ ranges from 0 to 9) and the fermionic sector $\psi_\alpha$ (where $\alpha$ ranges from 1 to 16) are both $N \times N$ traceless Hermitian matrices.

The ten gamma matrices, denoted as $\Gamma^\mu$, serve as generators of the Clifford algebra $\text{C}\ell(9,1)$. 
Contraction of Lorentz indices is achieved using the Minkowski metric $\eta = \text{diag}(+, -, -, -)$.

The IKKT model is proposed as a constructive formulation of type IIB superstring theory. To illustrate this connection, we start with the Green-Schwarz action for the IIB superstring in the Schild gauge
\beq
\label{chap6:eq-GS}
S_{\rm Schild} = \int d^2 \sigma \left[ \sqrt{g} \alpha \left( \frac{1}{4} \{ X^\mu, X^\nu \}^2 - \frac{i}{2} \psib \Gamma^\mu \{X^\mu, \psi\} \right) + \beta\sqrt{g} \right],
\eeq
where $\alpha$ and $\beta$ are constants related to the string tension. 
The worldsheet coordinates are denoted by $\sigma^a = (\tau, \sigma)$, with $a = 0,~1$ corresponding to the worldsheet coordinates, and $\sigma \in [0,~2\pi)$. 
The target space coordinates are represented by $X^\mu (\tau, \sigma)$, where $\mu = 0, 1, \cdots, 9$. 
For a closed string, the periodic boundary condition is $X^\mu(\tau, \sigma) = X^\mu(\tau, \sigma + 2\pi)$.

It is important to clarify that the worldsheet refers to the two-dimensional surface traced by a moving string, while the target space denotes the spacetime through which the string moves. 
The Poisson bracket on the worldsheet, denoted by $\{~,~\}$, is defined as
\begin{equation}
\{X(\sigma), Y(\sigma)\} = \frac{1}{\sqrt{g}}g^{ab} \partial_a X \partial_b Y,
\end{equation}
where $g^{ab}$ is the worldsheet metric and $g = \det(g^{ab})$.

To transition from the continuum description to a matrix formulation, we use the Goldstone-Hoppe regularization scheme. 
In the large $N$ limit, where $N$ represents the size of the matrices and acts as a regulator in this scheme, the correspondence between the continuum and matrix-regularized models is given by
\beq
\label{GHcorr}
-i [~,~] \longleftrightarrow \{~,~\} \quad \text{and} \quad \text{Tr} \longleftrightarrow \int d^2\sigma \sqrt{g}.
\eeq

Applying this correspondence to the Green-Schwarz action in the Schild gauge, the matrix version of the action becomes
\beq
\label{SMat}
S_{\rm matrix}(\beta) = \alpha \left( -\frac{1}{4} \text{Tr} [X_\mu, X_\nu]^2 - \frac{1}{2} \text{Tr} (\psib \Gamma^\mu[X_\mu, \psi]) \right) + \beta \text{Tr} \mathbbm{1}_N,
\eeq
where $\mathbbm{1}_N$ denotes the $N \times N$ identity matrix. The partition function for this matrix model is
\beq
Z = \sum_{N = 0}^\infty \int dX d\psi \, \text{e}^{iS_{\rm matrix}(\beta)}.
\eeq

This formulation captures the essence of the ten-dimensional type IIB superstring theory in a matrix regularization framework.

The action presented in Eq. \eqref{SMat} is identical to that in Eq. \eqref{Actikkt}, except for the $\beta N$ term. 
In Ref. [\refcite{Ishibashi:1996xs}], the parameter $\beta$ in the matrix-regularized Schild action was interpreted as the `chemical potential' associated with the $N$ eigenvalues of $X_{\mu}$. 
The IKKT matrix model, given in Eq. \eqref{Actikkt}, provides a comprehensive framework for describing a varying number of D-objects -- such as instantons, strings, branes, and their combinations -- represented by block-diagonal matrices. 
The off-diagonal blocks in this setup correspond to the interactions between these distinct objects.

To illustrate how these objects are naturally included in the model, consider a background with a block-diagonal structure:
\beq
X_\mu = 
\begin{pmatrix}
    X_\mu^{(1)} & 0 \\
    0 & X_\mu^{(2)}
\end{pmatrix}.
\eeq
The classical equations of motion for $X_\mu$ then separate into equations for two independent blocks:
\beq
[X_\nu^{(1)},[X^{\mu^{(1)}}, X^{\nu^{(1)}}]] = 0 \quad \text{and} \quad [X_\nu^{(2)},[X^{\mu^{(2)}}, X^{\nu^{(2)}}]] = 0.
\eeq

This indicates the presence of two separate entities, $X_\mu^{(1)}$ and $X_\mu^{(2)}$. 
These entities can be further decomposed as:
\begin{align}
X_\mu^{(1)} &= x_\mu^{(1)}\mathbbm{1} + \widetilde{X_\mu^{(1)}}, \quad x_\mu^{(1)} = \frac{1}{N^{(1)}} \text{Tr}(X_\mu^{(1)}), \quad \text{Tr}(\widetilde{X_\mu^{(1)}}) = 0; \\
X_\mu^{(2)} &= x_\mu^{(2)}\mathbbm{1} + \widetilde{X_\mu^{(2)}}, \quad x_\mu^{(2)} = \frac{1}{N^{(2)}} \text{Tr}(X_\mu^{(2)}), \quad \text{Tr}(\widetilde{X_\mu^{(2)}}) = 0.
\end{align}
Here, $x_\mu^{(1)}$ and $x_\mu^{(2)}$ denote the centers of mass for the matrices $X_\mu^{(1)}$ and $X_\mu^{(2)}$, respectively. 
Extending this to an arbitrary number of objects is straightforward, similar to a many-body configuration in second quantization. 
Interactions between these blocks are analyzed by computing the effective action for the submatrices and integrating over the remaining degrees of freedom. 
The appearance of the $\beta$ term in the one-loop effective action of a single D-string in flat spacetime points to the fact that the matrix-regularized Schild action, as presented in Eq. \eqref{SMat}, represents a low-energy effective theory of a single D-string. 
Thus, the IKKT model, encoding multiple objects rather than a single D-string, strongly supports its role as a constructive definition of type IIB superstring theory.

Let us briefly explore the symmetries inherent in the IKKT model:

\begin{itemize}
\item {\it SU($N$) Gauge Symmetry:} The model, derived from ten-dimensional super Yang-Mills theory, is invariant under SU($N$) gauge transformations. 
Specifically, the gauge transformations are given by
\begin{equation}
\label{gaugetransfo}
X_\mu \rightarrow U^\dagger X_\mu U, \quad \psi_\alpha \rightarrow U^\dagger \psi_\alpha U,
\end{equation}
where $U$ is an element of SU($N$).

\item {\it SO(9,1) Lorentz Symmetry:} The IKKT model exhibits explicit Lorentz symmetry. 
The invariance of both the bosonic and fermionic components of the action under Lorentz transformations is straightforward to verify.

\item {\it $\mathcal{N} = 2$ Supersymmetry:} The model has invariance under specific supersymmetric transformations. 
These are:
\begin{align}
\delta^{(1)} X_\mu &= i \bar{\epsilon}_1 \Gamma_\mu \psi, \\
\delta^{(1)} \psi &= \frac{i}{2} \Gamma^{\mu \nu} [X_\mu, X_\nu] \epsilon_1,
\end{align}
and
\begin{align}
\delta^{(2)} X_\mu &= 0, \\
\delta^{(2)} \psi &= \epsilon_2,
\end{align}
where $\epsilon_1$ and $\epsilon_2$ are constant Majorana-Weyl spinors representing the supersymmetric transformation parameters. 
The transformations $\delta^{(1)}$ are remnants of the $\mathcal{N} = 1$ supersymmetry in ten-dimensional theory, after dimensional reduction. 
However, due to the reduction to a point, the commutation relation of $\delta^{(1)}$ is trivial, and thus it does not correspond to spacetime supersymmetry. 

Nonetheless, an additional supersymmetry, $\delta^{(2)}$, along with a bosonic symmetry emerges after reduction. 
By combining $\delta^{(1)}$ and $\delta^{(2)}$ as follows:
\begin{align}
\tilde{\delta}^{(1)} &= \delta^{(1)} + \delta^{(2)}, \\
\tilde{\delta}^{(2)} &= i (\delta^{(1)} - \delta^{(2)}),
\end{align}
we obtain the commutators:
\begin{align}
[\tilde{\delta}^{(1)}_\epsilon, \tilde{\delta}^{(2)}_\xi] X_\mu &= 2 i \bar{\epsilon} \Gamma_\mu \xi \delta^{ij}, \\
[\tilde{\delta}^{(1)}_\epsilon, \tilde{\delta}^{(2)}_\xi] \psi &= 0,
\end{align}
which form the $\mathcal{N} = 2$ supersymmetry algebra. 
The commutator of two supersymmetry transformations corresponds to a translation, which can be interpreted as a spacetime translation, with the eigenvalues of the bosonic matrices representing points in spacetime. 
Thus, the transformations $\tilde{\delta}^{(1)}$ and $\tilde{\delta}^{(2)}$ manifest ten-dimensional $\mathcal{N} = 2$ spacetime supersymmetry.
\end{itemize}

\subsection{IKKT model with Euclidean signature}

In the Euclidean IKKT model, obtained by Wick rotation from the Lorentzian version, the transformation for the gamma matrices and coordinates is:
\beq
\Gamma^0 \rightarrow -i \Gamma^{10}, \quad X_0 \rightarrow i X_{10},
\eeq
and the contractions are now performed using the Euclidean metric $\delta^{\mu \nu}$ instead of the Minkowski metric.

To work out the explicit representation of the $\Gamma^{\mu} (\mu = 1, 2, \cdots, 10)$ matrices in a ten-dimensional Euclidean spacetime, we use the Clifford algebra:
\beq
\{ \Gamma^\mu, \Gamma^\nu \} = 2 \delta^{\mu \nu} \mathbbm{1}_{32}.
\eeq

In even dimensions $d$, the gamma matrices of size $2^{\frac{d}{2}} \times 2^{\frac{d}{2}}$ can be chosen to be off-block diagonal:
\beq
\Gamma^\mu =    
\begin{pmatrix}
    0 & \gamma^\mu \\
    \widetilde{\gamma}^\mu & 0
\end{pmatrix},
\eeq
where $\gamma^\mu$ are $2^{\frac{d}{2} - 1} \times 2^{\frac{d}{2} - 1}$ matrices, and $\widetilde{\gamma}^\mu = (\gamma^\mu)^\dagger$.

For $d = 10$, we have the chiral matrix $\Gamma^{11}$ and the charge conjugation matrix $\mathcal{C}$ given by:
\beq
\Gamma^{11} = 
\begin{pmatrix}
    \mathbbm{1}_{16} & 0\\
    0 & -\mathbbm{1}_{16}
\end{pmatrix}, \quad 
\mathcal{C} = \begin{pmatrix}
    0 & \mathbbm{1}_{16}\\
    -\mathbbm{1}_{16} & 0
\end{pmatrix}.
\eeq

Ref. [\refcite{Ambjorn:2000dx}] provides an explicit representation of the ten $16 \times 16$ gamma matrices:
\beq
\begin{aligned}
\Gamma^1 &= i \sigma^2 \otimes \sigma^2 \otimes \sigma^2 \otimes \sigma^2, & \quad \Gamma^2 &= i \sigma^2 \otimes \sigma^2 \otimes \bf{1} \otimes \sigma^1, \\
\Gamma^3 &= i \sigma^2 \otimes \sigma^2 \otimes \bf{1} \otimes \sigma^3, & \Gamma^4 &= i \sigma^2 \otimes \sigma^1 \otimes \sigma^2 \otimes \bf{1}, \\
\Gamma^5 &= i \sigma^2 \otimes \sigma^3 \otimes \sigma^2 \otimes \bf{1}, & \Gamma^6 &= i \sigma^2 \otimes \bf{1} \otimes \sigma^1 \otimes \sigma^2, \\
\Gamma^7 &= i \sigma^2 \otimes \bf{1} \otimes \sigma^3 \otimes \sigma^2, & \Gamma^8 &= i \sigma^1 \otimes \bf{1} \otimes \bf{1} \otimes \bf{1}, \\
\Gamma^9 &= i \sigma^3 \otimes \bf{1} \otimes \bf{1} \otimes \bf{1}, & \Gamma^{10} &=  \bf{1} \otimes \bf{1} \otimes \bf{1} \otimes \bf{1}.
\end{aligned}
\eeq

In the above, $\sigma^{i} (i = 1, 2, 3)$ are the Pauli matrices and $\bf{1}$ represents the $2 \times 2$ identity matrix. 

In the Euclidean representation of the IKKT model, the charge conjugation matrix $\mathcal{C}$ satisfies:
\beq
\mathcal{C} \Gamma^\mu \mathcal{C}^\dagger = (\Gamma^\mu)^T \quad \text{and} \quad \mathcal{C}^T = \mathcal{C}.
\eeq

In this specific representation, $\mathcal{C}$ becomes the identity matrix.

The structure of gamma matrices in even-dimensional spacetime always allows for an off-diagonal block form. In the Euclidean case, the Weyl projection corresponds to applying the chirality projection operator, which is determined by the highest-rank gamma matrix. 
Consequently, the SO($9,1$) Lorentz symmetry in the Lorentzian version of the IKKT model transforms into the SO(10) rotational symmetry in the Euclidean signature.

For numerical studies of the IKKT matrix model, we use the relation between the Yang-Mills coupling $g$, the number of colors $N$, and the t'Hooft coupling $\lambda$:
\beq
\lambda = g^2 N
\eeq

Rewriting the action, Eq. \eqref{Actikkt}, in terms of $\lambda$ we get
\beq
S = -\frac{N}{4\lambda} \text{tr} \Big( [X_\mu, X_\nu][X^\mu, X^\nu] \Big) - \frac{N}{2\lambda} \text{tr} \Big( \psib \Gamma^\mu[X_\mu, \psi] \Big).
\eeq

To achieve a finite and well-defined partition function, we perform the following field redefinitions:
\beq
X_\mu \to \lambda^{\frac{1}{4}} X_\mu, \quad \psi \to \lambda^{\frac{3}{8}} \psi.
\eeq

This leads to the action:
\beq
S_{\rm IKKT} = S_\text{b} + S_\text{f},
\eeq
where
\beq
S_\text{b} = - \frac{1}{4} N \text{tr} \left([X_\mu, X_\nu]^2 \right),
\eeq
and
\beq
S_\text{f} = - \frac{1}{2} N \text{tr} \left(\psi_\alpha (\mathcal{C} \Gamma^\mu)_{\alpha\beta}[X_\mu, \psi_\beta] \right).
\eeq

In this action, the $N \times N$ traceless Hermitian matrices $X_\mu$ (for $\mu = 1, 2, \cdots, 10$) and the 16-component Majorana-Weyl spinors $\psi_\alpha$ (for $\alpha = 1, 2, \cdots, 16$) transform as vectors and spinors under SO(10) symmetry, respectively. 
The action is invariant under SU($N$) gauge symmetry, extended $\mathcal{N} = 2$ supersymmetry, and SO(10) rotational symmetry.

It can be shown that the partition function
\beq
\label{eqn:ikkt-action}
Z = \int dX d\psi e^{-S_{\rm IKKT}} 
\eeq
is finite and well-defined \cite{Krauth:1998xh,Austing:2001pk}.

After integrating out the fermionic degrees of freedom, the partition function, is given by:
\beq
Z = \int dX \, {\rm Pf} \mathcal{M} \, e^{-S_{\rm b}} = \int dX \, e^{-S_{\rm eff}},
\eeq
where the effective action is:
\beq
S_{\rm eff} = S_\text{b} - \ln {\rm Pf} \mathcal{M}.
\eeq

Here, $\mathcal{M}$ is the fermionic operator, a $16(N^2 - 1) \times 16(N^2 - 1)$ antisymmetric matrix. 
To determine the explicit form of $\mathcal{M}$, we expand $X_\mu$ and $\psi_\alpha$ in terms of the $N^2-1$ generators $ \{t^a\} $ of SU($N$):
\beq
X_\mu = \sum_{a = 1}^{N^2-1} X_\mu^a t^a \quad \text{and} \quad \psi_\alpha = \sum_{b = 1}^{N^2-1} \psi_\alpha^b t^b,
\eeq
where $X_\mu^a$ are real and $\psi_\alpha^b$ are Grassmann variables. 
The traceless, Hermitian generators are normalized by $\text{tr} (t^a t^b) = \delta^{ab}$.

Using the properties of SU($N$) structure constants, the matrix $\mathcal{M}$ is given by:
\beq
\mathcal{M}_{\alpha a, \beta b} = \frac{N}{2} \Gamma_{\alpha \beta}^\mu \, \text{tr} \left( X_\mu [t^a, t^b] \right).
\eeq

The eigenvalues of the $X_\mu$ matrices can be interpreted as defining the `radial extent' of spacetime in each direction:
\beq
\langle \lambda_\mu \rangle \equiv \left\langle \frac{1}{N} \text{tr} \left( X_\mu^2 \right) \right\rangle.
\eeq

This quantity $\lambda_\mu$ serves as an order parameter for SSB. 
In the large-$N$ limit, if the extents $\lambda_\mu$ are not equivalent, meaning they vary along different directions, then SO(10) symmetry is said to spontaneously break down to SO($d$), where $d < 10$.

Numerical studies of the bosonic IKKT model using Monte Carlo methods and $1/D$ expansions have not observed any SSB \cite{Hotta:1998en}. 
Subsequent phase-quenched Monte Carlo studies also failed to find evidence for SSB \cite{Ambjorn:2000dx,Anagnostopoulos:2012ib}. 
These findings suggest that the complex phase of the Pfaffian might be crucial in SSB, with its fluctuations indicating a severe sign problem. 
Therefore, phase-quenched approximations may be inadequate. 

\subsection{Complex Langevin simulations of the IKKT model}

This section explores the application of the complex Langevin method to the Euclidean IKKT model. 
The update equation for the bosonic matrices $X_\mu$ at fictitious Langevin time $\theta$ is given by:
\beq
\frac{d(X_{\mu})_{ij}}{d\theta} = - \frac{\partial S_{\rm eff}}{\partial (X_{\mu})_{ji}} + (\eta_{\mu})_{ij}(\theta),
\eeq
where $\eta_{\mu}(\theta)$ is a Hermitian Gaussian noise term following the probability distribution:
\beq
P[\eta_{\mu}(\theta)] \sim \exp \left(-\frac{1}{2} \int {\rm tr} \left( \eta^2_{\mu}(\theta) \right) \right).
\eeq

The gradient of the effective action $S_{\rm eff}$ can be computed using the following expressions:
\beq
\frac{\partial S_{\rm eff}}{\partial (X_\mu)_{ji}} = \frac{\partial S_\text{b}}{\partial (X_\mu)_{ji}} - \frac{1}{2} \frac{\partial \left( \text{tr} \left[ \ln \mathcal{M} \right] \right)}{\partial (X_\mu)_{ji}},
\eeq
where the contributions are:
\beq
\frac{\partial S_\text{b}}{\partial (X_\mu)_{ji}} = -N \left( \left[ X_\nu, \left[X_\mu, X_\nu\right] \right] \right)_{ij},
\eeq
and
\beq
- \frac{1}{2} \frac{\partial \left( \text{tr} \left[ \ln \mathcal{M} \right] \right)}{\partial (X_\mu)_{ji}} = -\frac{1}{2} \text{tr} \left[ \mathcal{M}^{-1} \frac{\partial \mathcal{M}}{\partial (X_\mu)_{ji}} \right].
\eeq

The derivative of the fermionic operator $\mathcal{M}$ with respect to $ (X_\mu)_{ji} $ is
\beq
\frac{\partial \mathcal{M}}{\partial (X_\mu)_{ji}} = \frac{N}{2} \Gamma_{\alpha \beta}^{\mu} \left( \left[ t^a, t^b \right] \right)_{ij}.
\eeq

These equations provide the necessary components to implement the complex Langevin method for numerical simulations of the Euclidean IKKT model.

To ensure the correctness of the complex Langevin method, it is important to monitor the distribution of the magnitude of the drift term. 
The magnitude of the drift term $u$ is defined as:
\beq
u \equiv \sqrt{\frac{1}{10N^2} \sum_{\mu=1}^{10} \sum_{i,j=1}^{N} \left| \frac{\partial S_{\rm eff}}{\partial (X_{\mu})_{ji}} \right|^2}.
\eeq

\begin{figure}[htbp]
	\centering
	\includegraphics[width=.7\textwidth,origin=c,angle=0]{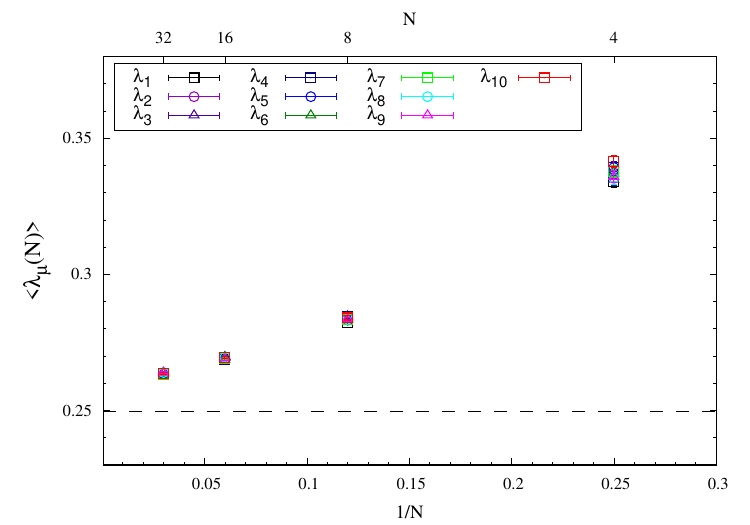}
	\includegraphics[width=.65\textwidth,origin=c,angle=0]{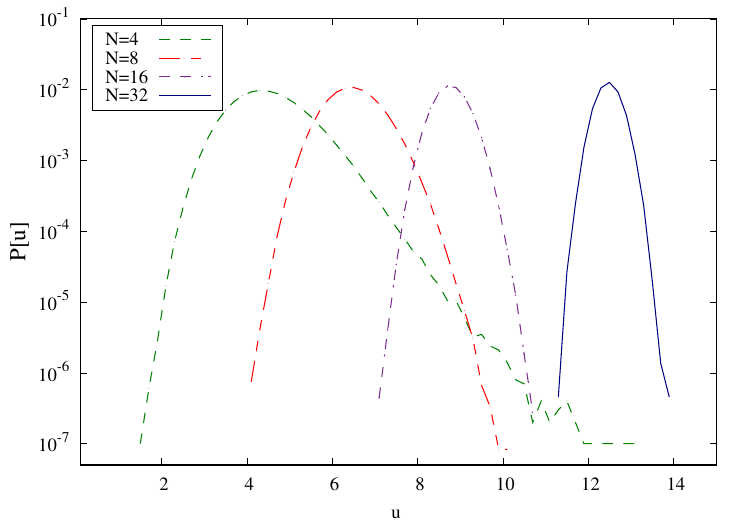}
	\caption{The top panel shows the expectation value of order parameter $\lambda_{\mu}$ for the bosonic IKKT model. The bottom panel shows the probability of magnitude of drift term for various $N$.} 
	\label{fig:bos}
\end{figure}

Figure \ref{fig:bos} shows the results from complex Langevin simulations for the bosonic IKKT model. 
The top panel illustrates that the SO(10) symmetry remains unbroken even for finite $N$ and approaches the expected analytical result in the large-$N$ limit \cite{Hotta:1998en}. 
This suggests that the simulation is correctly capturing the symmetry properties of the model. 
The bottom panel demonstrates that the probability of drift decreases exponentially or faster, indicating that the simulations are reliable.

However, while applying the complex Langevin method to the Euclidean IKKT model, two significant issues are encountered\cite{Kumar:2023nya,Kumar:2022giw}:
\begin{itemize}
\item[1.] \textit{Excursion Problem:} This occurs when the field values move far from their typical region of configuration space, potentially leading to non-physical results or instability in the simulations. 
This problem often manifests as fields taking on unreasonably large values, which can affect the reliability of the simulation.

\item[2.] \textit{Singular-Drift Problem:} This problem arises when the drift term becomes very large or singular, causing difficulties in the numerical integration of the Langevin equation. 
A large drift term can lead to numerical instabilities and inaccurate results.

\end{itemize}

In the following subsections, we discuss these issues in more detail and explore strategies to address them effectively.

\subsubsection{Taming the excursion problem with gauge cooling}
\label{sec:Taming_the_excursion_problem_with_gauge_cooling}

The Pfaffian of the IKKT model is complex in general. 
In complex Langevin simulations this can lead to excursions of the bosonic matrices $X_\mu$ into anti-Hermitian directions, expanding the gauge group from SU($N$) to $\text{SL}(N, \mathbb{C})$. 
This results in the {\it excursion problem}, where $X_\mu$ moves away from its expected SU($N$) configuration.

A proposed solution to mitigate this issue is \textit{gauge cooling} \cite{Seiler:2012wz}. 
Gauge cooling helps to keep the matrices close to the Hermitian subspace by correcting the deviations caused by the complex nature of the Pfaffian.

To monitor the deviation of $X_\mu$ from Hermitian configurations, we use the \textit{Hermiticity norm} \cite{Nagata:2016vkn}:
\beq
\mathcal{N}_{\rm H} \equiv - \frac{1}{10N} \sum_{\mu} \text{tr} \left( \left[ X_\mu - X_\mu^\dagger \right]^2 \right).
\eeq
This norm quantifies how much $X_\mu$ deviates from being Hermitian. 

The matrix fields $X_\mu$ show invariance under an enlarged gauge symmetry:
\beq
X_\mu \rightarrow g X_\mu g^{-1}, \quad g \in \text{SL}(N, \mathbb{C}),
\eeq
where
\beq
g = \text{e}^{-\alpha \delta \mathcal{N}_{\rm H}}, \quad \delta \mathcal{N}_{\rm H} = \frac{1}{N} \sum_\mu \left[ X_\mu, X_\mu^\dagger \right], \quad \alpha \in \mathbb{R}^+.
\eeq
The norm $\mathcal{N}_{\rm H}$ is not invariant under this transformation.
Thus, applying the gauge transformation iteratively at each Langevin step minimizes $\mathcal{N}_{\rm H}$ and helps to keep $X_\mu$ close to Hermitian configurations. 
Complex Langevin simulations show that after applying gauge cooling, the Hermiticity norm $\mathcal{N}_{\rm H}$ is well-controlled, indicating that the issue of excursions is effectively managed.

\subsubsection{Taming the singular-drift problem with mass deformations}

The computation of the gradient of the effective fermionic action involves $\mathcal{M}^{-1}$, the inverse of the fermion operator. 
The singular-drift problem occurs when the eigenvalues of $\mathcal{M}$ cluster near zero. 
To address this issue, one approach is to shift the eigenvalues of the fermion operator away from zero. 
This can be achieved by introducing fermion bilinear mass deformation terms into the action \cite{Ito:2016efb}. 

Generally, these deformations take the form:
\beq
\Delta S = \frac{N}{2}\epsilon m_\mu \text{tr}(X_\mu^2) + \frac{N}{2} \text{tr} \left( \psi_\alpha (\mathcal{CA})_{\alpha\beta} \psi_\beta \right),
\eeq
where $m_\mu$ represents the mass vector and $\mathcal{A}$ is a complex anti-symmetric $16 \times 16$ matrix. 
Due to the constraints imposed by Majorana-Weyl spinors in ten dimensions, only bilinears of rank three and seven tensors (which are equivalent due to duality relations) are viable \cite{Wetterich:1982eh}. 
Specifically, $\mathcal{A}$ can be represented as $i m_{\rm f} \epsilon_{\mu\nu\sigma} \Gamma_{\mu} \Gamma_{\nu}^\dagger \Gamma_\sigma$, where $\epsilon_{\mu \nu \sigma}$ is a totally anti-symmetric 3-form. 
Here, $\epsilon$ and $m_{\rm f}$ are the deformation parameters.

These deformations not only break the SO(10) symmetry but also lead to SUSY breaking. 
The extended $\mathcal{N} = 2$ SUSY is crucial for incorporating gravity into the model. 
A recent study \cite{Anagnostopoulos:2020xai} examined similar deformations and found that the SO(10) symmetry was spontaneously broken down to SO(3), consistent with results from the Gaussian expansion method \cite{Nishimura:2011xy}. 
Investigating SSB with these deformations requires a three-step extrapolation process ($N \to \infty$, $\epsilon \to 0$, and $m_{\rm f} \to 0$), which introduces systematic errors. 
An alternative approach is to use SUSY-preserving deformations, which reduces the number of extrapolation steps to just two.

\subsection{SUSY preserving mass deformations}

It is possible to introduce supersymmetry preserving deformations \cite{Metsaev:2001bj,Blau:2001ne,Bonelli:2002mb,Cvetic:2002si, Austing:2003kd}, including a Myers term, to the original IKKT model $S_{\rm IKKT}$. 
The modified action is given by:
\bea
\label{eqn:susy-deformed-ikkt-action}
S &=& S_{\rm IKKT} + S_\Omega, \nn \\
S_{\rm  \Omega} &=& N~ \text{tr} \left( M^{\mu \nu} X_{\mu} X_{\nu} + i N^{\mu \nu \gamma} X_\mu \left[X_{\nu}, X_{\gamma} \right] + \frac{i}{8} \psib N_3 \psi \right),
\eea
where $N_3 = \Gamma^{\mu \nu \gamma} N_{\mu \nu \gamma}$, with $N^{\mu \nu \gamma}$ denoting a totally anti-symmetric tensor, and $M_{\mu \nu}$ is the mass matrix. 
This deformed model was first introduced in Ref. [\refcite{Bonelli:2002mb}]. 
Recently it became the subject of a recent study given in Ref. [\refcite{Hartnoll:2024csr}]. There, this model was termed as the polarized IKKT model.

The action of the the model remains invariant under the following SUSY transformations:
\bea
\label{eqn:susy-deformed-ikkt-transformations}
\delta X^{\mu} &=& - \frac{1}{2} \overline{\varepsilon} \Gamma^{\mu} \psi, \\
\delta \psi &=& \frac{1}{4} \left[ X^{\mu}, X^{\nu} \right] \Gamma_{\mu \nu} \varepsilon - \frac{i}{16} X^{\mu} \left( \Gamma_{\mu} N_3 + 2 N_3 \Gamma_{\mu} \right) \varepsilon,
\eea
subject to the mass/flux constraint:
\beq
\left[ N_3 (\Gamma^{\mu} N_3 + 2 N_3 \Gamma^{\mu}) + 4^3 M^{\mu \nu} \Gamma_{\nu} \right] \varepsilon = 0.
\eeq

To verify the invariance of the action, we first note that the variation arising from the term $[X_\mu, X_\nu]^2$ is canceled by the variation of the term $\psib \Gamma^\mu [X_\mu, \psi]$ due to the invariance of the undeformed IKKT model. 
We then examine the remaining terms, which split into two components: one linear in $X_\mu$ and the other proportional to the commutator of the bosonic fields. 
Both parts must vanish, leading to constraint equations.

The linear part of the variation is:
\bea
\delta S_{\text{linear}} &=& \frac{i}{4} \psib N_3 \left( - \frac{i}{16} X_\mu \left( \Gamma^\mu N_3 + 2 N_3 \Gamma^\mu \right) \varepsilon \right) + 2 M^{\mu \nu} X_\mu \left( \frac{1}{2} \psib \Gamma_\nu \varepsilon \right) \nn \\
&=& \psib \left( \frac{1}{4^3} N_3 (\Gamma^\mu N_3 + 2 N_3 \Gamma^\mu) + M^{\mu \nu} \Gamma_\nu \right) \varepsilon X_\mu,
\eea
which yields the constraint equation:
\beq
N_3 (\Gamma^\mu N_3 + 2 N_3 \Gamma^\mu) + 4^3 M^{\mu \nu} \Gamma_\nu = 0.
\eeq

The commutator part of the variation is:
\beq
\delta S_{\text{comm}} = i \psib \left( \frac{1}{16} N_3 \Gamma^{\mu \nu} + \frac{1}{16} \Gamma^\mu (\Gamma^\nu N_3 + 2 N_3 \Gamma^\nu) + \frac{3}{2} N^{\mu \nu \rho} \Gamma_\rho \right) \varepsilon [X_\mu, X_\nu].
\eeq

A particular choice of $N_3$ and $M$ that ensures the invariance of the action under the SUSY transformations is:
\bea
N_3 &=& \mu \Gamma^{789}, \\
M &=& \frac{\mu^2}{4^3} (\mathbbm{1}_7 \oplus 3\mathbbm{1}_3).
\eea

A specific solution to the constraint equations is given by:
\bea 
\label{eqn:susy-deformed-ikkt-solution}
N_{3} &=& -\Omega\Gamma^{8}{\Gamma^{9}}^{\dagger}\Gamma^{10}, \\
N^{\mu \nu \gamma} &=&  \frac{\Omega}{3!} \sum_{\mu,\nu,\gamma=8}^{10}{\epsilon^{\mu \nu \gamma}}, \\
M &=&  \frac{\Omega^2}{4^3} \left( \mathbb{I}_7 \oplus 3\mathbb{I}_3 \right).
\eea

This choice explicitly breaks the ten-dimensional SO(10) rotational symmetry to SO(7) $\times$ SO(3). 
To recover the original IKKT model and study spontaneous symmetry breaking, one can take the limit $\Omega \to 0$ in the large-$N$ regime.

For the supersymmetric deformed model, the gradient of the bosonic action includes additional contributions as follows:
\bea
\label{eqn:bos-mass-matrix}
-\frac{N M ^{\mu \nu} \partial \left( X_{\mu} X_{\nu}\right) }{\partial X_{\sigma}} &=& -2N M ^{\sigma \nu}X_{\nu}^{T} \\
&=& \frac{2\Omega^2 N}{4^3}  
\begin{cases}
    X_{\sigma}^{T} & \text{for } \sigma = 1, 2, \cdots, 7, \\
    3X_{\sigma}^{T} & \text{for } \sigma = 8, 9, 10,
\end{cases} \\
\frac{i N \partial \left( N^{\mu \nu \gamma} X_\mu \left[X_{\nu}, X_{\gamma} \right] \right) }{\partial X_{\sigma}} &=& i \Omega N \sum_{\nu, \gamma = 8}^{10} \left( \epsilon^{\sigma \nu \gamma} X_{\nu} X_{\gamma} \right) \\
&=& i \Omega N
\begin{cases}
    [X_{9}, X_{10}]^{T} & \text{for } \sigma = 8, \\
    [X_{10}, X_{8}]^{T} & \text{for } \sigma = 9, \\
    [X_{8}, X_{9}]^{T} & \text{for } \sigma = 10.
\end{cases}
\eea

The fermion bilinear deformation alters the fermion operator as follows:
\beq
\mathcal{M}_{\alpha a \beta b} \to \tilde{\mathcal{M}}_{\alpha a \beta b} = \frac{N}{2} \Gamma_{\alpha \beta}^\mu \text{tr} \left( X_\mu \left[t^a, t^b \right] \right) - \frac{i \Omega N}{8} \left( \Gamma^{8} {\Gamma^{9}}^{\dagger} \Gamma^{10} \right)_{\alpha \beta} \delta_{ab}.   
\eeq

\begin{figure}[htbp]
	\centering
	\includegraphics[width=.32\textwidth,origin=c,angle=270]{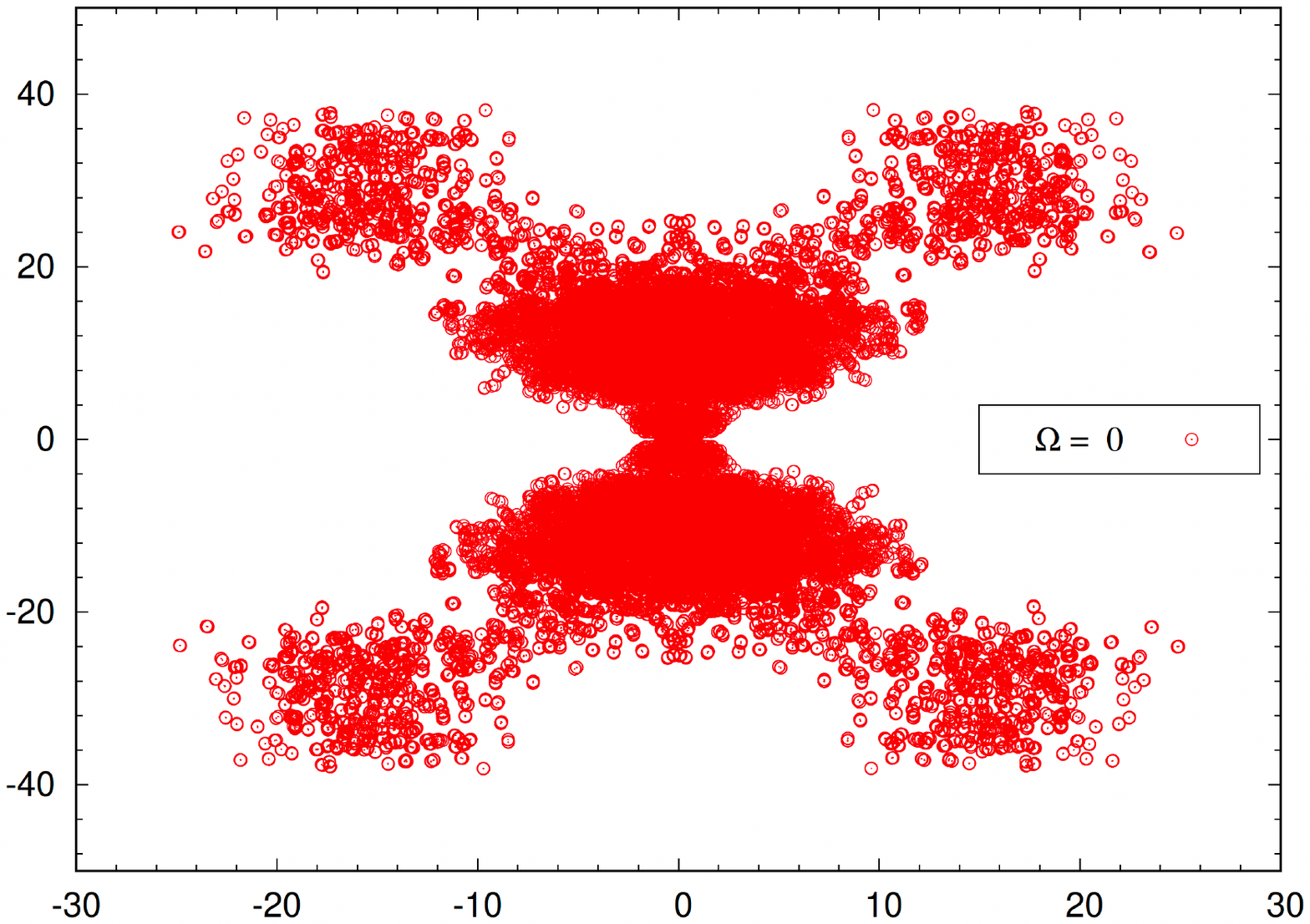}
	\includegraphics[width=.32\textwidth,origin=c,angle=270]{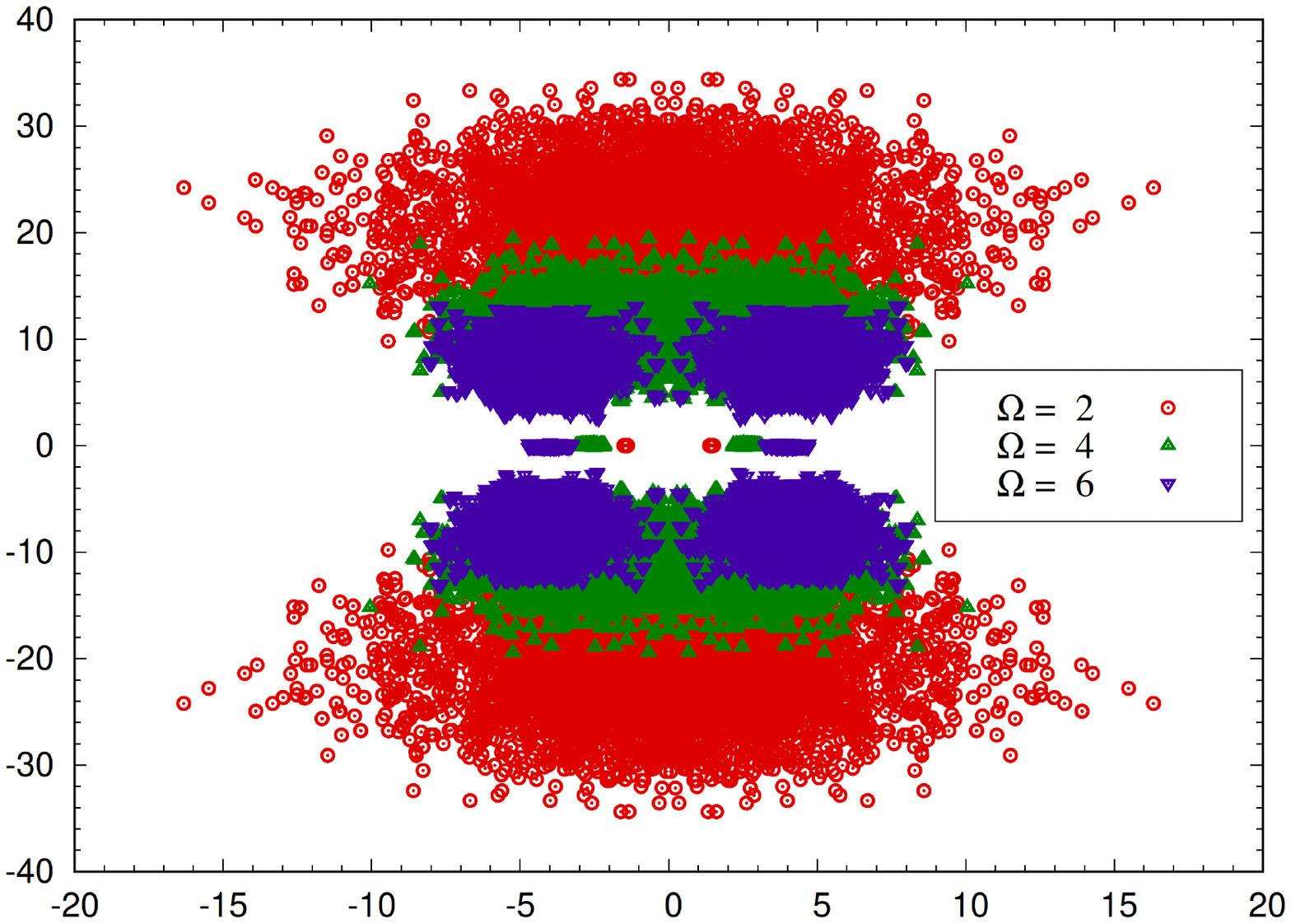}
	\includegraphics[width=.32\textwidth,origin=c,angle=270]{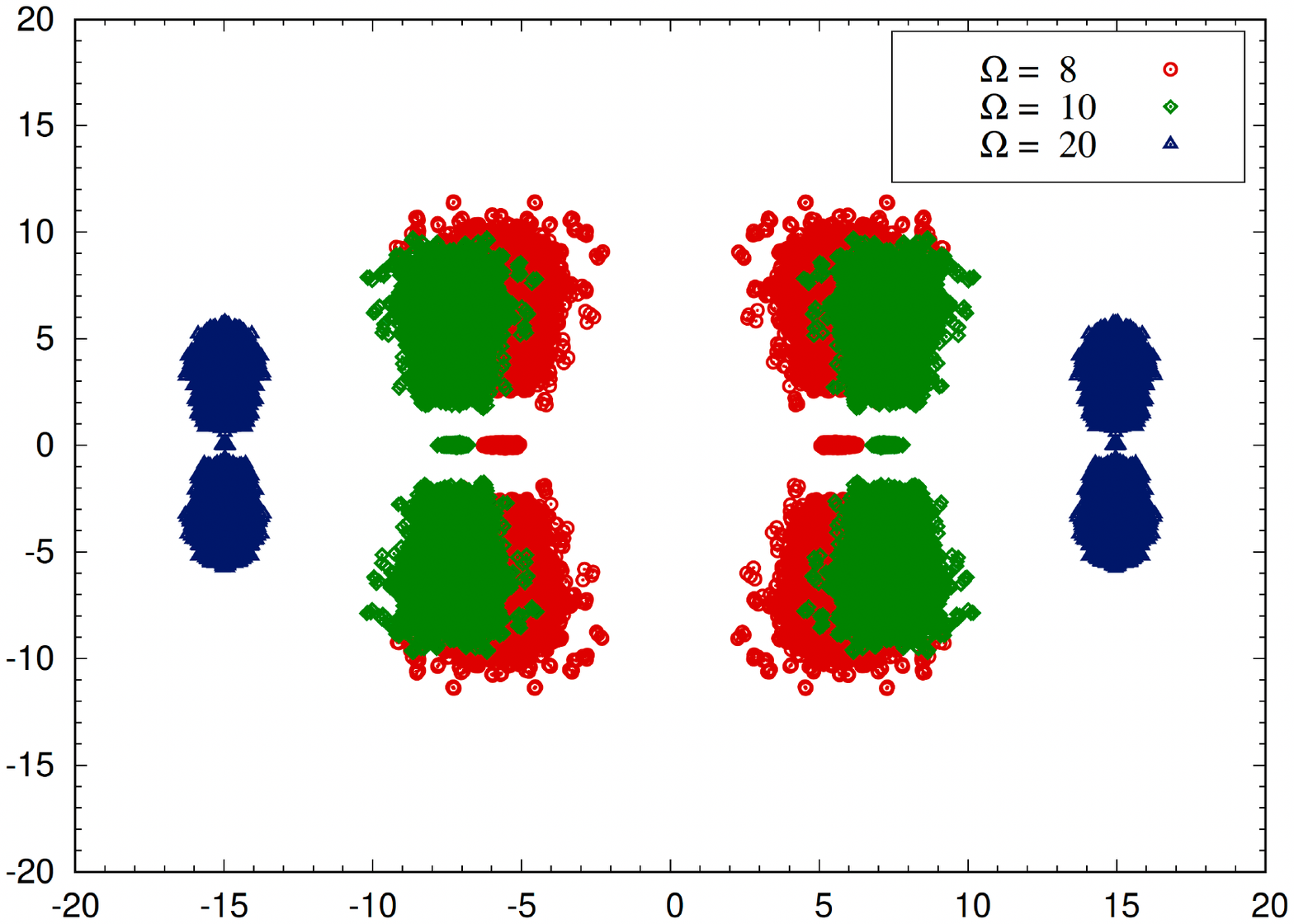}
	\caption{The scatter plot of real versus imaginary part of the eigenvalues of the fermion operator $\mathcal{M}$ for the IKKT matrix model with SUSY preserving mass deformations. The mass deformation parameter $\Omega$ takes various values from 0 to 20. The simulations were performed for $N = 6$.} 
	\label{fig-eigen-nc6.png}
\end{figure}

The eigenvalue distribution of the fermion operator \(\mathcal{M}\) for the SUSY-preserving mass-deformed IKKT model are shown in Fig. \ref{fig-eigen-nc6.png} (taken from Refs. [\refcite{Kumar:2022giw,Kumar:2023nya}]).
For the case where \(\Omega = 0\), corresponding to the original IKKT model, we observe the singular-drift problem. 
However, as \(\Omega\) increases, the eigenvalues of \(\mathcal{M}\) shift further from the origin. 
These observations suggest that the SUSY-preserving mass deformations effectively mitigate the singular-drift problem.

\subsubsection{Bosonic IKKT model with mass deformations}

Introduction of bosonic Gaussian mass deformation terms and a Myers term into the bosonic IKKT matrix model results in the following deformed model action
\beq
S_{\rm b} = S_{\rm bIKKT} + S_{\rm G} + S_{\rm Myers},
\eeq
where
\beq
S_{\rm G} = \frac{\Omega^2 N}{4^3}~ {\rm tr} \left( \sum_{i = 1}^7 X_i^2 + 3 \sum_{a = 8}^{10}  X_a^2 \right),
\eeq
and
\beq
S_{\rm Myers} = \frac{i \Omega N}{3!}~ {\rm tr} \left( \sum_{a, b, c = 8}^{10} X_a \left[X_b, X_c \right] \right).
\eeq

The complex Langevin method is applied for various values of the mass deformation parameter $\Omega$ to investigate whether the ten-dimensional rotational symmetry remains intact in the limit $\Omega \to 0$. 
The order parameter $\lambda_\mu(\Omega)$ shows an inverse dependence on $\Omega$, leading to $\lambda_\mu (\Omega)$ diverging as $\Omega$ approaches zero.

To address this, one can define the normalized extent values as
\beq
\label{eqn:normalized_extent}
\langle \rho_\mu (\Omega) \rangle \equiv \left \langle \frac{\lambda_\mu (\Omega)}{\sum_\mu \lambda_\mu (\Omega)} \right\rangle.
\eeq

The normalized extents $\rho_\mu$ reduce the dependence on the deformation parameter $\Omega$, making it easier to identify whether the SO(10) symmetry is broken. 
If SO(10) symmetry is indeed broken, the normalized extents will differ across directions.

\begin{figure}[htbp]
	\centering
	\includegraphics[width=.7\textwidth,origin=c,angle=0]{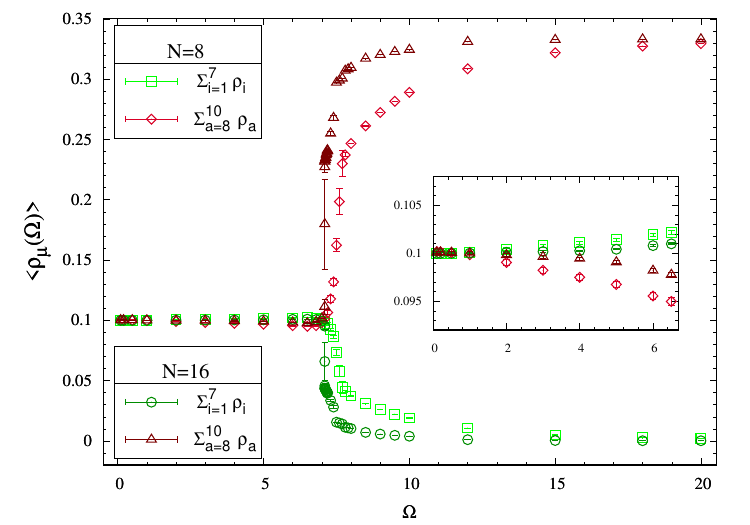}
	\includegraphics[width=.7\textwidth,origin=c,angle=0]{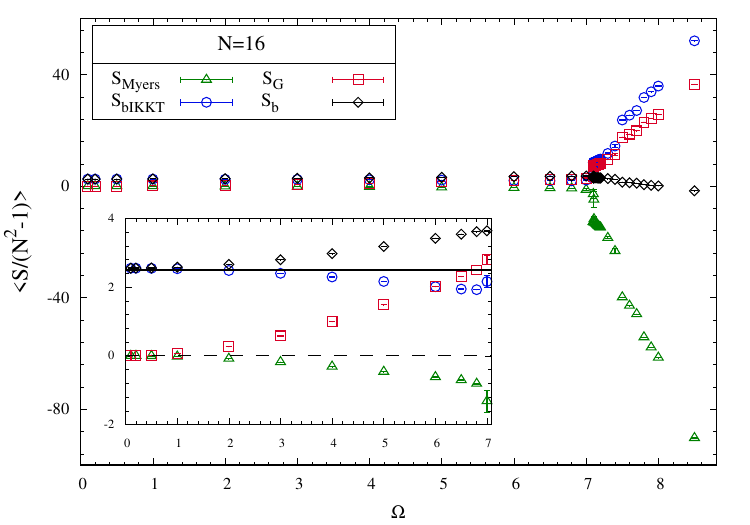}
	\caption{The complex Langevin simulations results for the bosonic IKKT model with mass deformations and a Myers term. On the top panel, the averaged extents, $\frac{1}{7}\sum_{i = 1}^7 \rho_i (\Omega)$ and $\frac{1}{3} \sum_{a = 8}^{10} \rho_a (\Omega)$ versus the mass deformation parameter $\Omega$ are shown for $N = 8, 16$. On the bottom panel, the bosonic action terms against the mass deformation parameter $\Omega$ are shown for $N = 16$.} 
	\label{fig:bos-myers}
\end{figure}

In this model, explicit symmetry breaking from SO(10) to SO(7) $\times$ SO(3) is observed for sufficiently large values of $\Omega$. 
To quantify this, one can use the averaged extents as order parameters:
\beq
\frac{1}{7} \sum_{i = 1}^7 \rho_i (\Omega) \quad \text{and} \quad \frac{1}{3} \sum_{a = 8}^{10} \rho_a (\Omega).
\eeq

The top panel of Fig. \ref{fig:bos-myers} shows the results for these averaged extents. 
The two averaged extents converge as $\Omega \to 0$, indicating that the SO(10) symmetry of the original bosonic IKKT model is restored. 
This suggests that the bosonic mass deformation and the Myers term do not contribute to the spontaneous symmetry breaking of SO(10).

Additionally, there is a phase transition around $\Omega \sim 7.1$ for $N = 16$. 
This transition is likely due to changes in the saddle point configurations introduced by the Myers term. 
The bottom panel of Fig. \ref{fig:bos-myers} illustrates the dominance of the Myers term beyond $\Omega \sim 7.1$. 
As seen from the inset plot, as $\Omega \to 0$, both the contributions from the Gaussian deformation and the Myers term vanish, restoring the bosonic IKKT model.

\subsubsection{IKKT model with mass deformations}

This subsection presents results from complex Langevin simulations of the IKKT model with supersymmetry preserving mass deformations\cite{Kumar:2023nya,Kumar:2022giw}. 
Due to the complex nature of the action, the expectation values of the extent observables become complex. 
To mitigate this issue and maintain the Hermitian nature of the matrices as closely as possible, gauge cooling algorithm (outlined in Sec. \ref{sec:Taming_the_excursion_problem_with_gauge_cooling}) can be employed.

Note that while discussing extents (or normalized extents), we focus exclusively on their real parts, as the imaginary components are negligible.

\begin{figure}[htbp]
	\centering
	\includegraphics[width=.7\textwidth,origin=c,angle=0]{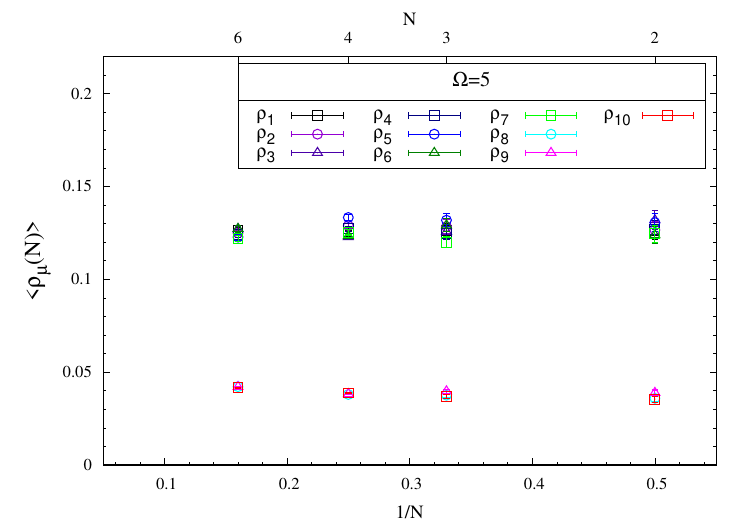}
	\includegraphics[width=.7\textwidth,origin=c,angle=0]{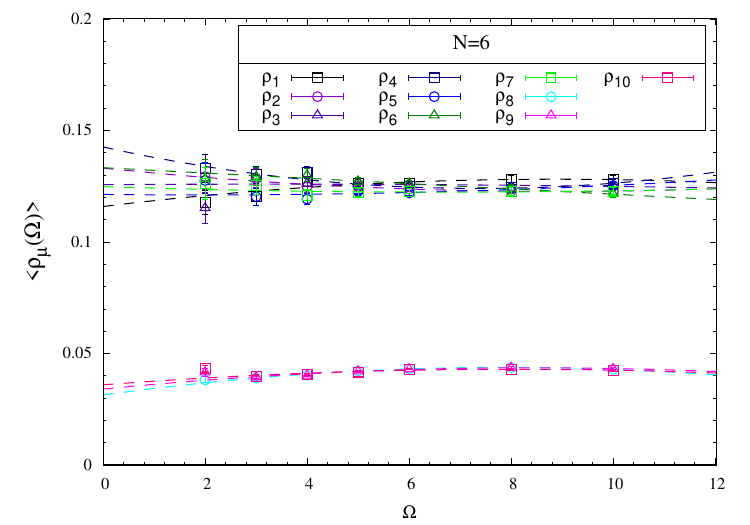}
	\caption{Complex Langevin simulation results for the IKKT matrix model with supersymmetry preserving mass deformations. The top panel shows the normalized extents (order parameter) $\rho_\mu$ against $N$ for fixed $\Omega = 5$. The bottom panel displays the normalized extents $\rho_\mu$ against $\Omega$ for fixed $N = 6$.} 
	\label{fig:rho-N-Omega}
\end{figure}

The top panel of Fig. \ref{fig:rho-N-Omega} shows the normalized extents $\rho_\mu$ for a fixed mass deformation parameter $\Omega = 5$ and various matrix sizes $N$. 
For sufficiently large $\Omega$, explicit SO(7) $\times$ SO(3) symmetry breaking is evident. 
These finite-$N$ results suggest that the extents $\rho_\mu$ are nearly independent of $N$, though larger $N$ computations are necessary for a definitive analysis of the behavior.

Estimating $\mathcal{M}^{-1}$ involves a computational time complexity of $O(N^6)$, which poses a bottleneck for the algorithm. 
In this study, $N = 6$ is used as a proxy for the large-$N$ limit. 
The $\Omega \to 0$ limit is examined on the bottom panel of Fig. \ref{fig:rho-N-Omega}. 
The complex Langevin simulations are unreliable for $\Omega < 2$. 
As $\Omega \to 0$, the original IKKT matrix model is recovered. 
Even for $N = 6$, spontaneous breaking of SO(10) $\to$ SO(7) $\times$ SO(3) symmetry is evident. 
Notably, SO(7) symmetry seems to further break down into smaller subgroups as $\Omega \to 0$, suggesting a SO($d$) symmetric vacuum with $d < 7$. 
To fully understand the symmetric vacuum structure of the IKKT matrix model, large-$N$ extrapolations are required.

\section{Conclusions and future directions}
\label{sec:Conclusions_and_future_directions}

In this review, we have focused on using the complex Langevin method as a tool to investigate systems with complex actions, especially supersymmetric theories in various dimensions. 
Lattice regularization of field theories provides a robust foundation for numerical investigations. 
However, in systems with complex actions, its effectiveness is constrained by the sign problem.
This limitation motivates the exploration of advanced methodologies and complex Langevin method is one of them. 

We have explored several examples that make use of the complex Langevin method. They are supersymmetric theories in zero, one, and two dimensions, including supersymmetric quantum mechanics and the IKKT matrix model.  

After giving a brief overview of lattice regularized path integrals in Sec. 2 and details on the sign problem in complex actions and methods to cure it in Sec. 3, we gave a brief review and current status of the CLM in Sec. 4.
  
In Section 5, we demonstrated the utility of the CLM in addressing the sign problem within the controlled environment of zero-dimensional quantum field theory models. 
These simplified systems provide an ideal testing ground for the CLM.
The section explores various models, including bosonic systems, ${\cal PT}$-symmetric theories, and scenarios involving spontaneous supersymmetry (SUSY) breaking.
Through these examples, the section underscores the robustness of the CLM in overcoming severe sign problems and producing reliable results, validated using criteria like drift term decay and consistency with Fokker-Planck dynamics. 
The absence of spatial and temporal dimensions allows detailed analysis of convergence properties, stability, and numerical performance, establishing confidence in the foundational principles of this method.

While zero-dimensional models simplify the mathematical and computational challenges, they highlight key aspects of CLM’s behavior that inform its application to higher-dimensional systems. 
The findings from these studies validate CLM’s potential to address complex systems with oscillatory integrals and emphasize the importance of stabilization techniques in maintaining accuracy and convergence.

Section 6 highlighted the effectiveness of the CLM in addressing the sign problem and exploring supersymmetry-breaking dynamics in supersymmetric quantum mechanics (SUSY QM). 
By extending the foundational work in zero-dimensional models, this section demonstrated the CLM’s ability to tackle one-dimensional systems where the complexity of interactions and observables increases.
The section emphasizes the method’s utility in simulating lattice-discretized SUSY QM, with key observables like correlation functions and Ward identities, providing critical insights into dynamical SUSY breaking. 
${\cal PT}$-symmetric theories are also explored, offering intriguing results on non-Hermitian dynamics and their interplay with SUSY.

Reliability tests, such as analyzing drift term behavior and verifying consistency with Langevin dynamics, validate the accuracy of CLM in this setting. 
Despite numerical challenges, CLM proves capable of managing oscillatory integrals and stabilizing complexified field trajectories, underscoring its robustness for supersymmetric systems.

In section 7, we demonstrated the capability of the CLM to tackle the challenges of simulating two-dimensional quantum field theories, including scalar field models and supersymmetric systems. 
By expanding the method’s application from simpler zero- and one-dimensional cases, this section highlights its effectiveness in addressing more intricate interactions and dynamics characteristic of higher-dimensional systems.

Key results include the successful application of CLM to scalar field theories with $\phi^4$ terms and ${\cal PT}$-symmetric interactions and the ${\cal N} = 1$ Wess-Zumino model, a cornerstone of supersymmetric field theory. 
These examples showcase CLM’s ability to overcome the sign problem, handle oscillatory integrals, and produce meaningful results in systems with complex actions.

The section also emphasizes the robustness of CLM in managing numerical challenges such as unstable trajectories and divergent field configurations, which become more pronounced in higher dimensions. 
Techniques for stabilizing simulations, combined with reliability checks such as drift term analysis, ensure the accuracy of results and bolster confidence in the method’s applicability.

Section 8 highlighted the application of the CLM to the IKKT matrix model, a key candidate for the non-perturbative formulation of superstring theory. 
The IKKT model presents significant challenges, including a complex fermion determinant and computational difficulties related to large number of degrees of freedom. 
CLM proves to be a promising tool for addressing these challenges, enabling simulations in otherwise inaccessible regimes.

The section emphasizes the innovative techniques employed to stabilize CLM simulations for the IKKT model. 
These include {\it gauge cooling}, which regulates large excursions in field configurations, and mass deformations, which mitigate singular drift terms while preserving supersymmetry. 
These advancements ensure the reliability and stability of simulations, allowing for accurate exploration of the rich structure of this theory.
Applications of CLM to the IKKT model demonstrate its ability to investigate spontaneous SUSY breaking and dynamics related to emergent spacetime. 
These results highlight the method’s potential to provide insights into the non-perturbative behavior of quantum gravity and string theory.

This section underscores the transformative role of CLM in tackling the computationally demanding IKKT model. 
The success of these applications not only advances the understanding of the IKKT model but also establishes CLM as a powerful tool for exploring the frontiers of string theory and non-perturbative quantum field theory. 
This work sets the stage for future studies into even more complex models and broader applications in theoretical physics.

There are several future directions to address.

{\it Improving convergence and stability:} Developing rigorous mathematical frameworks to ensure convergence of the CLM for complex actions remains an important goal. 
This includes addressing issues with boundary terms, non-self-adjoint operators, and unbounded trajectories that can lead to incorrect equilibrium distributions.

{\it Higher-dimensional applications:} Extending CLM to higher-dimensional quantum field theories, particularly in the context of lattice gauge theories and supersymmetric models, offers immense potential. 
Exploring its utility in strongly coupled systems, finite-density QCD, and condensed matter physics could provide groundbreaking insights.

{\it Benchmarking against quantum simulations:} With advances in quantum computing, benchmarking CLM results against quantum simulation techniques could validate its reliability and highlight complementary strengths. 
This collaboration between classical and quantum computational methods may unlock new avenues for tackling the sign problem.

{\it Refining stabilization techniques:} Enhancing existing stabilization methods, such as gauge cooling, adaptive step sizes, and mass deformations, will be crucial for improving the robustness of CLM. 
Tailored approaches for specific models, especially those with severe sign problems, could further expand its applicability.

{\it Exploring non-Hermitian systems:} Investigating CLM’s application to non-Hermitian quantum mechanics and ${\cal PT}$-symmetric systems could open new research frontiers, offering a deeper understanding of exotic phenomena and symmetry-breaking dynamics.

{\it Integrating machine learning:} Employing machine learning techniques to optimize paths in the complexified field space or enhance sampling efficiency could significantly improve the accuracy and scalability of CLM.

{\it Applications in cosmology and quantum gravity:} Applying CLM to theories of quantum gravity, such as the IKKT matrix model, and exploring its implications for early universe cosmology or black hole physics could bridge computational tools with fundamental theoretical questions.

{\it Cross-disciplinary applications:} Extending CLM beyond particle physics to areas like condensed matter systems, financial modeling, and network theory could highlight its versatility in addressing complex systems with oscillatory integrals.

Addressing these directions will advance the theoretical and practical scope of CLM and contribute broadly to fundamental physics, computational physics and interdisciplinary problem-solving.

\section{Acknowledgments}

We extend our gratitude to Ujjwal Basumatary, Gaurav Dadwal, Navdeep Singh Dhindsa, Piyush Kumar, Vamika Longia, Ashutosh Tripathi, and Kunal Verma for their invaluable discussions.
The work of AJ was supported in part by the Start-up Research Grant from the University of the Witwatersrand. The work of AK was  partly supported by the National Natural Science Foundation of China under Grants No. 12293064, No. 12293060, and No. 12325508, as well as the National Key Research and Development Program of China under Contract No. 2022YFA1604900.


\end{document}